\documentclass{aa}

\usepackage[varg]{txfonts}
\usepackage{graphicx}
\usepackage{algorithm}
\usepackage{algorithmic}
\usepackage{color}
\usepackage{placeins}
\usepackage{tabularx}
\usepackage{makecell}

\usepackage{CJK}

\usepackage{mathtools}
\newcommand{\noarrow}{\relbar\mkern-5.5mu\relbar}

\usepackage{natbib,twoopt}
\usepackage[breaklinks=true]{hyperref}
\bibpunct{(}{)}{;}{a}{}{,}

\authorrunning{Zhang et al.}
\titlerunning{Interpreting deep learning-based stellar mass estimation}

\begin{document}

\begin{CJK*}{UTF8}{gbsn}

\title{Interpreting deep learning-based stellar mass estimation via causal analysis and mutual information decomposition}

\author{Wei Zhang (张伟)\inst{\ref{inst1}, \ref{inst2}}
\and Qiufan Lin (林秋帆)\inst{\ref{inst1}}\thanks{Corresponding author; linqf@pcl.ac.cn}
\and Yuan-Sen Ting (丁源森)\inst{\ref{inst3}, \ref{inst4}}
\and Shupei Chen (陈树沛)\inst{\ref{inst1}}
\and Hengxin Ruan (阮恒心)\inst{\ref{inst1}}
\and Song Li (李松)\inst{\ref{inst1}}
\and Yifan Wang (王一凡)\inst{\ref{inst1}}}

\institute{Pengcheng Laboratory, Nanshan District, Shenzhen, Guangdong 518000, P.R. China\label{inst1}
\and Harbin Institute of Technology, Nanshan District, Shenzhen, Guangdong 518000, P.R. China\label{inst2}
\and Department of Astronomy, The Ohio State University, Columbus, OH 43210, USA\label{inst3}
\and Center for Cosmology and AstroParticle Physics (CCAPP), The Ohio State University, Columbus, OH 43210, USA\label{inst4}
}

\date{Received; accepted}

\abstract{End-to-end deep learning models fed with multi-band galaxy images are powerful data-driven tools used to estimate galaxy physical properties in the absence of spectroscopy. However, due to a lack of interpretability and the associational nature of such models, it is difficult to understand how the information that is included in addition to integrated photometry (e.g., morphology) contributes to the estimation task. Improving our understanding in this field would enable further advances into unraveling the physical connections among galaxy properties and optimizing data exploitation. Therefore, our work is aimed at interpreting the deep learning-based estimation of stellar mass via two interpretability techniques: causal analysis and mutual information decomposition. The former reveals the causal paths between multiple variables beyond nondirectional statistical associations, while the latter quantifies the multicomponent contributions (i.e., redundant, unique, and synergistic) of different input data to the stellar mass estimation. We leveraged data from the Sloan Digital Sky Survey (SDSS) and the Wide-field Infrared Survey Explorer (WISE). With the causal analysis, meaningful causal structures were found between stellar mass, photometry, redshift, and various intra- and cross-band morphological features. The causal relations between stellar mass and morphological features not covered by photometry indicate contributions coming from images that are complementary to the photometry. With respect to the mutual information decomposition, we found that the total information provided by the SDSS optical images is effectively more than what can be obtained via a simple concatenation of photometry and morphology, since having the images separated into these two parts would dilute the intrinsic synergistic information. A considerable degree of synergy also exists between the $g$ band and other bands. In addition, the use of the SDSS optical images may essentially obviate the incremental contribution of the WISE infrared photometry, even if infrared information is not fully covered by the optical bands available. Taken altogether, these results provide physical interpretations for image-based models. Our work demonstrates the gains from combining deep learning with interpretability techniques, and holds promise in promoting more data-driven astrophysical research (e.g., astrophysical parameter estimations and investigations on complex multivariate physical processes).
}

\keywords{methods: data analysis -- methods: statistical -- techniques: image processing -- surveys -- galaxies: evolution}

\maketitle

\section{Introduction} \label{sec:intro}

The observed physical properties of galaxies, such as the stellar mass ($M_{\ast}$), star formation rate (SFR), and metallicity, offer valuable insights into galaxy formation and evolution. Next-generation imaging surveys, including the \textit{Euclid} survey \citep{Laureijs2011}, \textit{Nancy Grace Roman} Space Telescope \citep{Spergel2015}, \textit{Vera C. Rubin} Observatory Legacy Survey of Space and Time \citep[LSST;][]{Ivezic2019}, and China Space Station Telescope \citep[CSST;][]{Zhan2018}, will produce an unprecedented amount of astronomical data and bring new opportunities to revolutionize our understanding of the evolution of galaxies and the Universe. Estimating the physical properties of galaxies is one of the crucial areas comprised by the science goals of the next-generation surveys.

The estimation of galaxy physical properties leverages the notion that the observed spectral or photometric features of a galaxy are tied to its physical processes and can therefore be used to infer its physical properties. While spectroscopy is a powerful way to estimate physical properties, obtaining spectroscopic follow-up measurements for individual galaxies is too expensive to match the rapid growth of imaging data envisioned by future surveys. Regarding non-spectroscopy approaches, traditional methods typically rely on the modeling of spectral energy distributions (SEDs) and fit with observed integrated photometry. Such SED fittings provide strong physical constraints, but lack flexibility and may be unreliable for new galaxy types. Furthermore, these methods usually require a huge amount of computing time, making them impractical for processing large datasets.

Data-driven methods, empowered by the fast development of machine learning, have emerged as promising approaches to large-scale photometric data analysis. A flexible machine learning model, in particular, a deep learning neural network with good predictive power, can automatically learn complex nonlinear dependence relations from data and establish a mapping between the input data and target physical properties. Compared with traditional SED fitting methods, data-driven methods reduce the reliance on physical priors, offer the potential to make more accurate predictions, and greatly improve computational efficiency. These advantages have been demonstrated by a few studies that adopted different machine learning methods, such as self-organizing maps \citep{Davidzon2022, LaTorre2024}, random forests \citep{Acquaviva2016, Bonjean2019, DelliVeneri2019, Mucesh2021}, CatBoost \citep{Zeraatgari2024}, deep learning neural networks \citep{DelliVeneri2019, Dobbels2019, WuBoada2019, Surana2020, BuckWolf2021, Euclid_Bisigello2023, Chu2024, Zeraatgari2024, Zhong2024}, and other machine learning methods \citep{Acquaviva2016}.

In particular, some works \citep[e.g.,][]{WuBoada2019, BuckWolf2021, Euclid_Bisigello2023, Zhong2024} have directly utilized multi-band galaxy images labeled with physical properties or combined galaxy images with integrated photometry to train end-to-end deep learning models. These image-based models tend to perform better in terms of their estimation accuracy than other machine learning methods that only use integrated photometry as their input data. This is because galaxy images contain more information on physical properties than integrated photometry does, and such information can be automatically extracted by a deep learning model. In a word, image-based deep learning models have shown great promise in handling future large-scale data with high efficiency and accuracy.

Despite the merits, the ``black box'' nature of deep learning makes it hard to interpret how a trained end-to-end model reaches a certain prediction given the input data, especially for image-based models that exhibit a high level of complexity. The lack of model interpretability would not only hinder our understanding or discovery of the underlying mechanisms in the determination of physical properties, but also pose difficulties in developing better models to optimize data exploitation in data-driven applications. In principle, using a simple model (e.g., a linear model) could help regain the model transparency, but this would compromise the model expressivity and the predictive power; thus, it would limit the exploitation of information from large-scale data and lead to disfavored results. Therefore, unraveling the black box while retaining the predictive power is a challenging but crucial task.

Interpreting image-based models amounts to understanding how the variables or features encoded in galaxy images in addition to integrated photometry (e.g., morphology) contribute to the estimation of physical properties. Similar insights can also be gained by investigating how the information on redshift ($z$) is encoded in galaxy images, since physical properties and redshift both affect observed photometry and they are largely degenerate given a certain set of photometric data. There have been a number of studies investigating the impact of galaxy morphology on the estimation of physical properties or photometric redshift (photo-$z$); however, no strict consensus has been reached. 

For example, early works such as those of \citet{Way2009}, \citet{Singal2011}, and \citet{JonesSingal2017} have suggested that combining morphological parameters with photometry does not offer a statistically significant improvement on the photo-$z$ estimation, probably due to the influence of morphology-introduced noise. \citet{Chu2024} included morphological parameters such as the S\'{e}rsic index and the inclination when analyzing the impacts of input features on the estimation of stellar, dark matter, and total masses, as well as the stellar mass-to-light ratio ($M_{\ast}/L$), while the impacts of these morphological parameters are overwhelmed by other parameters such as galaxy luminosity. On the contrary, \citet{Yip2011} found a trend between the inclination and the overestimation of photo-$z$, which was re-investigated by \citet{Pasquet2019}. \citet{Soo2018} suggested that galaxy morphology would improve the photo-$z$ estimation, especially when photometric bands are insufficient. This conclusion is similar to the one drawn by \citet{Dobbels2019} on the estimation of $M_{\ast}/L$. In addition, \citet{WuBoada2019}, \citet{BuckWolf2021}, and \citet{Zhong2024} have found that downgrading the image resolution would have a negative impact on the estimation of physical properties, implying that extra information is conveyed by morphology additional to photometry. \citet{Euclid_Bisigello2023} found that the inclusion of simulated \textit{Euclid} $H_E$-band images may be helpful for the estimation of stellar mass, although it is not very useful for photo-$z$ or SFR.

Furthermore, it is still not conclusive how much each photometric band may contribute to the estimation of physical properties or photo-$z$, and whether the contributions of different bands are overlapping, unique, or in synergy. \citet{WuBoada2019}, \citet{BuckWolf2021}, and \citet{Zhong2024} have found that dropping out certain photometric bands would degrade the estimation results, compared to using all available bands; whereas the question of how different bands interact in terms of their contributions to the target prediction warrants further investigation.

Several studies, such as those of \citet{Hoyle2015}, \citet{Acquaviva2016}, \citet{Bonjean2019}, \citet{DelliVeneri2019}, \citet{Lu2024}, and \citet{Zeraatgari2024} have analyzed the contributions of input features from different photometric bands using ``feature importance'' produced by machine learning models. However, as noted by \citet{Euclid_Humphrey2023} and \citet{Lu2024}, feature importance can only display the ``net'' contributions that may sometimes be misleading, because the importance shared between co-linear or dependent features would be absorbed by the dominant ones, leaving the impression that the remaining features are less important (or even not at all). In other words, it is difficult to tell apart the dependence or independence relations between the input features using feature importance.

More importantly, the shortcomings of feature importance point to the limitations of the associational nature of most machine learning methods, which is tightly connected to the interpretability issue. Although machine learning is capable of capturing complex and even unknown dependences between variables, in most cases, such predictive dependences are simply nondirectional statistical associations, rather than causal relations. They only show the overall ``appearance" of the entangled mechanisms behind a machine learning model, but cannot reveal causal structures that indicate the way correlated variables influence each other and the direction of influence between variables. Thus, they do not allow us to differentiate among disparate causal mechanisms that produce the same associational behaviors. Furthermore, overall associations between input and target variables do not elaborate on how the multivariate information on the target is distributed among different input variables, making it hard to uncover their mutual relations and their individual or synergistic contributions to the target prediction. Therefore, any further advancement of research into interpreting deep learning models requires statistical and analytical methods that break through the barriers of nondirectional associations and the entanglement of multivariate information.

In this work, we have sought to interpret the deep learning-based estimation of physical properties by resorting to the principles of causal learning and information theory, exploring their potentials for gaining interpretability beyond pure statistical associations. As a case study, we analyzed the estimation of stellar mass for galaxies with redshift up to $z \sim 0.33$, using multi-band optical images, photometry, morphology, spectroscopic redshift (spec-$z$), and other catalog data from the Sloan Digital Sky Survey \citep[SDSS;][]{York2000} and infrared photometry from the Wide-field Infrared Survey Explorer \citep[WISE;][]{Wright2010}. For the purposes of a causal analysis beyond nondirectional associations, we applied a framework with supervised contrastive learning and $k$-nearest neighbors (KNNs) to investigate the causal paths between stellar mass, photometry, spec-$z$, and various intra- and cross-band morphological features. In particular, the causal relations between stellar mass and morphological features cannot be entirely accounted for by integrated photometry; thus, they suggest certain implications regarding the extra contributions of multi-band images (in addition to photometry) to the stellar mass estimation that can be captured by image-based models. For the information-theoretical analysis, we used deep learning neural networks to estimate the mutual information between stellar mass and various sets of input data, then decomposed the mutual information into redundant, unique, and synergistic components using the method from \citet{WilliamsBeer2010}. This offers a quantification of the multicomponent contributions of different input sets to the stellar mass estimation, showing how the information on stellar mass is distributed across the input data. We analyzed the decomposed contributions of different data modalities, involving photometry, morphology, images, and spec-$z$, as well as separate photometric bands.

In a nutshell, the causal analysis tells how multiple variables are connected and form causal structures, while the mutual information decomposition quantifies the ``interactions'' between variables that are not described by the causal structures (e.g., synergy). Both techniques are performed in a data-driven manner and provide complementary views on the data structures and mechanisms behind the stellar mass-predicting process. Thus, they help draw up physical interpretations of image-based stellar mass estimation models. This work underscores the benefits of combining the predictive power of deep learning and the interpretability of causal and information-theoretical analysis techniques, with a great deal of promise for optimizing data exploitation and promoting more data-driven scientific research.

In addition, we note that there have been many visual explanation methods in computer vision that can be used to understand image-based deep learning models, such as saliency maps \citep{Simonyan2013}, local interpretable model-agnostic explanations \citep[LIME;][]{Ribeiro2016}, class activation map \citep[CAM;][]{Zhou2016}, deep Taylor decomposition \citep{Montavon2017}, deep learning important features \cite[DeepLIFT;][]{Shrikumar2017}, and deep dream\footnote{\url{https://research.google/blog/inceptionism-going-deeper-into-neural-networks}}. We did not adopt these methods, since they would mainly highlight pixel-level features relevant to the prediction task but lack physical meanings at the object level. For example, the highlighted pixels enclosing a galaxy would not reveal the curvature of its light profile as a predictive feature. We also did not use association-based interpretability tools, such as feature importance and Shapley additive explanations \citep[SHAP;][]{LundbergLee2017}, because they do not serve our purpose of uncovering causal structures between galaxy properties.

This paper organized is as follows. Section~\ref{sec:data} describes the data used in this work. Section~\ref{sec:methods} presents our causal analysis and mutual information decomposition methods. Sections~\ref{sec:results_causal} and \ref{sec:results_info} present the results. A comparison between our work and other studies is given in Sect.~\ref{sec:comparison}. Section~\ref{sec:discussions} discusses the limitations of this work and offers suggestions for future work. Finally, we summarize our findings and give concluding remarks in Sect.~\ref{sec:conclusion}. The details of training and validating the models used for our causal analysis and mutual information estimation are presented in Appendix~\ref{sec:training_dev}. A discussion on KNN is presented in Appendix~\ref{sec:assess_knn}. More results on our causal analysis and mutual information decomposition are provided in Appendix~\ref{sec:more_res}.

\section{Data} \label{sec:data}

\begin{figure*}
\begin{center}
\centerline{\includegraphics[width=1.0\linewidth]{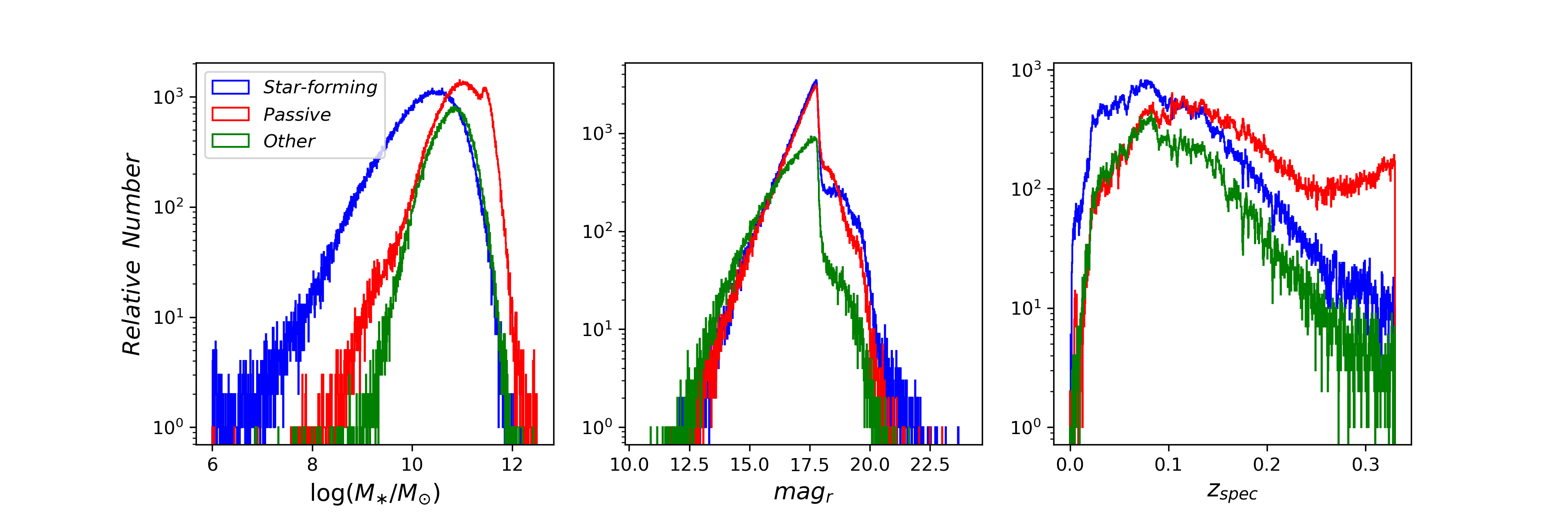}}
\caption{Distributions of stellar mass, $r$-band magnitude and spec-$z$ for the SDSS data used in our work, shown for star-forming, passive, and other galaxies.}
\label{fig:logm_r_z_dist}
\end{center}
\end{figure*}

\begin{table*}
\caption{List of parameters used for interpreting the stellar mass estimation.} \label{tab:data}
\centering
\begin{tabular}{l | l}
\hline
Parameter  &  Description \\
\hline
$mag_u$, $mag_g$, $mag_r$, $mag_i$, $mag_z$, $mag_{W1}$, $mag_{W2}$, $mag_{W3}$  &  Optical and infrared magnitudes \\
$mag_u$ \textit{error}, $mag_g$ \textit{error}, $mag_r$ \textit{error}, $mag_i$ \textit{error}, $mag_z$ \textit{error}  &  Optical magnitude errors \\
$mag_{W1}$ \textit{error}, $mag_{W2}$ \textit{error}  &  Infrared magnitude errors (excluding the $W3$ band) \\
$u-g$, $g-r$, $r-i$, $i-z$, $z-W1$, $W1-W2$, $W2-W3$  &  Optical and infrared colors \\
$n_u$, $n_g$, $n_r$, $n_i$, $n_z$  &  Optical-band S\'{e}rsic indices \\
$n_u / n_g$, $n_g / n_r$, $n_r / n_i$, $n_i / n_z$  &  Cross-band ratios of S\'{e}rsic indices \\
$[b/a]_u$, $[b/a]_g$, $[b/a]_r$, $[b/a]_i$, $[b/a]_z$  &  Optical-band inclinations based on exponential fits \\
$[b/a]_u / [b/a]_g$, $[b/a]_g / [b/a]_r$, $[b/a]_r / [b/a]_i$, $[b/a]_i / [b/a]_z$  &  Cross-band ratios of inclinations \\
$R_{50,u}$, $R_{50,g}$, $R_{50,r}$, $R_{50,i}$, $R_{50,z}$  &  Optical-band Petrosian radii containing 50\% of Petrosian fluxes \\
$R_{90,u}$, $R_{90,g}$, $R_{90,r}$, $R_{90,i}$, $R_{90,z}$  &  Optical-band Petrosian radii containing 90\% of Petrosian fluxes \\
$R_{90,u} / R_{50,u}$, $R_{90,g} / R_{50,g}$, $R_{90,r} / R_{50,r}$, $R_{90,i} / R_{50,i}$, $R_{90,z} / R_{50,z}$  &  Intra-band ratios of Petrosian radii \\
$R_{90,u} - R_{50,u}$, $R_{90,g} - R_{50,g}$, $R_{90,r} - R_{50,r}$, $R_{90,i} - R_{50,i}$, $R_{90,z} - R_{50,z}$  &  Intra-band differences of Petrosian radii \\
$R_{50,u} / R_{50,g}$, $R_{50,g} / R_{50,r}$, $R_{50,r} / R_{50,i}$, $R_{50,i} / R_{50,z}$  &  Cross-band ratios of Petrosian radii \\
$R_{90,u} / R_{90,g}$, $R_{90,g} / R_{90,r}$, $R_{90,r} / R_{90,i}$, $R_{90,i} / R_{90,z}$  &  \\
$R_{90,u} / R_{50,g}$, $R_{90,g} / R_{50,r}$, $R_{90,r} / R_{50,i}$, $R_{90,i} / R_{50,z}$  &  \\
$z_{spec}$  &  Spectroscopic redshift (spec-$z$) \\
$\log$(\textit{SFR})  &  Star formation rate\tablefootmark{*} \\
$M_{\ast}/L_u$, $M_{\ast}/L_g$, $M_{\ast}/L_r$, $M_{\ast}/L_i$, $M_{\ast}/L_z$  &  Optical-band mass-to-light ratios\tablefootmark{**} \\
$\log$(\textit{sSFR})  &  Specific star formation rate\tablefootmark{**} \\
\textit{Metallicity}  &  Metallicity\tablefootmark{**} \\
\textit{Age}  &  Stellar age\tablefootmark{**} \\
\textit{Dust} $\tau_1$  &  Optical depth of dust attenuation around young stars\tablefootmark{**} \\
$T_{look-back}$  &  Look-back formation time\tablefootmark{**} \\
\textit{PSF}$_u$, \textit{PSF}$_g$, \textit{PSF}$_r$, \textit{PSF}$_i$, \textit{PSF}$_z$  &  FWHMs of optical-band point spread functions \\
$E(B-V)$  &  Galactic reddening\tablefootmark{***} \\
\hline
\end{tabular}
\tablefoot{
\tablefoottext{*}{Following the method from \citet{Brinchmann2004} with aperture corrections as described in \citet{Salim2007}.}
\tablefoottext{**}{Following the method from \citet{Conroy2009}.}
\tablefoottext{***}{Based on the dust map from \citet{Schlegel1998}.}
}
\end{table*}

This work makes use of a sample of galaxies from the Sloan Digital Sky Survey \citep[SDSS;][]{York2000} and the Wide-field Infrared Survey Explorer \citep[WISE;][]{Wright2010}. In particular, we mainly used the SDSS optical images, the SDSS optical photometry and the WISE infrared photometry as input data for the estimation of stellar mass. The SDSS and WISE photometry, morphological parameters in optical bands, spec-$z$, physical properties, and other data were used as features or variables for our causal analysis or mutual information decomposition.

The SDSS catalog data were selected from the SDSS Data Release 12 \citep{Alam2015}, available on the SDSS CasJobs website\footnote{\url{https://skyserver.sdss.org/CasJobs}}. Since we were interested in how spec-$z$ acts in the stellar mass estimation, we collected galaxies that have spectroscopically measured redshifts, flagged as \texttt{z}. We retrieved the SDSS photometric data composed of Petrosian magnitudes, fluxes, and magnitude errors of each galaxy in the five optical bands $u,g,r,i,z$, flagged as \texttt{petroMag\_?}, \texttt{petroFlux\_?}, and \texttt{petroMagErr\_?}, respectively, where \texttt{?} stands for a band from $u,g,r,i,z$. The morphological parameters we retrieved include five-band inclinations $b/a$ based on exponential fits and Petrosian radii containing 50\% or 90\% of Petrosian fluxes, flagged as \texttt{expAB\_?}, \texttt{petroR50\_?}, and \texttt{petroR90\_?}, respectively. We also took S\'{e}rsic indices $n$ in the five bands from New York University Value-Added Galaxy Catalog \citep[NYU-VAGC;][]{Blanton2005}\footnote{\url{http://sdss.physics.nyu.edu/vagc/sersic.html}}.

For physical properties, we retrieved stellar mass and SFR estimates from the MPA-JHU DR8 catalog. The stellar mass estimates are expressed as $\log(M_{\ast} / M_{\odot})$, where $M_{\odot}$ refers to the solar mass. They were estimated via stellar population synthesis (SPS) modeling, using the method presented by \citet{Kauffmann2003}. For each galaxy, we took the median estimate from the probability density function (PDF) of log total stellar mass, flagged as \texttt{lgm\_tot\_p50}. The SFR estimates were obtained using emission line measurements within the SDSS 3 arcsec-diameter fiber aperture when available \citep[described in][]{Brinchmann2004}.\ Aperture corrections were performed to estimate the SFR outside the fiber \citep[described in][]{Salim2007}. The total SFR is the sum of SFR within and outside the fiber. For those galaxies with weak emission lines within the fiber, the SFR was instead estimated by fitting integrated photometry. For each galaxy, we took the median estimate from the PDF of log total SFR, flagged as \texttt{sfr\_tot\_p50}. Furthermore, we retrieved other physical properties estimated using the flexible SPS code described by \citet{Conroy2009}. These properties comprise the mean estimates (from the PDFs) of specific star formation rate (sSFR), metallicity, stellar age, optical depth for dust attenuation around young stellar populations, look-back formation time, and five-band mass-to-light ratios $M_{\ast}/L$, flagged as \texttt{ssfr\_mean}, \texttt{metallicity\_mean}, \texttt{age\_mean}, \texttt{dust1\_mean}, \texttt{t\_age\_mean}, and \texttt{m2l\_?}, respectively, where \texttt{?} stands for a band from $u,g,r,i,z$.

In addition, for each galaxy, we retrieved the full widths at half maximum (FWHMs) of five-band point spread functions (PSFs), flagged as \texttt{psffwhm\_?}, and obtained galactic reddening $E(B-V)$ along the line of sight based on the dust map from \citet{Schlegel1998}.

From the WISE All-Sky Data Release catalog, we retrieved magnitudes in infrared bands $W1$, $W2$, and $W3$, flagged as \texttt{w?mpro}, where \texttt{?} represents 1, 2, or 3. These magnitudes were measured with profile-fit photometry if the flux measurement has S/N > 2, or replaced with the 95\% confidence brightness upper limits if S/N < 2. Magnitude errors in the $W1$ and $W2$ bands were also retrieved, flagged as \texttt{w?sigmpro}, where \texttt{?} represents 1 or 2. We did not take $W3$-band magnitude errors due to a considerable portion of null values that correspond to the magnitudes of brightness upper limits. Similar to \citet{Bonjean2019}, we did not use photometry in the $W4$ band due to its poorer quality.

We selected galaxies using the criteria \texttt{survey=sdss}, \texttt{targetType=SCIENCE}, and \texttt{class=GALAXY}. We set \texttt{insideMask=0}, \texttt{clean=1}, \texttt{z>0}, and \texttt{zWarning=0} to retain galaxies that are not in a mask, have clean optical photometry, and have positive and reliable spec-$z$. We also filtered out those with null values in stellar mass and infrared photometry, by \texttt{lgm\_tot\_p50$\not\eq$-9999}, \texttt{w?mpro$\not\eq$9999} where \texttt{?} represents 1, 2, or 3, and \texttt{w?sigmpro$\not\eq$9999} where \texttt{?} represents 1 or 2. Since no null value remains in the $W1$-band and $W2$-band magnitude errors, all the magnitudes in these two bands are profile-fit magnitudes, while the $W3$ band also contains the magnitudes of brightness upper limits. These selections resulted in a total of 581\,805 galaxies with redshift up to $z \sim 0.33$ and stellar mass in a range between 6.0 and 12.5 dex. We randomly selected 431\,805 galaxies as a training sample, 50\,000 galaxies as a validation sample, and 100\,000 galaxies as a test sample. The training sample was used to train the neural networks for the causal analysis and the mutual information estimation. The validation sample was used to monitor all the training processes. The test sample was used to present all our results.

To better illustrate the behaviors of different galaxy populations, we considered a division of galaxy populations based on the Baldwin-Phillips-Terlevich (BPT) diagram \citep{Baldwin1981} as described in \citet{Brinchmann2004}. In specific, \texttt{bptclass=-1} corresponds to unclassifiable galaxies that have no or very weak emission lines; \texttt{bptclass=1} corresponds to star-forming galaxies; \texttt{bptclass=2} corresponds to low S/N star-forming galaxies; \texttt{bptclass=3} corresponds to composite galaxies; \texttt{bptclass=4} corresponds to active galactic nuclei (AGNs), excluding low ionization nuclear emission line regions (LINERs); \texttt{bptclass=5} corresponds to low S/N LINERs. In our analysis, the galaxies with \texttt{bptclass=-1} were referred to as ``passive'' galaxies; those with \texttt{bptclass=1} and \texttt{bptclass=2} were referred to as ``star-forming'' galaxies; all the remaining galaxies were set together and referred to as ``other'' galaxies. The distributions of stellar mass, $r$-band magnitude and spec-$z$ for these three galaxy populations are shown in Fig.~\ref{fig:logm_r_z_dist}.

We took stamp galaxy images processed by \citet{Pasquet2019}. Each stamp image is a cutout from one of the five optical bands and encompasses $64\times64$ pixels in spatial dimensions with a galaxy at the center. The pixel scale is 0.396 arcsec. The five stamp images of a galaxy in the $u,g,r,i,z$ bands in sequence constitute a data instance with $64\times64\times5$ dimensions. To reduce the contrast between the peak flux intensities of different galaxies, all the images were rescaled using the formula
\begin{equation}
I = \begin{cases}
-\log_e (-I_0 + 1.0), \quad I_0 < 0,\\
\log_e (I_0 + 1.0), \quad I_0 > 0,
\end{cases}
\label{eq:rescaling}
\end{equation}
where $I$ and $I_0$ refer to the rescaled and the original pixel intensities, respectively.

The aforementioned photometric, morphological, physical, and other parameters, as well as multiple intra- and cross-band combinations of the original morphological features, constitute a long list of parameters (summarized in Table~\ref{tab:data}), a non-exhaustive but rich search base we used to interpret the stellar mass estimation. We note that all the parameters including stellar mass are measured or derived quantities rather than the ground-truth values, thus there would be measurements biases and errors. Furthermore, any deep learning models in principle cannot reach beyond the accuracy set by the training sample. Therefore, the aim of this work is not to predict the ground-truth stellar mass values. Rather, we regarded the stellar mass estimates as reference values and investigated how and to what extent images and other measured features can provide information to reproduce the stellar mass measurements, bearing in mind that the reference values may be biased or erroneous.

\section{Methods} \label{sec:methods}

\subsection{Overview} \label{sec:overview}

This section presents our methods for interpreting deep learning-based stellar mass estimation models, including the causal analysis and the mutual information decomposition. 
For the causal analysis, we established a causal graph to depict the stellar mass-predicting process of end-to-end deep learning models, based on a framework with supervised contrastive learning and KNN. This framework projects potentially complex input data to low-dimensional latent vectors that encode the information on stellar mass, to which KNN can be applied. This offers a highly efficient avenue to find out whether any given variable is external, namely, containing the information on stellar mass but missing in the input data of a deep learning model. In other words, using the known stellar mass, photometric, morphological, physical, and other parameters of the nearest neighbors of each test galaxy in the latent space, we can check the local (conditional) independence between stellar mass and all the other parameters. Any parameter that shows statistically significant association with stellar mass is deemed as an external variable.

Focusing on a photometry-only model that is only fed with integrated optical photometry, we investigated the causal structures between its external variables and stellar mass. In particular, morphological features are the external variables for this photometry-only model. Since all morphological features are measured on the basis of images and can be captured by image-based models, interpretations for image-based models can be made based on the findings regarding how the unaccounted-for morphological parameters for this photometry-only model are causally linked to stellar mass.

We then estimated the mutual information between stellar mass and different sets of input data using deep learning models, and applied the mutual information decomposition to quantify the redundant, unique, and synergistic information components. This, complementary to the causal analysis, describes how the information on stellar mass is distributed across different data modalities and photometric bands, revealing their multicomponent contributions to the stellar mass estimation from the information-theoretical perspective.
The details of the causal analysis and the mutual information decomposition are elaborated in the following two subsections, respectively.

\begin{figure}
\begin{center}
\centerline{\includegraphics[width=1.0\linewidth]{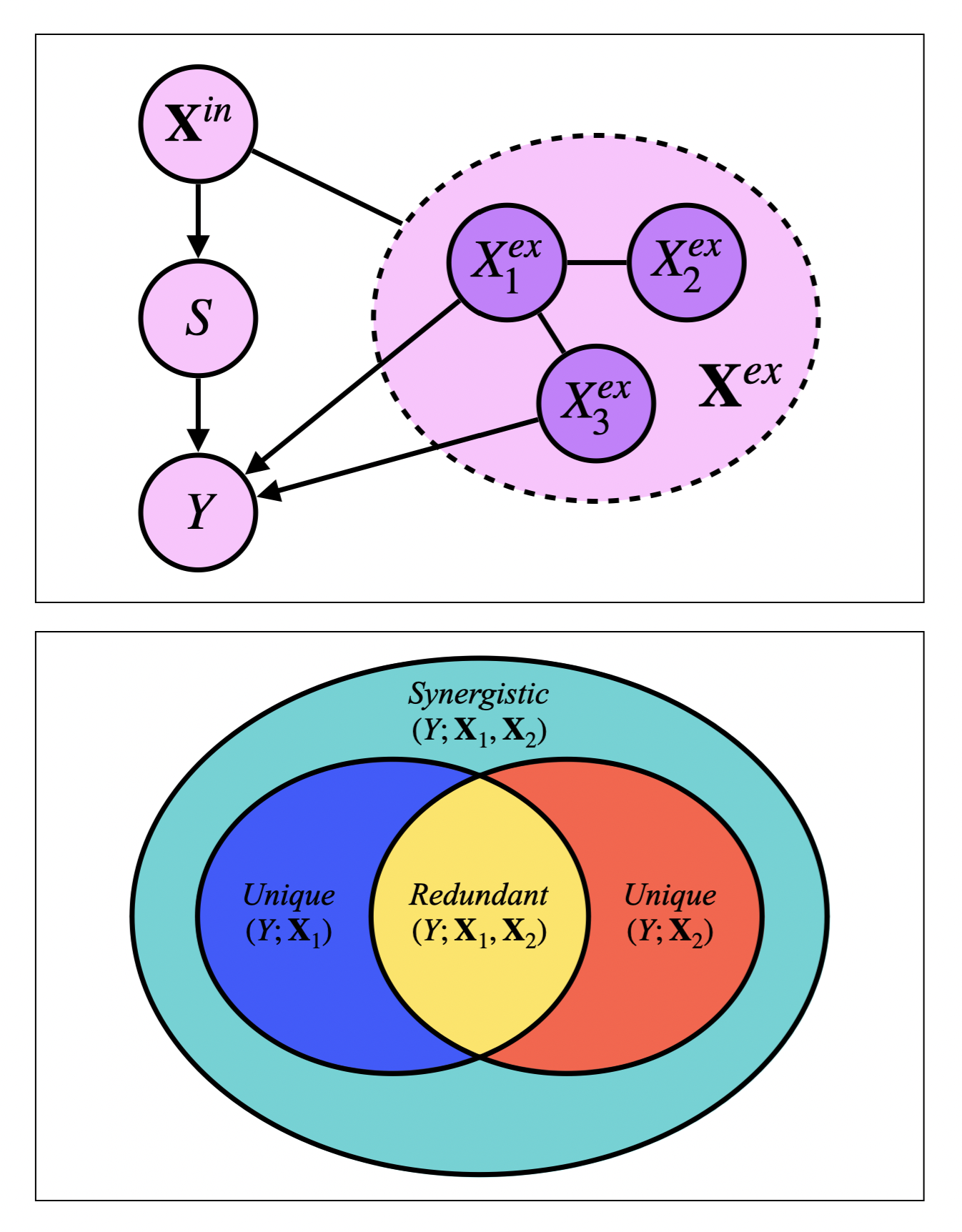}}
\caption{Causal analysis and mutual information decomposition methods adopted in this work. \textit{Upper panel:} Causal graph that represents the stellar mass-predicting process of an end-to-end deep learning model. Each node refers to a variable or a set of variables. Each arrow represents a causal link. $\mathbf{X}^{in}$ refers to the set of input data to the model. $Y$ refers to the target variable (i.e., stellar mass in this work). $\mathbf{X}^{ex}$ refers to the set of external variables that contain the information on stellar mass but missing in the input data. $S$ refers to the low-dimensional latent vector that encodes the information on stellar mass extracted from the input data, the intermediary variable between $\mathbf{X}^{in}$ and $Y$. The line between $\mathbf{X}^{ex}$ and $\mathbf{X}^{in}$ that has no direction specified refers to their possible dependence, which is not necessarily a direct causal link. There may be inner structures between the individual variables in the set $\mathbf{X}^{ex}$ and $Y$, shown by the exemplar variables $X^{ex}_1$, $X^{ex}_2$, and $X^{ex}_3$. The undirected lines between $X^{ex}_1$ and $X^{ex}_2$ and between $X^{ex}_1$ and $X^{ex}_3$ refer to their possible undirected dependences. \textit{Lower panel:} Diagram of the decomposition of mutual information between the target $Y$ and two sets of input data $\mathbf{X}_1$, $\mathbf{X}_2$. $Redundant(Y; \mathbf{X}_1, \mathbf{X}_2)$ refers to the redundant information on $Y$ that both $\mathbf{X}_1$ and $\mathbf{X}_2$ can provide. $Unique(Y; \mathbf{X}_1)$ and $Unique(Y; \mathbf{X}_2)$ refer to the unique information that only $\mathbf{X}_1$ or $\mathbf{X}_2$ can provide. $Synergistic(Y; \mathbf{X}_1, \mathbf{X}_2)$ refers to the synergistic information that exists only when both $\mathbf{X}_1$ and $\mathbf{X}_2$ are available.}
\label{fig:method}
\end{center}
\end{figure}

\subsection{Causal analysis} \label{sec:causal}

\subsubsection{General idea} \label{sec:general_causal}

The first interpretability technique we leveraged in this work is based on the principle of causal learning. Causal learning \citep{Pearl2009} aims to identify the causal paths between different variables, uncover the causal structures of multivariable systems, and infer the causal effects of some variables on other variables. Therefore, causal learning can provide not only predictive but also interpretable insights into the data-generating mechanisms behind multivariate systems, bridging the gap between statistical associations and domain knowledge. In many situations where interventional or controlled experiments are infeasible, researchers have to discover causal structures solely using observational data (e.g., astronomical data). Causal learning offers the possibility to achieve this goal. While not possible to identify the causal direction between only two variables without prior knowledge, causal learning leverages the presence of other variables to infer directionality and establish causal links; thus, the whole process of causal discovery can be conducted in a purely data-driven manner. We refer to \citet{HeinzeDeml2018} and \citet{Yao2021} for in-depth introductions.

Causal learning has been widely used in many scientific fields, whereas there is still vast room for exploitation in astronomy. Although the ideas involved in ​​causal learning may be recognized in commonly used analysis techniques (e.g., matching, variable control, and randomization tests), only a small number of studies in astronomy have explicitly referred to causal learning. For example, \citet{Scholkopf2016}, \citet{Wang2016}, and \citet{Gebhard2022} relied on causal learning to reduce instrumental and systematic effects to facilitate exoplanet detection. \citet{Pasquato2023} and \citet{Jin2025} applied causal learning to analyze the relationship between supermassive black holes and host galaxies. \citet{Mucesh2024} used causal learning to analyze the effect of environment on star formation, providing a new perspective on this long-debated topic of galaxy formation and evolution. In short, there would be broad prospects for the application of causal learning in astronomy. The combination of causal learning and machine learning would further exploit their advantages and make deeper impacts on data mining and scientific discovery \citep[e.g.,][]{Scholkopf2021, Deng2022, Kaddour2022, Berrevoets2023}.

To represent the stellar mass-predicting process of an end-to-end deep learning model, we established a causal graph that illustrates how the set of input data $\mathbf{X}^{in}$ and the set of external variables $\mathbf{X}^{ex}$ causally affect the prediction of the target $Y$ (i.e., stellar mass in this work), shown in the upper panel of Fig.~\ref{fig:method}. This is in the form of a directed acyclic graph (DAG) that contains no directed cycles between variables. The causal path $\mathbf{X}^{in} \rightarrow S \rightarrow Y$ describes the flow of information inside the model. The input set $\mathbf{X}^{in}$ may comprise galaxy images, integrated photometry, or additional input data (e.g., galactic reddening $E(B-V)$). The low-dimensional representation $S$ encodes the information on $Y$ extracted from $\mathbf{X}^{in}$. As $\mathbf{X}^{in}$ cannot provide all the information on $Y$, the external set $\mathbf{X}^{ex}$, containing information unaccounted for by $\mathbf{X}^{in}$, have causal links to $Y$ outside the model. $\mathbf{X}^{ex}$ may contain spectroscopic properties (e.g., spec-$z$) or morphological features for a photometry-only model. There may be inner structures between the individual variables in $\mathbf{X}^{ex}$ and $Y$, as exemplified by the variables $X^{ex}_1$, $X^{ex}_2$, and $X^{ex}_3$ that may have direct or indirect causal paths to $Y$. The variables in $\mathbf{X}^{ex}$ would have contributions to the estimation of $Y$ in addition to $\mathbf{X}^{in}$ if input into the model. $\mathbf{X}^{ex}$ and $\mathbf{X}^{in}$ may be dependent, as represented by the undirected line in between, namely, $\mathbf{X}^{ex} \noarrow \mathbf{X}^{in}$. Any shared information between them that contributes to the estimation of $Y$ can be captured by $S$.

We note that our causal graph represents the stellar mass-predicting process of an end-to-end model (for the purpose of interpretation) rather than the actual physical mechanisms. As we elaborate in our work, this provides a universal way to tell apart the input and external variables with respect to any given model whatever the input data are, needing not to modify the causal graph to accord with different models that have various input data. Furthermore, the stellar mass estimates are derived quantities subject to the impacts of various variables, which we do not regard as independent ground-truth values. Correspondingly, the target $Y$, considered as a dependent variable, acts as the descendant of the other nodes.

We can identify three basic components that generate our causal graph: chains, confounders, and colliders. For any variables $A$, $B,$ and $C$, a chain $A \rightarrow B \rightarrow C$ refers to a causal path from $A$ to $C$ in which $B$ acts as a mediator to transmits the information or the impact of $A$ to $C$ (e.g., $\mathbf{X}^{in} \rightarrow S \rightarrow Y$). A confounder in $A \leftarrow B \rightarrow C$ refers to the variable $B$ that causally affects both $A$ and $C$, which introduces a statistical association between $A$ and $C$ even if no direct causal link exists between them. A collider in $A \rightarrow B \leftarrow C$ refers to the variable $B$ that both $A$ and $C$ have causal effects on; $A$ and $C$ may not necessarily be independent, and they may affect $B$ individually or synergistically (e.g., $S \rightarrow Y \leftarrow \mathbf{X}^{ex}$). These three components are the simplest causal structures that act as basic building blocks for capturing and describing the causal interactions in complex systems.

In the cases with a chain or a confounder, $A$ and $C$ are not marginally independent in general, namely, $A \not\perp C$, or $p(A, C) \neq p(A) p(C)$ where $p$ refers to the probability density. However, they are conditionally independent, or ``$d$-separated'', once conditioned on $B$; namely, we obtain $A \perp C | B$, which means that no information can flow between $A$ and $C$ when the causal path is blocked. It can be proven by simply observing that $p(C | A, B) = p(C | B)$ since $C$ does not directly depend on $A$; thus, it leads to $p(A, C | B) = p(A | B) p(C | A, B) = p(A | B) p(C | B)$, or $A \perp C | B$. As a simple illustration, the season affects the rainfall frequency, and a rainfall causes the grass to be wet and a pavement to be slippery, making the season, the average wetness of the grass, and the average slipperiness of the pavement be associated; given a certain rainfall, how wet the grass is would no longer rely on the season, and would not depend on how slippery the pavement is.

This idea of $d$-separation lays the foundation for our causal analysis. Based on our causal graph, we propose the following expressions, \begin{equation}
\mathbf{X}^{in} \perp Y | S, \,\,\,\,\, \mathbf{X}^{ex} \not\perp Y | S,
\label{eq:prop1}
\end{equation}
and
\begin{equation}
X^{ex}_2 \perp Y | S, X^{ex}_1, \,\,\,\,\, X^{ex}_3 \not\perp Y | S, X^{ex}_1.
\label{eq:prop2}
\end{equation}
$\mathbf{X}^{in}$ and $Y$ are independent once conditioned on $S$, but $\mathbf{X}^{ex}$ and $Y$ cannot be $d$-separated as there are direct links in between. Therefore, we can conduct local independence tests by conditioning on $S$ to identify the constituents of $\mathbf{X}^{ex}$: any variable that exhibits statistically significant association with $Y$ belongs to $\mathbf{X}^{ex}$, while any variable that shows no significant association with $Y$ either has already been included in $\mathbf{X}^{in}$ or has no (statistically significant) contributory information on $Y$. Similarly, conditional independence tests can be applied to analyze the causal structures between the individual variables in $\mathbf{X}^{ex}$ and $Y$. Once conditioned on both $S$ and the given conditional variable $X^{ex}_1$, the indirect paths through $S$ or $X^{ex}_1$ to $Y$ are blocked; any variable that shows no significant association with $Y$ can possibly be linked to $Y$ only through $X^{ex}_1$, represented by $X^{ex}_2 \noarrow X^{ex}_1 \rightarrow Y$, regardless of the directionality of the link between $X^{ex}_1$ and $X^{ex}_2$. On the other hand, any variable that shows statistically significant association with $Y$ has a causal link to $Y$ not through $X^{ex}_1$, represented by $X^{ex}_3 \rightarrow Y \leftarrow X^{ex}_1$, no matter whether $X^{ex}_1$ and $X^{ex}_3$ are dependent or not. In other words, $X^{ex}_1$ can explain the overall contribution of $X^{ex}_2$ to the estimation of $Y$ if input into the model in addition to $\mathbf{X}^{in}$, but not the contribution of $X^{ex}_3$. It would be unnecessary to use $X^{ex}_2$ anymore once $X^{ex}_1$ is added to $\mathbf{X}^{in}$, whereas $X^{ex}_3$ would still be useful. We note that we took the faithfulness assumption: $A \perp_{data} C | B \Rightarrow{} A \perp_{graph} C | B$, meaning that the statistical independencies observed in data can deduce $d$-separations in the causal graph.

\subsubsection{Outline of our analysis} \label{sec:outline_causal}

We established the causal path $\mathbf{X}^{in} \rightarrow S \rightarrow Y$ for several photometry-only and image-based models individually using a framework based on supervised contrastive learning and KNN. These models are denoted as ``$\mathbf{M}_{ugriz}$,'' ``$\mathbf{M}_{ugrizW123}$,'' ``$\mathbf{I}_{ugriz}$,'' ``$\mathbf{I}_{ugriz} \cup \mathbf{M}_{W123}$,'' and ``$\mathbf{I}_{ugriz} \cup \mathbf{M}_{W123} \cup z_{spec}$.'' They are different in terms of input data (summarized in Table~\ref{tab:models_causal}). We included galactic reddening $E(B-V)$ in the input data as it impacts the photo-$z$ estimation \citep{Pasquet2019} and also the stellar mass estimation, though we did not explicitly show its contribution in this work. For each model, we detected the external variables in $\mathbf{X}^{ex}$ with high efficiency from the long list of photometric, morphological, physical, and other parameters presented in Table~\ref{tab:data}. Then using the detected external variables for the photometry-only model $\mathbf{M}_{ugriz}$, we investigated the causal structures in $\mathbf{X}^{ex} \rightarrow Y$, which give hints on the causal mechanisms behind image-based models.

\begin{table}[!ht]
\caption{Summary of the models in the causal analysis.} \label{tab:models_causal}
\centering
\begin{tabularx}{\columnwidth}{l | X}
\hline
$\mathbf{M}_{ugriz}$  &  A photometry-only model fed with magnitudes and errors in the five optical bands $u,g,r,i,z$ and galactic reddening $E(B-V)$.  \\
\hline
$\mathbf{M}_{ugrizW123}$  &  A photometry-only model same as $\mathbf{M}_{ugriz}$, but also fed with magnitudes and errors in the three infrared bands $W1,W2,W3$ (except errors in the $W3$ band).  \\
\hline
$\mathbf{I}_{ugriz}$  &  An image-based model fed with images in the five optical bands $u,g,r,i,z$ and galactic reddening $E(B-V)$.  \\
\hline
$\mathbf{I}_{ugriz} \cup \mathbf{M}_{W123}$  &  An image-based model same as $\mathbf{I}_{ugriz}$, but also fed with magnitudes and errors in the three infrared bands $W1,W2,W3$ (except errors in the $W3$ band).  \\
\hline
$\mathbf{I}_{ugriz} \cup \mathbf{M}_{W123} \cup z_{spec}$  &  An image-based model same as $\mathbf{I}_{ugriz} \cup \mathbf{M}_{W123}$, but also fed with spec-$z$.  \\
\hline
\end{tabularx}
\end{table}

The detailed procedures for establishing the causal path $\mathbf{X}^{in} \rightarrow S \rightarrow Y$, detecting external variables in $\mathbf{X}^{ex}$, and investigating the causal structures in $\mathbf{X}^{ex} \rightarrow Y$ are presented below.

\subsubsection{Establishing the causal path $\mathbf{X}^{in} \rightarrow S \rightarrow Y$} \label{sec:causal_path_XY}

The establishment of the causal path $\mathbf{X}^{in} \rightarrow S \rightarrow Y$ essentially amounts to having the input data $\mathbf{X}^{in}$ projected to the low-dimensional and stellar mass-sensitive representation $S$, a crucial step in our analysis. This can be realized using the framework initially developed by \citet{Lin2024} for the photo-$z$ estimation, which includes supervised contrastive learning and KNN procedures. Conditioning on $S$ is equivalent to obtaining a group of nearest neighbors in the $S$ space that constitute a local (conditional) multivariate distribution, allowing conditional independent tests to be performed. In principle, KNN cannot always be directly preformed on $\mathbf{X}^{in}$ because the input data may be high-dimensional and contain irrelevant information. Therefore, supervised contrastive learning should be performed first to compress $\mathbf{X}^{in}$ and obtain the low-dimensional representation $S$.

The supervised contrastive learning procedure involves an encoder, an estimator, and a decoder, which are all neural networks. The encoder is fed with $\mathbf{X}^{in}$ and outputs two low-dimensional vectors. One vector is $S$ that encodes the information on stellar mass while the other vector encodes other information on $\mathbf{X}^{in}$. Then, $S$ is fed to the estimator with the softmax function applied on its last layer to give a probability density estimate of stellar mass, regarding each stellar mass bin as a class. This softmax output is constrained with a cross-entropy loss using one-hot labels converted from the known stellar mass values. The discretization of the stellar mass estimates is conducted primarily in order to be coherent with the estimation of mutual information described in Sect.~\ref{sec:implement_info}. The supervision by the stellar mass labels is intended to let meaningful stellar mass information extracted from $\mathbf{X}^{in}$ and propagate to $S$. The two vectors from the encoder are concatenated and fed to the decoder, reconstructing the main input data in $\mathbf{X}^{in}$ (i.e., images and magnitudes) constrained by a mean square error (MSE) loss. The reconstructed main input data and the original additional input data (i.e., magnitude errors and galactic reddening $E(B-V)$) are then concatenated and re-input into the encoder, repeating the process above. This second pass is intended to produce positive pairs for contrastive learning, in which a contrastive loss is adopted to minimize the contrast of $S$ between positive pairs and maximize between negative pairs (i.e., different galaxies), with the contrast characterized by the Euclidean distance. Once trained, the encoder can be used to obtain $S$ given $\mathbf{X}^{in}$ (i.e., $\mathbf{X}^{in} \rightarrow S$), then stellar mass can be seen as one axis of the local multivariate distribution constructed by KNN in the $S$ space (i.e., $S \rightarrow Y$). We refer to \citet{Lin2024} for the details of the supervised contrastive learning procedure.

We conducted supervised contrastive learning individually for each of the five cases listed in Table~\ref{tab:models_causal}. When $\mathbf{X}^{in}$ contains images, we used the same encoder, estimator, and decoder networks from \citet{Lin2024}, except with different inputs and outputs. The encoder network is a modified version of the inception network developed by \citet{Treyer2024} that has six inception modules. We took 16 dimensions for $S$ and 512 dimensions for the other vector output by the encoder. The decoder network mainly consists of convolutional layers and bilinear interpolation for upsampling. The estimator network has two fully connected layers, with 1024 nodes in the first layer. 

When $\mathbf{X}^{in}$ does not involve images, we used simpler networks instead. The simpler encoder network consists of 20 consecutive fully connected layers and two additional parallel fully connected layers that produce the two vectors. The two vectors both have eight dimensions. The decoder network has 20 fully connected layers. The estimator network has two fully connected layers. The last layers of these networks all have no activation. Each of the remaining layers has 128 nodes and each layer is activated by the Leaky Rectified Linear Unit (Leaky ReLU) with a leaky rate of 0.2.

We used 520 bins to express the distribution of stellar mass in a range between 6.0 and 12.5 dex (Fig.~\ref{fig:logm_r_z_dist}). The same stellar mass bin width was used for the softmax outputs of both the two versions of estimators and the one-hot labels, as well as for estimating mutual information (Sect.~\ref{sec:implement_info}). The networks were trained from scratch using the mini-batch gradient descent. We have validated that this implementation is robust for the qualitative analysis that we conducted in this work. The details of model training and validation are presented in Appendix~\ref{sec:training_dev}.

\subsubsection{Detecting external variables in $\mathbf{X}^{ex}$} \label{sec:detect_X}

The detection of external variables in $\mathbf{X}^{ex}$ relies on KNN applied on the $S$ space established by supervised contrastive learning. For each of the five models, we obtained the latent vectors $S$ for all the galaxies in the test sample and the training sample. For each galaxy in the test sample, we searched for its nearest neighbors from the training sample in the $S$ space, using the same distance measure (i.e., the Euclidean distance) applied in supervised contrastive learning. Within each group of nearest neighbors, local independence tests can be performed between stellar mass and any other variables. We assume that the various features of the nearest neighbors of each test galaxy can be regarded as random realizations from a local multivariate distribution conditioned on $S$. Any statistically significant associations with stellar mass are due to information not fully accounted for by $S$, thus included in $\mathbf{X}^{ex}$ but not in $\mathbf{X}^{in}$; otherwise, the information that can be captured by $S$ would be further utilized by the model to improve the stellar mass estimation until no association is left.

For all the five models, we took a constant value $k=100$ as the number of selected nearest neighbors for each test galaxy. This $k$ value ensures the localization in the $S$ space, but is not too small to introduce strong shot noise (Appendix~\ref{sec:assess_knn}). With the known photometric, morphological, physical, and other parameters at hand (listed in Table~\ref{tab:data}), we can easily tell which variables belong to $\mathbf{X}^{ex}$ given a model. In fact, by training the model only once, any known features can be tested. This is much more efficient and computationally cheaper than other feature selection approaches that require retraining models or re-implementing estimation methods each time the features are changed \citep[e.g.,][]{DIsanto2018}.

We used two metrics to describe the association between stellar mass and any query variable within each group of nearest neighbors. The first metric is the Pearson correlation coefficient. The second metric, named the predictive efficiency, is defined as
\begin{equation}
\epsilon(X_q) = 1 - \frac{\sigma^2_{S}(Y|X_q)}{\sigma^2_{S}(Y)},
\label{eq:pred_efficiency}
\end{equation}
where $X_q$ is the query variable; $\sigma^2_{S}(Y)$ is the variance of the target $Y$ (i.e., stellar mass) estimated using the nearest neighbors (i.e., conditioned on $S$); $\sigma^2_{S}(Y|X_q)$ is the same as $\sigma^2_{S}(Y)$ but with $Y$ quadratically regressed on $X_q$. The predictive efficiency describes the reduction in the variance of $Y$ given $X_q$. A more impactful $X_q$ leads to more reduction in the variance of $Y$ and an $\epsilon(X_q)$ closer to unity.

We note that these two metrics were adopted with the observation that the multivariate distribution conditioned on $S$ is already ``local'' and does not have strong nonlinear behaviors. The predictive efficiency is equivalent to the squared correlation in the case where $X_q$ and $Y$ are linearly associated, meaning that the predictive efficiency would grow faster than the correlation for strong associations but more slowly for weak associations. These two metrics provide two angles to view the causal links: the correlation is capable of indicating whether a trend between $X_q$ and $Y$ is averagely ascending or descending, while the predictive efficiency is an indicator of the ``strength'' of causal effects and more relevant to the outcome of the stellar mass estimation. We analyzed the results on the detection of external variables using both the correlation and the predictive efficiency.

\subsubsection{Investigating causal structures in $\mathbf{X}^{ex} \rightarrow Y$} \label{sec:str_causal_effects_XY}

Moving forward with the detected external variables in $\mathbf{X}^{ex}$, the causal structures in $\mathbf{X}^{ex} \rightarrow Y$ would reveal how different variables beyond $\mathbf{X}^{in}$ are causally contributory to the stellar mass estimation in addition to $\mathbf{X}^{in}$. In particular, the causal structures in $\mathbf{X}^{ex} \rightarrow Y$ for the photometry-only model $\mathbf{M}_{ugriz}$ would involve variables not covered by optical photometry such as morphology, infrared photometry, and spec-$z$. Since all morphological parameters are measured on the basis of images, the causal paths between morphology and stellar mass (possibly through other variables) would provide meaningful insights into the causal mechanisms behind image-based models.

We performed conditional independence tests to analyze the causal structures in $\mathbf{X}^{ex} \rightarrow Y$ for the model $\mathbf{M}_{ugriz}$. For this, each individual test should involve stellar mass and two variables from $\mathbf{X}^{ex}$, one of which is used as a query variable and the other is a conditional variable. Using the nearest neighbors of each test galaxy in the $S$ space (i.e., conditioned on $S$), we define the conditional predictive efficiency to describe the association between stellar mass and any query variable conditioned on a third variable, as
\begin{equation}
\epsilon(X_q|X_c) = 1 - \frac{\sigma^2_{S}[(Y|X_c)|(X_q|X_c)]}{\sigma^2_{S}(Y|X_c)},
\label{eq:pred_efficiency_cond}
\end{equation}
where $X_q$ and $X_c$ are the query variable and the conditional variable, respectively, both from $\mathbf{X}^{ex}$; $\sigma^2_{S}(Y|X_c)$ is the variance of the target $Y$ estimated using the nearest neighbors after quadratically regressed on $X_c$; $\sigma^2_{S}[(Y|X_c)|(X_q|X_c)]$ is the same as $\sigma^2_{S}(Y|X_c)$ but with $X_q$ and $Y$ both quadratically regressed on $X_c$, and then $Y|X_c$ quadratically regressed on $X_q|X_c$. A near-zero $\epsilon(X_q|X_c)$ would imply that $X_q$ does not have any significant contribution to the estimation of $Y$ unexplained by $X_c$. This is exemplified by $X^{ex}_2 \noarrow X^{ex}_1 \rightarrow Y$ in Fig.~\ref{fig:method} where $X^{ex}_1$ is $X_c$ and $X^{ex}_2$ is $X_q$. On the other hand, a positive $\epsilon(X_q|X_c)$ would imply that $X_q$ has a contribution to the estimation of $Y$ not covered by $X_c$, exemplified by $X^{ex}_3 \rightarrow Y \leftarrow X^{ex}_1$ where $X^{ex}_1$ is $X_c$ and $X^{ex}_3$ is $X_q$.

For the practical purpose of stellar mass estimation, we are more interested in to what extent a variable can impact the outcome. Therefore, we analyzed the causal structures in $\mathbf{X}^{ex} \rightarrow Y$ primarily using the conditional predictive efficiency rather than the correlation metric. For the model $\mathbf{M}_{ugriz}$, using the external morphological features as the query $X_q$ and spec-$z$ or infrared photometry as the condition $X_c$, we can determine whether the contributions of the morphological features to the stellar mass estimation in addition to optical photometry can be explained by spec-$z$ or infrared photometry; if instead using the external morphological features as the condition $X_c$, we can determine how much the impact of spec-$z$ or infrared photometry can be recovered by providing the morphological features.

\subsection{Mutual information decomposition} \label{sec:info}

\subsubsection{General idea} \label{sec:general_info}

The second interpretability technique we used in this work relies on the principle of information theory. 
Information theory provides an elegant way to define and quantify information, and describe the transmission or processing of information. Rooted in Shannon's pioneer work \citep{Shannon1948}, information theory has developed into an interdisciplinary field and become especially appealing for the analysis of complex systems. Information theory is also a rather novel approach to astronomy. It has been adopted in applications regarding compact stars \citep[e.g.,][]{deAvellarHorvath2012}, cosmology \citep[e.g.,][]{Pandey2016, PandeySarkar2016, Vazza2017, GarciaAlvarado2020, MartaPinho2020}, and exoplanet characterization \citep[e.g.,][]{GilbertFabrycky2020, Bartlett2022, Segal2024, Vannah2024}. 

Out of the most commonly used concepts from information theory, mutual information gives a measure of statistical dependences between variables even in the presence of complex nonlinearity. Furthermore, the method proposed by \citet{WilliamsBeer2010} offers a possibility to probe the ``structure'' of mutual information, namely, decomposing multivariate mutual information into redundant, unique, and synergistic components, which leads to a deeper understanding of how mutual information breaks down across different variables and how a complex system functions. It is also possible to combine such mutual information decomposition with causal insights to quantify complex causal relations in multivariate systems \citep{MartinezSanchez2024}.

In the context of the stellar mass estimation, mutual information offers a way to characterize the contributions of input data by measuring the reduction in the entropy of stellar mass given the input data, irrespective of the functional relationship between them. We followed the idea from \citet{WilliamsBeer2010} to decompose the mutual information between any given tuple of two input datasets and stellar mass, as illustrated in the lower panel of Fig.~\ref{fig:method}.

We first define the instance-wise mutual information for any single data instance (i.e., a galaxy), as
\begin{equation}
i(y; x) = \log_e p(y|x) - \log_e p(y),
\label{eq:mutual_info}
\end{equation}
where $x$ and $y$ refer to an input data instance and the corresponding target, respectively; $p(y)$ refers to the prior probability density of $y$, equivalent to the normalized stellar mass distribution shown in Fig.~\ref{fig:logm_r_z_dist}; $p(y|x)$ refers to the probability density of $y$ given $x$; thus $-\log_e p(y)$ and $-\log_e p(y|x)$ give the instance-wise entropy and conditional entropy (in units of nats), respectively. This equation describes the information gain relative to the prior once the input $x$ is given.

Using the instance-wise mutual information, we now discuss the different components of mutual information between the target $Y$ and two input datasets $\mathbf{X}_1$, $\mathbf{X}_2$. The redundant information (denoted as ``rdn'') is the amount of information on $Y$ that can be provided by either $\mathbf{X}_1$ or $\mathbf{X}_2$, indicating the overlapping contribution of $\mathbf{X}_1$ and $\mathbf{X}_2$. The redundant information within any given parameter space $\Omega$ is defined as
\begin{equation}
I_{\Omega}^{rdn} (Y; \mathbf{X}_1, \mathbf{X}_2) = \frac{1}{N_{\Omega}} \sum_{y,x_1,x_2 \in \Omega} \min \{i(y; x_1),i(y; x_2)\},
\label{eq:rdn}
\end{equation}
considered as the average of individual minimum instance-wise information estimates, where $N_{\Omega}$ refers to the number of data instances in $\Omega$. We note that our work regards the redundant information as the same amount of information on $Y$ carried by $\mathbf{X}_1$ or $\mathbf{X}_2$, not necessarily the same information.

The unique information (denoted as ``unq'') for $\mathbf{X}_1$ is the amount of information that can only be provided by $\mathbf{X}_1$ rather than $\mathbf{X}_2$, indicating the unique contribution of $\mathbf{X}_1$. It is defined as
\begin{equation}
I_{\Omega}^{unq} (Y; \mathbf{X}_1) = \frac{1}{N_{\Omega}} \sum_{y,x_1,x_2 \in \Omega} i(y; x_1) - I_{\Omega}^{rdn} (Y; \mathbf{X}_1, \mathbf{X}_2).
\label{eq:unq1}
\end{equation}
Similarly, the unique information for $\mathbf{X}_2$ is defined as
\begin{equation}
I_{\Omega}^{unq} (Y; \mathbf{X}_2) = \frac{1}{N_{\Omega}} \sum_{y,x_1,x_2 \in \Omega} i(y; x_2) - I_{\Omega}^{rdn} (Y; \mathbf{X}_1, \mathbf{X}_2).
\label{eq:unq2}
\end{equation}
Both the redundant information and the unique information can be provided by a set of data without reliance on the other set.

Finally, the synergistic information (denoted as ``syn'') is the amount of information jointly provided by $\mathbf{X}_1$ and $\mathbf{X}_2$, indicating the synergistic contribution of $\mathbf{X}_1$ and $\mathbf{X}_2$ that does not exist if either of the two datasets is unavailable. It is defined as
\begin{equation}
\begin{aligned}
I_{\Omega}^{syn} (Y; \mathbf{X}_1, \mathbf{X}_2) = &\frac{1}{N_{\Omega}} \sum_{y,x_1,x_2 \in \Omega} i(y; x_1 \cup x_2) - I_{\Omega}^{rdn} (Y; \mathbf{X}_1, \mathbf{X}_2)\\
&- I_{\Omega}^{unq} (Y; \mathbf{X}_1) - I_{\Omega}^{unq} (Y; \mathbf{X}_2),
\end{aligned}
\label{eq:syn}
\end{equation}
where $x_1 \cup x_2$ refers to the union of the two individual inputs.

As an intuitive illustration, $g$-band and $r$-band images would have a high level of redundancy in determining the $g$-band galaxy inclination, since the galaxy shapes manifested in these two bands are highly similar (i.e., Eq.~\ref{eq:rdn}). The excessive information on the $g$-band inclination provided by the $g$ band itself over the redundant part reflects its unique contribution (i.e., Eq.~\ref{eq:unq1} or \ref{eq:unq2}). On the contrary, the two bands would have to synergize to determine the $g-r$ color, since neither of them alone can perfectly provide color information (i.e., Eq.~\ref{eq:syn}).

We note that the predictive efficiency defined in Sect.~\ref{sec:detect_X} also provides a way to quantify the contributions of an external set of variables to the stellar mass estimation given another set as the input data of a model, but it cannot differentiate the redundant, unique, and synergistic components of their contributions. This necessitates the use of the mutual information decomposition.

\begin{table*}[!ht]
\caption{Summary of the cases with different data modalities for the mutual information decomposition. Galactic reddening $E(B-V)$ is included in all the datasets.} \label{tab:info_modalities}
\centering
\begin{tabularx}{\linewidth}{l | X}
\hline
<$\mathbf{M}_{ugriz} \,,\, \mathbf{I}^{norm}_{ugriz}$>  &  The set $\mathbf{M}_{ugriz}$ contains Petrosian magnitudes and errors in the five optical bands $u,g,r,i,z$.
$\mathbf{M}_{ugriz}$ corresponds to the photometry-only model defined in Table.~\ref{tab:models_causal}. The notations for models and datasets are used interchangeably. The set $\mathbf{I}^{norm}_{ugriz}$ contains the five-band images normalized by the corresponding Petrosian fluxes before rescaling the pixel intensities using Eq.~\ref{eq:rescaling}.
The use of $\mathbf{I}^{norm}_{ugriz}$ is intended to investigate the contribution of images in the absence of magnitude information. The overall morphological information is in principle encompassed in $\mathbf{I}^{norm}_{ugriz}$. Before normalizing the images, clipping is applied to the Petrosian fluxes in the five bands at the 2nd, 0.1th, 0.1th, 0.1th, and 0.1th percentiles of the per-band flux distributions for the whole sample, respectively, in order to avoid near-zero or negative fluxes in the devidend. This clipping operation is only concerned to a tiny fraction of the sample and does not affect our conclusion.  \\
\hline
<$\mathbf{M}_{ugriz} \,,\, \mathbf{Morph}$>  &  The set $\mathbf{M}_{ugriz}$ remains unchanged. The set $\mathbf{Morph}$ contains a few representative morphological parameters including S\'{e}rsic indices $n$, inclinations $b/a$, and Petrosian radii $R_{50}$ and $R_{90}$ in the five optical bands $u,g,r,i,z$ (20 parameters in total). 
These parameters (or their combinations) have been found to contribute to the stellar mass estimation in addition to integrated photometry by the causal analysis.  \\
\hline
<$\mathbf{M}_{ugriz} \,,\, \mathbf{M}_{W123}$>  &  The set $\mathbf{M}_{ugriz}$ remains unchanged. The set $\mathbf{M}_{W123}$ contains magnitudes and errors in the three infrared bands $W1,W2,W3$ (except errors in the $W3$ band). \\
\hline
<$\mathbf{I}_{ugriz} \,,\, \mathbf{M}_{W123}$>  &  The set $\mathbf{I}_{ugriz}$ contains the unnormalized images in the five optical bands.
The set $\mathbf{M}_{W123}$ remains unchanged.  \\
\hline
<$\mathbf{I}_{ugriz} \cup \mathbf{M}_{W123} \,,\, \{z_{spec}\}$>  &  The set $\mathbf{I}_{ugriz} \cup \mathbf{M}_{W123}$ contains the union of $\mathbf{I}_{ugriz}$ and $\mathbf{M}_{W123}$. The set $\{z_{spec}\}$ contains spec-$z$. \\
\hline
\end{tabularx}
\end{table*}

\begin{table*}[!ht]
\caption{Summary of the cases with different photometric bands for the mutual information decomposition. Galactic reddening $E(B-V)$ is included in all the datasets.} \label{tab:info_bands}
\centering
\begin{tabularx}{\linewidth}{l | X}
\hline
<$\mathbf{I}_{X} \,,\, \mathbf{I}_{\backslash X}$>  &  The set $\mathbf{I}_{X}$ contains images from one single optical band or two adjacent bands, and the set $\mathbf{I}_{\backslash X}$ contains images from the remaining bands, running over all the five optical bands $u,g,r,i,z$. For example, $<\mathbf{I}_{gr} \,,\, \mathbf{I}_{\backslash gr}>$ refers to the case in which one set contains images in the $g$ and $r$ bands, and the other set contains images in the $u$, $i$, and $z$ bands. In total, there are five possibilities of separating out one band and four possibilities of separating out two bands. \\
\hline
<$\mathbf{M}_{X} \,,\, \mathbf{M}_{\backslash X}$>  &  The set $\mathbf{M}_{X}$ contains magnitudes and errors from one single band or two adjacent bands, and the set $\mathbf{M}_{\backslash X}$ contains those from the remaining bands, running over all the five optical bands $u,g,r,i,z$ and the three infrared bands $W1,W2,W3$ (except errors in the $W3$ band). In total, there are eight possibilities of separating out one band and seven possibilities of separating out two bands. \\
\hline
\end{tabularx}
\end{table*}

\subsubsection{Outline of our analysis} \label{sec:outline_info}

We first applied the mutual information decomposition to analyze the contributions of different data modalities to the stellar mass estimation, including photometry, morphology, images, and spec-$z$. We considered five cases, each having a tuple of two input datasets. They are denoted as ``<$\mathbf{M}_{ugriz} \,,\, \mathbf{I}^{norm}_{ugriz}$>,'' ``<$\mathbf{M}_{ugriz} \,,\, \mathbf{Morph}$>,'' ``<$\mathbf{M}_{ugriz} \,,\, \mathbf{M}_{W123}$>,'' ``<$\mathbf{I}_{ugriz} \,,\, \mathbf{M}_{W123}$>,'' and ``<$\mathbf{I}_{ugriz} \cup \mathbf{M}_{W123} \,,\, \{z_{spec}\}$>'' (summarized in Table~\ref{tab:info_modalities}). We also investigated the contributions of different photometric bands, considering the image-based and photometry-only cases separately. They are denoted as ``<$\mathbf{I}_{X} \,,\, \mathbf{I}_{\backslash X}$>'' and ``<$\mathbf{M}_{X} \,,\, \mathbf{M}_{\backslash X}$>'' (summarized in Table~\ref{tab:info_bands}).

\subsubsection{Implementation} \label{sec:implement_info}

We resorted to deep learning neural networks to estimate mutual information. For each combination of $\mathbf{X}_1$ and $\mathbf{X}_2$, we trained three networks to obtain instance-wise mutual information estimates by inputting $\mathbf{X}_1$, $\mathbf{X}_2$, and the union of $\mathbf{X}_1$ and $\mathbf{X}_2$, respectively. The redundant, unique, and synergistic components within any parameter space can then be estimated using Eqs.~\ref{eq:rdn}, \ref{eq:unq1}, \ref{eq:unq2}, and \ref{eq:syn}.

We took the same encoder and estimator networks from supervised contrastive learning in Sect.~\ref{sec:causal_path_XY} to build end-to-end models for the mutual information estimation. The decoder networks were discarded. Similar to supervised contrastive learning, the two versions of networks were used depending on whether images were involved in the input data. The networks all have a softmax output with the same stellar mass bin width as in supervised contrastive learning, constrained with a cross-entropy loss using the one-hot stellar mass labels in a manner of supervised learning. They were trained using the same training procedure as in supervised contrastive learning. The cross-entropy loss gives direct estimates of the conditional entropy $-\log_e p(y|x)$. Using the prior probability density $p(y)$ produced with the same stellar mass binning, the estimates of the instance-wise mutual information $i(y; x)$ can be obtained (Eq.~\ref{eq:mutual_info}). While the values of $-\log_e p(y|x)$ and $-\log_e p(y)$ both rely on the stellar mass bin width, such reliance can be canceled out for $i(y; x)$.

We note that the mutual information estimates (and thus the decomposition) produced by neural networks are empirical rather than theoretical. Strictly speaking, such an estimate should be regarded as the lower bound of the ground truth. Primarily, we point out that mutual information would be diluted by noise in data. However, as we were interested in the impact of given data on the stellar mass estimation, we only considered the empirically measured mutual information in the presence of noise, rather than the intrinsic noise-free value. In addition, mutual information is subject to data distributions. We discuss the impacts of noise and data distributions when presenting the results on the mutual information decomposition (Sect.~\ref{sec:results_info}).

There are other factors that would affect the estimation of mutual information, such as the miscalibration of probability densities $p(y|x)$, the limit of model expressivity, the binning effect, and overfitting. Similar to supervised contrastive learning, we have validated that our implementation for the mutual information estimation exhibits qualitative robustness against these factors. We provide the details of model training and validation in Appendix~\ref{sec:training_dev}.

\begin{figure*}
\begin{center}
\centerline{\includegraphics[width=1.05\linewidth]{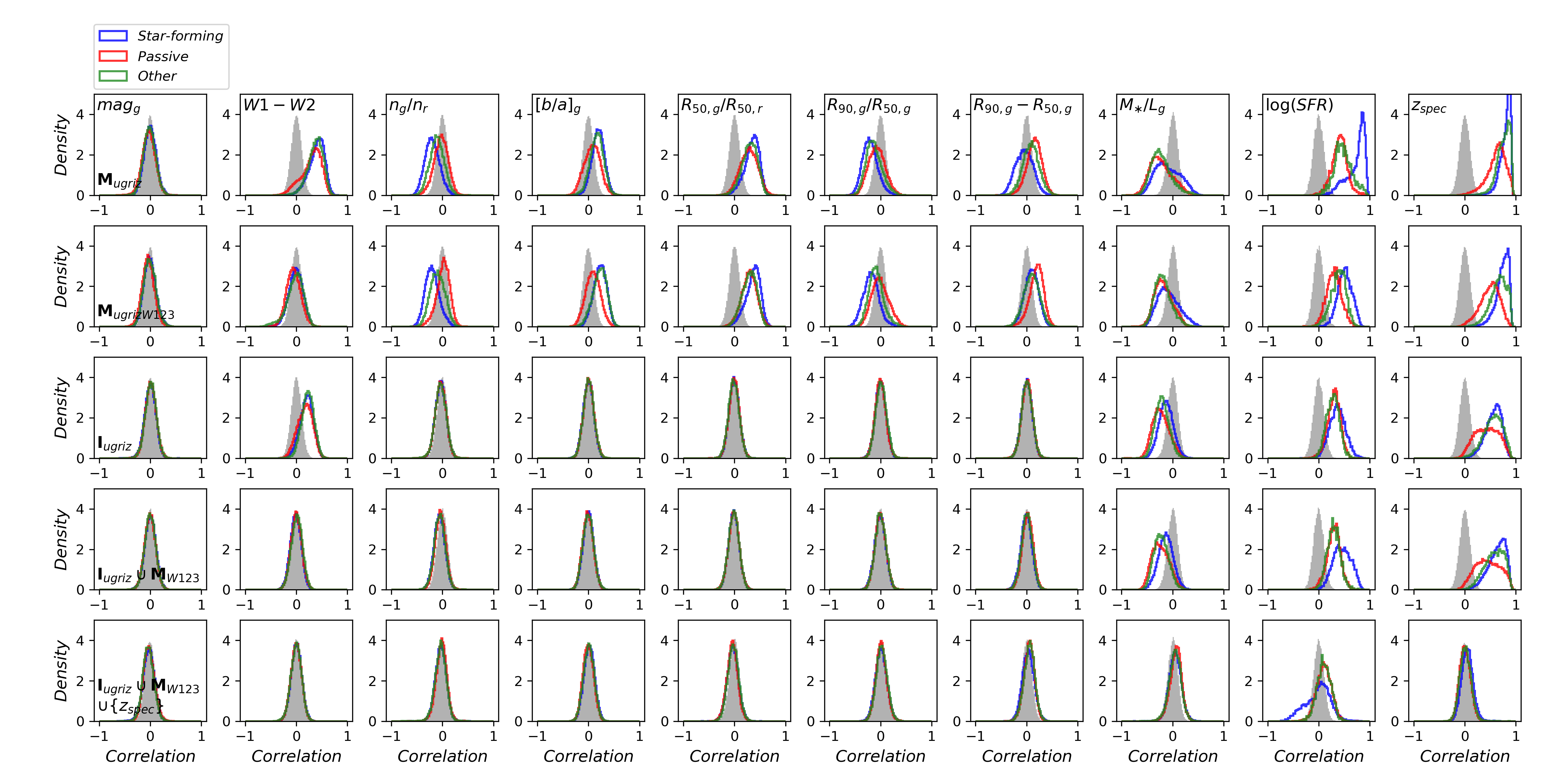}}
\caption{Distributions of local correlations between stellar mass and representative parameters for the five photometry-only or image-based models defined in Table~\ref{tab:models_causal}. Each column corresponds a parameter, and each row corresponds to a model. The original distributions are separately shown for star-forming, passive, and other galaxies from the test sample, illustrated as the colored curves. The distributions shown in grey are used as a contrast, produced by randomly permuting the stellar mass values within the nearest neighbors of each test galaxy. Deviations between the original and reference correlation distributions indicate external parameters for a given model. Primarily, optical photometry cannot entirely account for morphological information, infrared information, spec-$z$, and physical information related to stellar mass; while multi-band images can encompass intra- and cross-band morphological features that are both important for the stellar mass estimation.}
\label{fig:correlation_selected}
\end{center}
\end{figure*}

\begin{figure*}
\begin{center}
\centerline{\includegraphics[width=0.9\linewidth]{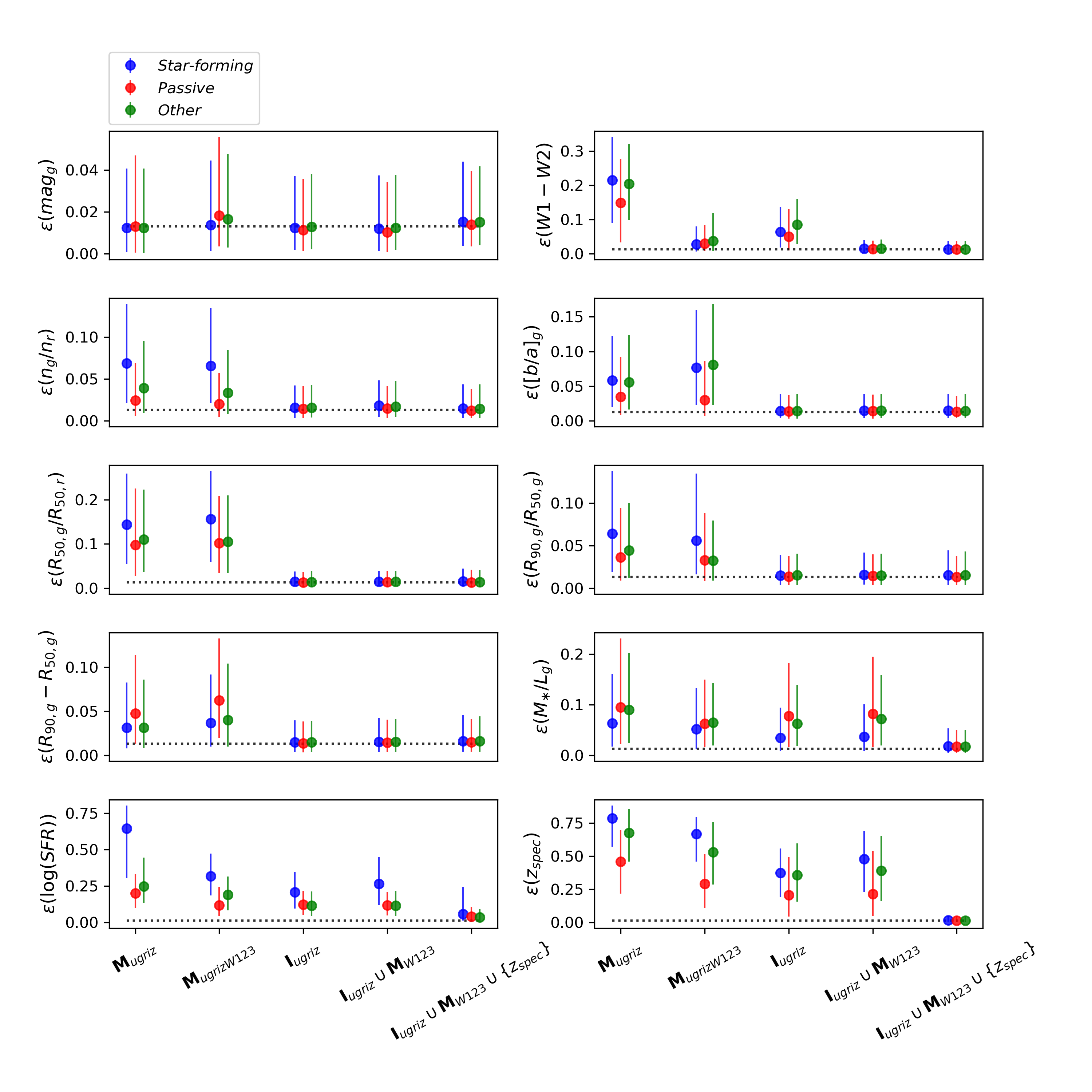}}
\caption{Predictive efficiency of representative parameters for the five photometry-only or image-based models defined in Table~\ref{tab:models_causal}. Each data point corresponds to the 50th of the predictive efficiency distribution over a galaxy population from the test sample (i.e., star-forming, passive, and other galaxies), and each error bar indicates the 16th and 84th percentiles. The black dotted lines indicate the reference value of the median predictive efficiency ($\sim 0.013$) estimated by randomly permuting the stellar mass values within the nearest neighbors of each test galaxy. The predictive efficiency reveals the same trends as in Fig.~\ref{fig:correlation_selected}, and is more indicative of the impact of each variable on the stellar mass estimation.}
\label{fig:pred_efficiency_selected}
\end{center}
\end{figure*}

\begin{figure*}
\begin{center}
\centerline{\includegraphics[width=0.8\linewidth]{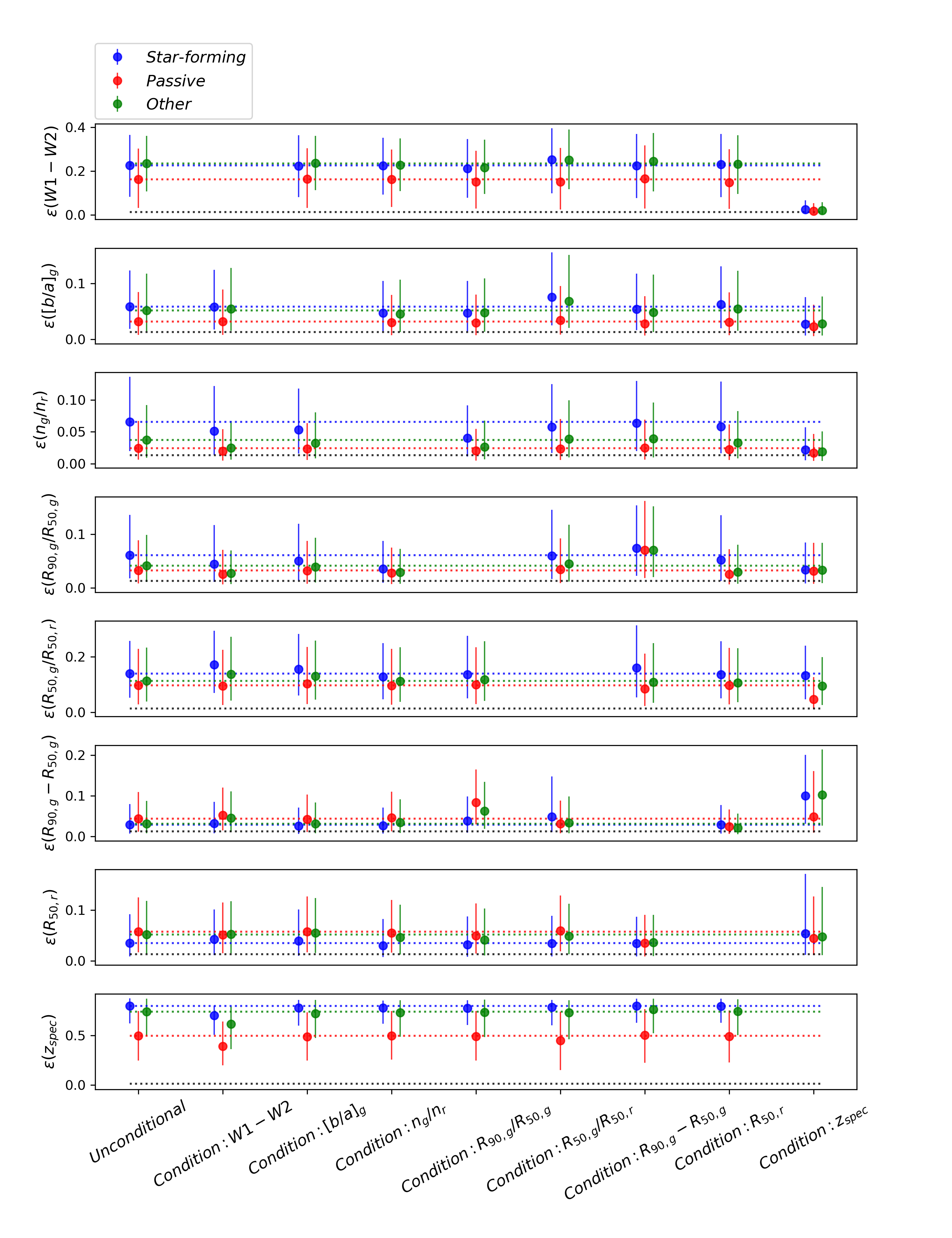}}
\caption{Conditional predictive efficiency of representative parameters for the photometry-only model $\mathbf{M}_{ugriz}$ defined in Table~\ref{tab:models_causal}. Each data point corresponds to the 50th of the (conditional) predictive efficiency distribution over a galaxy population from the test sample (i.e., star-forming, passive, and other galaxies), and each error bar indicates the 16th and 84th percentiles. In each row (corresponding to a parameter), the first triplet of data points shows the unconditional predictive efficiency to be compared with, and each of the remaining triplets shows the conditional predictive efficiency with the conditional variable labeled on the bottom. All the parameters including stellar mass are first conditioned on the $g-r$ color of the nearest neighbors of each test galaxy before computing the (conditional) predictive efficiency. The black dotted lines indicate the reference value of the median predictive efficiency ($\sim 0.013$) estimated by randomly permuting the stellar mass values within the nearest neighbors of each test galaxy. For better comparison, the blue, red, and green dotted lines indicate the three 50th percentiles (corresponding to the three galaxy populations) for the unconditional predictive efficiency. Based on the contrast between the conditional and unconditional predictive efficiency values, we mainly see that the contributions of $W1-W2$, $[b/a]_g$, $n_g/n_r$, and $R_{90,g}/R_{50,g}$ to the stellar mass estimation can be largely explained by spec-$z$, whereas the contributions of the other parameters such as $R_{50,g}/R_{50,r}$ for star-forming galaxies is essentially unexplained by spec-$z$. Furthermore, the contributions of all the morphological parameters cannot be fully explained by $W1-W2$, and vice versa.}
\label{fig:pred_efficiency_cond}
\end{center}
\end{figure*}

\section{Results on causal analysis} \label{sec:results_causal}

\subsection{Detection of external variables by local independence tests} \label{sec:res_ext_detect}

We present the results on the detection of external variables in $\mathbf{X}^{ex}$ for the five models $\mathbf{M}_{ugriz}$, $\mathbf{M}_{ugrizW123}$, $\mathbf{I}_{ugriz}$, $\mathbf{I}_{ugriz} \cup \mathbf{M}_{W123}$, and $\mathbf{I}_{ugriz} \cup \mathbf{M}_{W123} \cup z_{spec}$ defined in Table~\ref{tab:models_causal}, based on the long list of parameters in Table~\ref{tab:data}. Figure~\ref{fig:correlation_selected} illustrates the distributions of local correlations between stellar mass and a few representative parameters for the test sample. The correlation distributions for the full list of parameters are presented in Appendix~\ref{sec:more_corr}. Each correlation for a parameter is estimated using the nearest neighbors of a test galaxy. The distributions are separately shown as colored curves for the three galaxy populations from the test sample, namely, star-forming, passive, and other galaxies. We also randomly permuted the stellar mass values within each group of nearest neighbors to eliminate any possible dependences between stellar mass and other parameters, producing the correlation distributions shown in grey. These distributions are approximately Gaussian and centered at zero, whose spreads characterize the combined effects of intrinsic dispersions and noise in data, as well as the uncertainties introduced by the supervised contrastive learning and KNN implementation. They serve as a contrast for the original correlation distributions. In Fig.~\ref{fig:pred_efficiency_selected}, we further show the median predictive efficiency of the representative parameters over the three galaxy populations. Each predictive efficiency estimate is obtained using the nearest neighbors of a test galaxy as well.

The parameters that show significantly non-zero correlations with stellar mass or significantly non-zero predictive efficiency are external, which contain the information on stellar mass not encompassed in the input data of a certain model. We note that there may be undetected external variables for any model among the parameters investigated in our analysis, since all the parameters are measured or derived quantities whose noise would suppress the statistical significance of the correlation or the predictive efficiency. Despite this, there have been meaningful insights conveyed by these two metrics.

Firstly, for the optical magnitudes (exemplified by $g$-band magnitudes), there is good consistency between the original and reference correlation distributions for all the five models. The median predictive efficiency is also around the reference value $\sim 0.013$ estimated by randomly permuting the stellar mass values within the nearest neighbors of each test galaxy. This result is unsurprising since optical photometry is encompassed in the input data for every model.

In general, spec-$z$ and physical properties are the external variables for the models not involving spec-$z$. Spec-$z$ has the most significant dependence with stellar mass, and the physical properties generally show different levels of dependences, as indicated by the conspicuous deviations between the original and reference correlation distributions or the high predictive efficiency. This indicates that photometry or images in optical and infrared bands cannot provide complete information on spec-$z$ or physical properties. Adding spec-$z$ in the input data can remove the dependences between stellar mass and most physical properties including $M_{\ast}/L$, but there are still residual correlations with SFR, implying that spec-$z$ is a powerful predictor though not the only determinant of galaxy physical properties.

The contrast between the photometry-only models and the image-based models indicates that morphological features are external for the photometry-only models. In other words, integrated photometry in optical or infrared bands cannot convey morphological information as images do. Using images not only provides morphological information, but also reduces the dependences between stellar mass and spec-$z$ or other physical properties.

The contrast between the models with and without infrared photometry indicates that optical photometry or images cannot fully convey infrared information. Using infrared photometry also slightly reduces the dependences shown by spec-$z$ and the physical properties. However, it is interesting to notice that using optical images can greatly shrink the impact of infrared photometry on the stellar mass estimation even if optical bands do not fully cover infrared information, as implied by the much lower predictive efficiency $\epsilon(W1-W2)$ for the model $\mathbf{I}_{ugriz}$ compared to that for the model $\mathbf{M}_{ugriz}$. This is further confirmed in the aspect of mutual information (Sect.~\ref{sec:res_decompose1}).

For the photometry-only models, the external morphological features (i.e., S\'{e}rsic indices, inclinations, Petrosian radii, and their combinations) show disparate behaviors (illustrated more in Appendix~\ref{sec:more_corr}). The S\'{e}rsic indices $n$ do not have strong dependences with stellar mass in individual optical bands, while the cross-band ratio $n_g/n_r$ shows recognizable correlations. On the contrary, the inclinations $b/a$ in all the optical bands consistently show averagely positive correlations with stellar mass, while such dependences are canceled out for their cross-band ratios. The Petrosian radii $R_{50}$, $R_{90}$, and their multiple intra- and cross-band combinations generally have various dependence relations with stellar mass. These behaviors illustrate that the information on stellar mass conveyed by morphology is both intra- and cross-band, analogous to magnitudes and colors that may be both important for the stellar mass estimation. Various intra- and cross-band morphological parameters describe different aspects of the overall morphological information. Nonetheless, the contribution of a single morphological parameter to the stellar mass estimation is generally mild, as indicated by the low predictive efficiency. Among the shown morphological parameters, $R_{50,g} / R_{50,r}$ has the highest predictive efficiency.

It is also straightforward to identify disparate behaviors for different galaxy populations. For the morphological parameters, $n_g/n_r$, $[b/a]_g$, $R_{90,g}/R_{50,g}$, and $R_{50,g}/R_{50,r}$ seem to have stronger correlations with stellar mass or higher predictive efficiency for star-forming galaxies than for passive galaxies. On the contrary, the Petrosian radii such as $R_{50,g}$, $R_{50,r}$, $R_{90,g}$, and $R_{90,g}-R_{50,g}$ have stronger dependences with stellar mass for passive galaxies. $R_{90,g}/R_{50,g}$ and $R_{90,g}-R_{50,g}$, though similar, exhibit different dependence relations with stellar mass for star-forming and passive galaxies.
More discussions regarding these morphological parameters are presented in Sect.~\ref{sec:res_pred_efficiency_cond}. For the other parameters, we found distinctions between galaxy populations particularly in spec-$z$ and SFR, with the strongest dependences with stellar mass shown for star-forming galaxies.

Furthermore, it is possible to identify substructures in the correlation distributions for a galaxy population. For example, regarding passive galaxies, the correlations between stellar mass and spec-$z$ for the models not involving spec-$z$ may have more than one cluster. The correlations close to unity are mainly contributed by the passive galaxies that are relatively low-mass and located at low redshift, for which stellar mass has the strongest dependence on spec-$z$. Regarding star-forming galaxies, at least two groups can be detected in the correlations between stellar mass and $M_{\ast}/L$ for the model $\mathbf{M}_{ugriz}$, and between stellar mass and SFR for the model $\mathbf{I}_{ugriz} \cup \mathbf{M}_{W123} \cup z_{spec}$. In particular, low S/N star-forming galaxies (i.e., \texttt{bptclass=2}), having relatively red colors, high stellar mass and $M_{\ast}/L$, and located at relatively high redshift, mainly contribute to the peaks in the two correlation distributions, while other star-forming galaxies (i.e., \texttt{bptclass=1}), which may be more heterogeneous, contribute more widely spread correlations.

In summary, different groups of external variables have been found for the given models in our analysis. Morphological information that is related to stellar mass cannot be entirely accounted for by optical photometry. Infrared information cannot be entirely accounted for by optical photometry or images. Furthermore, spec-$z$ and physical information cannot be entirely accounted for by optical photometry, images, or infrared photometry, and spec-$z$ exhibits the strongest association with stellar mass. Both intra- and cross-band morphological features are contributory to the stellar mass estimation, though the contribution of a single morphological feature is generally insignificant (in the presence of photometry). These trends are not uniform for different galaxy populations or subsamples within a population.

\subsection{Causal structures between external variables and stellar mass revealed by conditional predictive efficiency} \label{sec:res_pred_efficiency_cond}

We identified meaningful causal structures using morphology, infrared photometry, and spec-$z$ for the photometry-only model $\mathbf{M}_{ugriz}$. These parameters are the detected external variables from Sect.~\ref{sec:res_ext_detect}, and the causal structures between these parameters and stellar mass would tell how they are causally linked to stellar mass and contribute in the stellar mass-predicting process complementary to optical photometry. Figure~\ref{fig:pred_efficiency_cond} illustrates the median conditional predictive efficiency of these variables for the model $\mathbf{M}_{ugriz}$ over the three galaxy populations in the test sample, compared to the unconditional predictive efficiency. The conditional predictive efficiency of every shown parameter is conditioned on every other parameter, estimated using the nearest neighbors of each test galaxy. In order to remove possible residual correlations existing between stellar mass and optical colors, all the parameters including stellar mass are first quadratically regressed on the $g-r$ color of the nearest neighbors of each test galaxy before computing the (conditional) predictive efficiency. Therefore, the unconditional predictive efficiency shown in Fig.~\ref{fig:pred_efficiency_cond} is slightly different from that in Fig.~\ref{fig:pred_efficiency_selected}. The results based on the correlation metric for the same parameters are presented in Appendix~\ref{sec:more_cond_corr}.

We found that the conditional predictive efficiency of $W1-W2$, $[b/a]_g$, $n_g/n_r$, and $R_{90,g}/R_{50,g}$ drops to low values once conditioned on spec-$z$. For star-forming galaxies, the median values decrease by 94.4\%, 68.0\%, 83.6\%, and 56.7\%, respectively, with reference to the unconditional cases (subtracting 0.013). This implies that the contribution of each of these parameters to the stellar mass estimation can be largely explained by spec-$z$, or largely ascribed to the strong dependence between spec-$z$ and stellar mass, as described by $X^{ex}_2 \noarrow X^{ex}_1 \rightarrow Y$ in Fig.~\ref{fig:method} where $X^{ex}_1$ represents spec-$z$ and $X^{ex}_2$ represents the other variables that have links to $Y$ through spec-$z$. On the other hand, all the shown morphological parameters have contributions to the stellar mass estimation largely unexplained by $W1-W2$, and vice versa, as described by $X^{ex}_3 \rightarrow Y \leftarrow X^{ex}_1$ in Fig.~\ref{fig:method} where $X^{ex}_1$ and $X^{ex}_3$ have separate links to $Y$. This means that combining infrared photometry with optical photometry would better constrain the SEDs and thus convey spec-$z$ information, but it cannot fully cover or be covered by optical morphological information.

The contributions of $n_g/n_r$ and $R_{90,g}/R_{50,g}$ for star-forming galaxies seem to be related, as indicated by the dropping conditional predictive efficiency when conditioned on each other. However, they do not seem to strongly overlap with the contribution of $[b/a]_g$. We speculate that the dependences between $n_g/n_r$, $R_{90,g}/R_{50,g}$, and spec-$z$ for star-forming galaxies may be mainly due to observational effects. Compared to galaxies at low redshift, high-redshift galaxies are heavily affected by the PSFs as they extend over fewer pixels on images. The PSF effect would lead to an overestimation of $R_{50,g}$ and make $R_{90,g}/R_{50,g}$ small for high-redshift galaxies, producing a trend between spec-$z$ and $R_{90,g}/R_{50,g}$. In the mean time, the PSFs in the $g$ band are on average wider than those in the $r$ band (whose median FWHMs are 1.25 and 1.13 arcsec, respectively), making $n_g$ more underestimated than $n_r$ for high-redshift galaxies, thus producing a trend between spec-$z$ and $n_g/n_r$. The PSF effect is more significant for star-forming galaxies than for passive galaxies in our sample, because star-forming galaxies at high redshift tend to be smaller. Therefore, for star-forming galaxies, $n_g/n_r$ and $R_{90,g}/R_{50,g}$ are mutually dependent and both dependent on spec-$z$. The dependence between $[b/a]_g$ and spec-$z$ for star-forming galaxies may be due to mixed effects. Since highly inclined galaxies cannot be well resolved at high redshift, small $[b/a]_g$ values tend to be associated with low spec-$z$. In addition, there are more passive galaxies at high redshift that have large $[b/a]_g$ values than other populations in our sample, enhancing the trend between $[b/a]_g$ and spec-$z$ captured by the nearest neighbors of each test galaxy. Another complication is that there may be a systematic bias in the stellar mass values estimated by \citet{Kauffmann2003}, as suggested by \citet{Maller2009}. No matter what the effects are, the dependence relation between $[b/a]_g$ and spec-$z$ appears to be different from those for $n_g/n_r$ and $R_{90,g}/R_{50,g}$. We point out that the amount of spec-$z$ information that can be recovered by a single morphological feature is generally insignificant, as implied by the almost unchanged conditional predictive efficiency of spec-$z$.

Meanwhile, the contribution of $R_{50,g}/R_{50,r}$ is essentially not covered by spec-$z$ for star-forming galaxies, with the median conditional predictive efficiency dropping by only 4.9\%. This reflects that the $g-r$ color gradients represented by $R_{50,g}/R_{50,r}$ may convey physical information unexplained by spec-$z$. The contribution of $R_{50,g}/R_{50,r}$ seems to be separate from those of all the other morphological parameters as well. However, $R_{50,g}/R_{50,r}$ has a large portion of contribution explained by spec-$z$ for passive galaxies, with the median conditional predictive efficiency dropping by 59.7\%. In addition, the contributions of $R_{90,g}-R_{50,g}$ and $R_{50,r}$ may be related (though with low statistical significance) but cannot be explained by spec-$z$, implying that $R_{90,g}-R_{50,g}$ may still essentially characterize the observed galaxy size, but the expected dependence between the average galaxy size and spec-$z$ has already been largely absorbed by optical photometry.

With these findings, we demonstrate that our causal analysis approach makes it efficient and straightforward to reveal data structures and provide insights for interpreting image-based models. In specific, we found that multiple intra- and cross-band morphological features may make use of the dependence between stellar mass and spec-$z$, or contribute to the stellar mass estimation in a way unexplained by spec-$z$, through either systematic or physical effects. These features characterize various aspects of morphology that are all encoded in images, thus image-based models can exploit the aforementioned trends to predict stellar mass.

\begin{figure*}
\begin{center}
\centerline{\includegraphics[width=1.0\linewidth]{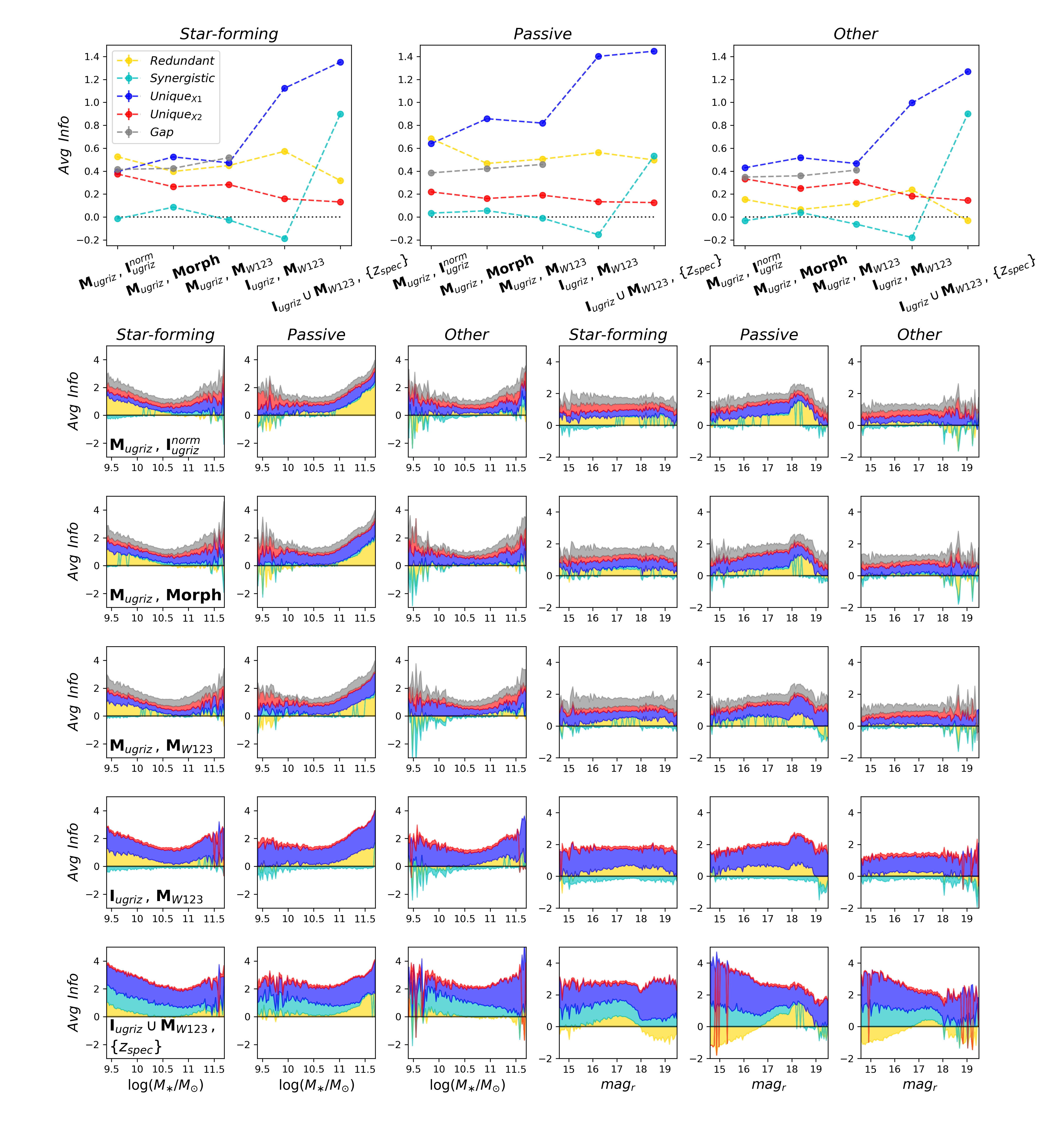}}
\caption{Mutual information decomposition for analyzing the contributions of different data modalities to the stellar mass estimation, involving photometry, morphology, images, and spec-$z$ (defined in Table~\ref{tab:info_modalities}). The redundant, unique, and synergistic components are all illustrated. In each tuple, the dataset labeled on the left is referred to as $X_1$, and the one on the right is referred to as $X_2$, distinguished for the unique information (shown in blue and red, respectively). \textit{First row:} Average information level of each component (in units of nats) for the five cases, separately shown for star-forming, passive, and other galaxies from the test sample. The information ``gap'' between each of the first three cases and $\mathbf{I}_{ugriz}$ is also illustrated, indicating the loss of synergy when separating photometry and morphology (i.e., <$\mathbf{M}_{ugriz} \,,\, \mathbf{I}^{norm}_{ugriz}$> and <$\mathbf{M}_{ugriz} \,,\, \mathbf{Morph}$>), or the outperformance of optical images over optical and infrared photometry combined (i.e., <$\mathbf{M}_{ugriz} \,,\, \mathbf{M}_{W123}$>). Furthermore, negative information is allowed as an indication of misinformation, which, particularly shown in the case <$\mathbf{I}_{ugriz} \,,\, \mathbf{M}_{W123}$>, shrinks the incremental contribution of infrared photometry in the presence of optical images. The black dotted lines indicate zero mutual information. \textit{Remaining rows:} Stack plots of different information components (in units of nats) as a function of stellar mass or $r$-band magnitude for the five cases, separately shown for star-forming, passive, and other galaxies from the test sample. The shown windows exclude the tails of the stellar mass and $r$-band magnitude distributions since they are heavily affected by low sample statistics. The vertical extents of the stacked areas above and below zero indicate the total amounts of positive and negative information, respectively, in any stellar mass or $r$-band magnitude bin. The varying information levels in the stack plots reflect the effects of data imbalance.}
\label{fig:info_globmean_mag_img_morph}
\end{center}
\end{figure*}

\begin{figure*}
\begin{center}
\centerline{\includegraphics[width=1.0\linewidth]{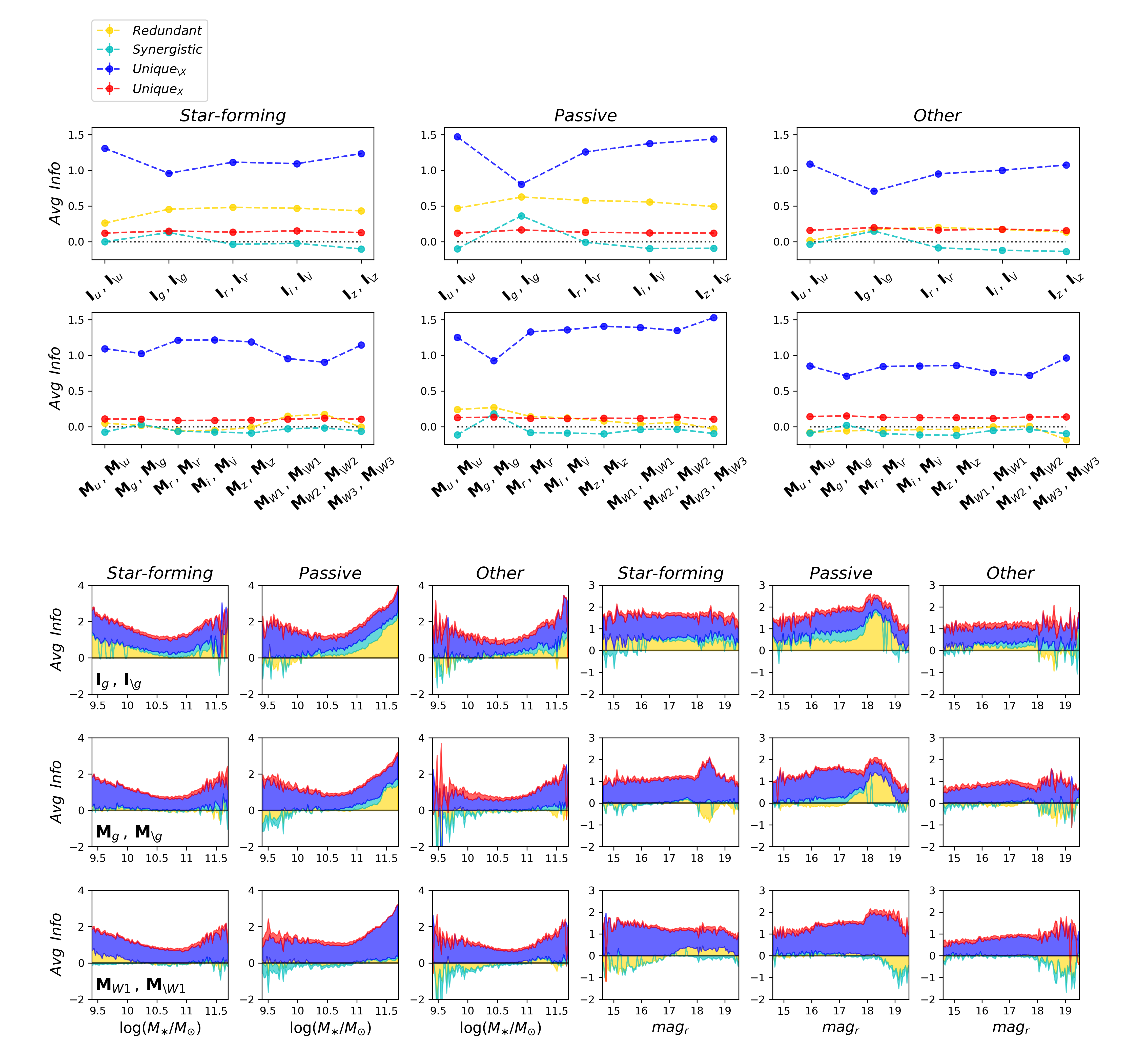}}
\caption{Mutual information decomposition for analyzing the contributions of different photometric bands to the stellar mass estimation (defined in Table~\ref{tab:info_bands}). The redundant, unique, and synergistic components are all illustrated. $\mathbf{I}$ ($\mathbf{M}$) refers to the datasets that contain images (photometry). The label $X$ refers to a single band that is separated out, and $\backslash X$ refers to the remaining bands, distinguished for the unique information (shown in red and blue, respectively). \textit{First row:} Average information level of each component (in units of nats) for the image-based cases in which $X$ runs over all the optical bands, separately shown for star-forming, passive, and other galaxies from the test sample. The black dotted lines indicate zero mutual information. \textit{Second row:} Same as the first row, but for the photometry-only cases in which $X$ runs over all the optical and infrared bands. No sharp contrast can be seen between different bands, except that the $g$ band has the largest incremental contribution in the presence of the other bands. The single-band images typically have larger contributions than the single-band photometry. \textit{Remaining rows:} Stack plots of different information components (in units of nats) as a function of stellar mass or $r$-band magnitude for the cases in which the $g$ band or the $W1$ band is separated out, shown for star-forming, passive, and other galaxies from the test sample. Similar to Fig.~\ref{fig:info_globmean_mag_img_morph}, the stack plots reveal the behaviors of data imbalance, which are not the same for optical and infrared bands.}
\label{fig:info_globmean_gw1drop}
\end{center}
\end{figure*}

\begin{figure*}
\begin{center}
\centerline{\includegraphics[width=1.0\linewidth]{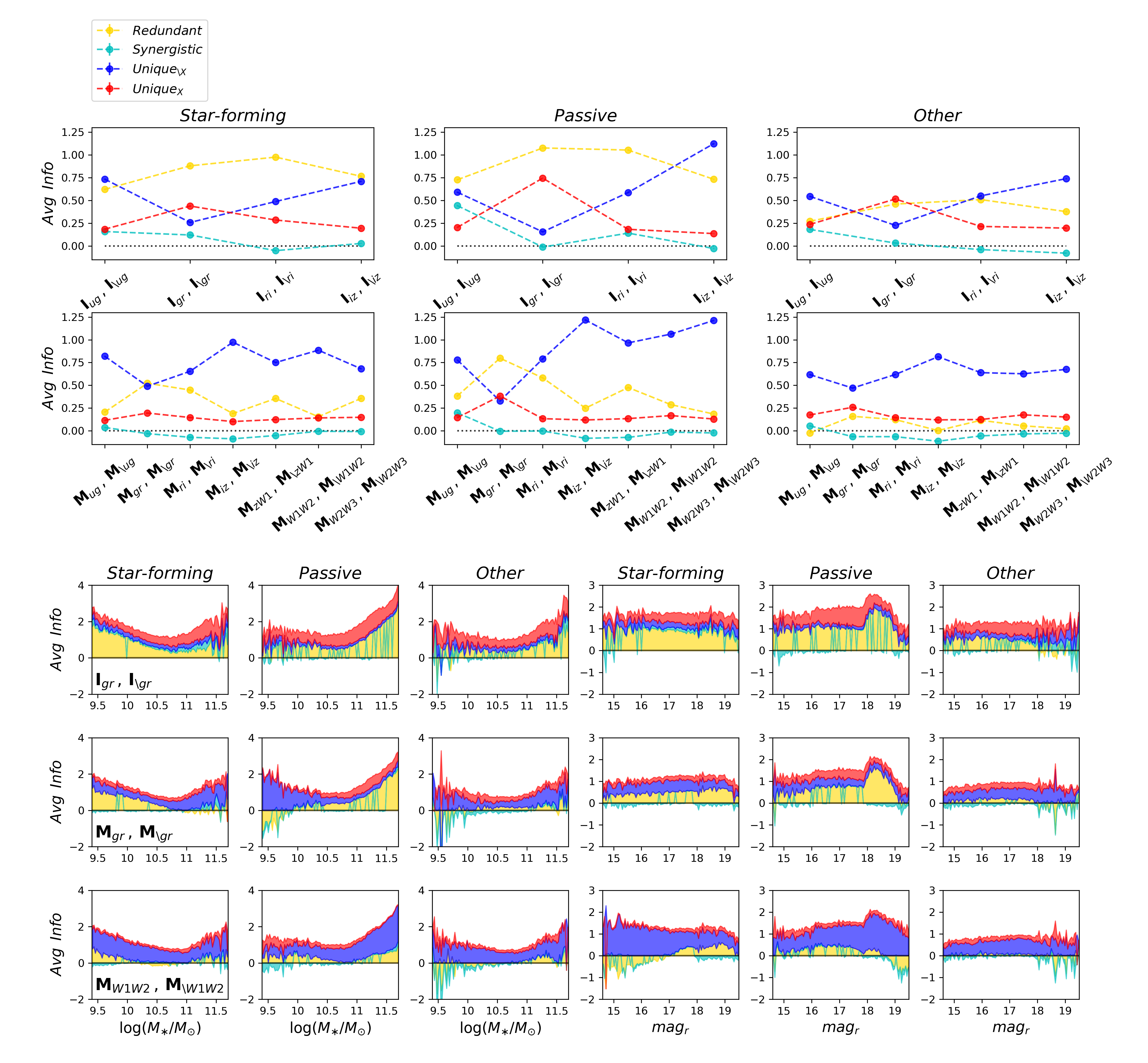}}
\caption{Same as Fig.~\ref{fig:info_globmean_gw1drop}, but with two adjacent bands separated out in each case. Compared to the cases with one single band, more information can generally be provided by two bands, in particular $g$-band and $r$-band images that have the strongest synergistic effect. The behaviors of data imbalance are revealed in the stack plots that are illustrated for separating out both the $g$ and $r$ bands, or both the $W1$ and $W2$ bands.}
\label{fig:info_globmean_grw1w2drop}
\end{center}
\end{figure*}

\section{Results on mutual information decomposition} \label{sec:results_info}

\subsection{Decomposing contributions of photometry, morphology, images, and redshift} \label{sec:res_decompose1}

Using the mutual information decomposition, we quantified the multicomponent contributions of photometry, morphology, images, and spec-$z$ to the stellar mass estimation. Figure~\ref{fig:info_globmean_mag_img_morph} illustrates the different information components for the five cases defined in Table~\ref{tab:info_modalities} that involve different data modalities, i.e., <$\mathbf{M}_{ugriz} \,,\, \mathbf{I}^{norm}_{ugriz}$>, <$\mathbf{M}_{ugriz} \,,\, \mathbf{Morph}$>, <$\mathbf{M}_{ugriz} \,,\, \mathbf{M}_{W123}$>, <$\mathbf{I}_{ugriz} \,,\, \mathbf{M}_{W123}$>, and <$\mathbf{I}_{ugriz} \cup \mathbf{M}_{W123} \,,\, \{z_{spec}\}$>. We also show the information ``gap'' between each of the first three cases and $\mathbf{I}_{ugriz}$ in the figure, quantifying the amount of missing mutual information with reference to that provided by the original optical images. We reiterate that our mutual information estimates are empirical rather than theoretical, characterizing the effectiveness of certain input data for the practical purpose of stellar mass estimation. We note that empirically measured negative information is allowed to exist as an indication of misinformation, which may be due to various factors such as noise in data, data imbalance, and estimation uncertainties introduced by the use of neural networks. We have checked that the uncertainties due to network implementation are negligible.

Firstly, it is noteworthy that the case <$\mathbf{M}_{ugriz} \,,\, \mathbf{I}^{norm}_{ugriz}$> cannot provide the total amount of mutual information given by $\mathbf{I}_{ugriz}$, implying that the separation of photometry and morphology results in information loss. In fact, the magnitudes and the normalized images in <$\mathbf{M}_{ugriz} \,,\, \mathbf{I}^{norm}_{ugriz}$> can be recombined to produce the original images in $\mathbf{I}_{ugriz}$. However, extracting photometry from images or measuring morphology would come along with intrinsic dispersions and measurement noise, making it difficult for neural networks to recognize the information gain from recombining photometry and morphology. Equivalently speaking, there is intrinsic synergy between photometry and morphology, but having them separately measured would empirically result in misinformation that offsets their synergistic information. This leads to a consequence that $\mathbf{M}_{ugriz}$ and $\mathbf{I}^{norm}_{ugriz}$ do not have apparently significant synergistic contributions to the stellar mass estimation. In addition, it can be seen that $\mathbf{M}_{ugriz}$ and $\mathbf{I}^{norm}_{ugriz}$ provide comparable redundant and unique information, meaning that the contribution of morphology alone may be considerable, though a large portion overlaps with the contribution of photometry.

Comparing the two cases <$\mathbf{M}_{ugriz} \,,\, \mathbf{I}^{norm}_{ugriz}$> and <$\mathbf{M}_{ugriz} \,,\, \mathbf{Morph}$>, we found that the five-band S\'{e}rsic indices $n$, inclinations $b/a$, and Petrosian radii $R_{50}$ and $R_{90}$ in $\mathbf{Morph}$ can make a fairly considerable contribution, though less significant than that by $\mathbf{I}^{norm}_{ugriz}$. This means that an essential portion of the overall morphological information useful for the stellar mass estimation can be conveyed by these parameters, despite that their individual contributions are generally insignificant. Nonetheless, $\mathbf{Morph}$ cannot recover the lost synergistic information as well. Based on these findings, we suggest that for tasks such as the stellar mass estimation that may rely on both photometric and morphological features, it would be preferable to use images each as a whole rather than a concatenation of photometry and morphology, because images can capture the intrinsic synergistic effect between photometry and morphology that may be unavailable when they are separate.

Regarding the two cases <$\mathbf{M}_{ugriz} \,,\, \mathbf{M}_{W123}$> and <$\mathbf{I}_{ugriz} \,,\, \mathbf{M}_{W123}$>, it is noticeable that $\mathbf{I}_{ugriz}$ produces a boost of unique information in comparison with the case with $\mathbf{M}_{ugriz}$. Even $\mathbf{M}_{ugriz}$ and $\mathbf{M}_{W123}$ combined cannot reach the level of information provided by $\mathbf{I}_{ugriz}$ alone, suggesting that use of images that contain information not limited to photometry may outperform the use of photometry over a broader band coverage. More interestingly, we found that the average synergistic information is negative in these two cases. The negative synergistic information comes from a portion of data instances for which the inclusion of both optical bands and infrared photometry does not improve but instead degrade the mutual information estimates compared to using only one input set. This may result from noise in infrared photometry or a change in the input data distribution due to the involvement of infrared photometry. In particular, for <$\mathbf{I}_{ugriz} \,,\, \mathbf{M}_{W123}$>, the negative synergistic information compensates the unique information provided by infrared photometry, leading to a result that the use of infrared photometry in addition to optical images has negligible improvement on the stellar mass estimation. This result is consistent with the finding by the predictive efficiency analysis in Sect.~\ref{sec:res_ext_detect}. In other words, for the SDSS and WISE data in our analysis, the use of optical images essentially relieves the requirement for infrared photometry even if the optical bands available do not fully cover infrared information. This marks another bonus of using images.

The case <$\mathbf{I}_{ugriz} \cup \mathbf{M}_{W123} \,,\, \{z_{spec}\}$> reveals that spec-$z$ contributes to the stellar mass estimation mainly in synergy with other input data. This is expected because spec-$z$ and physical properties jointly affect observed photometric data.

Furthermore, we point out that mutual information decomposition has the power to reveal interesting data structures and distributions. For example, the stacked plots in Fig.~\ref{fig:info_globmean_mag_img_morph} illustrate how different information components evolve with stellar mass and $r$-band magnitude. The U-shaped stacked areas as a function of stellar mass reflect the imbalance of the stellar mass distributions shown in Fig.~\ref{fig:logm_r_z_dist}. In other words, more information on stellar mass can be provided for the data instances on the two mass ends, since their instance-wise entropy is large as a result of low number densities; while the galaxies with medium masses have a strong prior due to high number densities and dominate the average information level over the sample. For the low-mass end, the redundant information for star-forming galaxies appears to be more than that for passive galaxies. This is because the low-mass end is populated with more star-forming galaxies than passive galaxies, where passive galaxies would require more informative input data to counteract the severe imbalance of the prior distribution and thus it is more difficult for both the two input sets in a tuple to have large and overlapping contributions. The trend is reversed for the high-mass end, which is populated with more passive galaxies than star-forming galaxies. In particular, the bump of redundant information at $r \sim 18.5$ for passive galaxies is contributed by those with the highest masses and at high redshift, whose masses are easier to predict relative to lower masses. We also notice that negative redundant information is produced by spec-$z$ at both low and high $r$-band magnitudes. This is due to data imbalance in magnitude, as the use of spec-$z$ alone introduces misinformation to dim and bright galaxies that are less frequent than medium-magnitude galaxies. Finally, optical bands are most informative for passive galaxies, signifying the data imbalance among galaxy populations.

\subsection{Decomposing contributions of different photometric bands} \label{sec:res_decompose2}

In Figs.~\ref{fig:info_globmean_gw1drop} and \ref{fig:info_globmean_grw1w2drop}, we illustrate the decomposed contributions of different photometric bands to the stellar mass estimation, by separating out one band or two adjacent bands each time (defined in Table~\ref{tab:info_bands}). The image-based and photometry-only cases are both shown.

When one band is separated out, we have found that the $g$ band has the largest incremental contribution among all the bands regardless of photometry or images, as indicated by an increase in the synergistic information and a drop in the unique information provided by the other bands, which is most significant for passive galaxies. For star-forming galaxies, there is a hint that the $W1$ and $W2$ bands have larger incremental contributions in addition to the $g$ band than the other bands in the photometry-only cases. The use of images enlarges the individual contributions of single bands, as indicated by an increase in redundant information. In particular, for passive galaxies, $g$-band images alone can provide around 40\% of the total amount of mutual information given by images in the five optical bands. Besides these differences, there is no sharp contrast between different bands. The unique information provided by any single band remains at a low level even for the $g$ band, as its contribution is essentially covered by the other bands. The low redundant information in the photometry-only cases further suggests that the individual contribution of single-band photometry is rather insignificant. Furthermore, there is insignificant and even negative synergistic information except for the cases in which the $g$ band is separated out. In fact, the inclusion of any single band except the $g$ band would not produce an incremental contribution over 0.1 nats in the presence of other bands, regardless of photometry or images. This in turn indicates that excluding a single band except the $g$ band would not heavily impact the prediction of stellar mass.

When two adjacent bands are separated out, we have found that two bands can generally provide more information than one band does, which is especially conspicuous in the image-based cases. This implies that a considerable amount of non-overlapping information may be conveyed by images from a second optical band in the presence of images from one band, though the incremental contribution of one single band would generally decrease when more bands are involved. In particular, the combination of the $g$ and $r$ bands has a noticeable difference from the other pairs of adjacent bands. Using images, the unique information provided by the $g$ and $r$ bands is even more than that provided by the remaining bands. This is mainly due to the fact that the strong synergistic effect between the $g$ band and other bands as shown in Fig.~\ref{fig:info_globmean_gw1drop} can be essentially absorbed by the $r$ band. In other words, the synergistic information between $\mathbf{I}_{X}$ and $\mathbf{I}_{\backslash X}$ shrinks when the $g$ and $r$ bands are grouped together in $\mathbf{I}_{X}$, in contrast to the case in which the $g$ and $u$ bands are grouped, for example. Such synergy may be due to the $4000 \r{A}$ break feature located at the $g$ band, and the cross-band morphological features that are related to stellar mass seen in Sects.~\ref{sec:res_ext_detect} and \ref{sec:res_pred_efficiency_cond}. The prominence of the $g$ band would also be owing to its relatively good S/N with which there is not too much misinformation to offset useful signal. In contrast, the grouping of any two adjacent bands not involving the $g$ band would produce a less significant incremental contribution in the presence of other bands, due to a low level of unique and synergistic information. These results highlight the uniqueness of the $g$ band for the stellar mass estimation with the SDSS data.

The stack plots in Figs.~\ref{fig:info_globmean_gw1drop} and \ref{fig:info_globmean_grw1w2drop} generally reveal the behaviors of data imbalance discussed in Sect.~\ref{sec:res_decompose1}. Different behaviors can be found by comparing the cases in which optical bands or infrared bands are separated out. For star-forming galaxies, the redundant information at the low-mass end almost vanishes when separating out $g$-band photometry, but remains positive when separating out $W1$-band photometry. This behavior is dominated by a group of low-mass dim star-forming galaxies, for which $W1$ or $W2$-band photometry is more informative than single-band optical photometry in predicting stellar mass, but at a cost of biasing the predictions for high-mass dim galaxies. The negative redundant information at low magnitudes when separating out $W1$-band and $W2$-band photometry is dominated by a group of low-mass bright star-forming galaxies, for which the predictions by $W1$-band and $W2$-band photometry are biased due to the influence of high-mass bright galaxies. For passive galaxies, when separating out $W1$-band and $W2$-band photometry, negative redundant information appears at high magnitudes, and the bump of redundant information at $r \sim 18.5$ that is present when separating out the $g$ and $r$ bands almost vanishes. This is due to the overall biases for high-mass dim galaxies when using $W1$-band and $W2$-band photometry. Such biases conversely improve the predictions for low-mass dim galaxies, thus contributing to the positive redundant information at low masses when separating out both the $W1$ and $W2$ bands. All the results above illustrate that the outcome of the mutual information decomposition heavily relies the data distributions, but it in turn can be used as a sensitive probe to analyze the structures within data. We present the stack plots for all the image-based and photometry-only cases in Appendix~\ref{sec:more_decompose2}.

\section{Comparison with other studies} \label{sec:comparison}

We compared our work with a few other representative studies on the estimation of physical properties or photo-$z$. Regarding the contributions of galaxy morphology, \citet{Way2009}, \citet{Singal2011}, and \citet{JonesSingal2017} suggested that the inclusion of several morphological parameters does not result in any statistically significant improvement on the photo-$z$ estimation, in which the measurement noise associated with the morphological parameters may play a negative role. We note that these studies analyzed the impacts of morphological parameters based on the reduction of estimation errors. Indeed, such dispersion-based metric directly indicates how much influence that certain variables can have on the outcome. Nonetheless, the correlation may be better suited for probing mild associations between the target and input parameters in the presence of noise \citep{TingWeinberg2022}. Furthermore, our results suggest that intra- and cross-band features are both contributory, whereas these studies only explored intra-band features. In fact, we found that only a group of morphological parameters that characterize various aspects of morphology may have considerable cumulative impact on the outcome.

There are other studies that gave supports for the usefulness of galaxy morphology. \citet{Soo2018} and \citet{Dobbels2019} found that morphology would noticeably improve the estimation of photo-$z$ or $M_{\ast}/L$ when photometry is suboptimal or partially lacking, while the improvement becomes insubstantial given high-quality photometry and sufficient band coverage. \citet{WuBoada2019}, \citet{BuckWolf2021}, and \citet{Zhong2024} found that for the image-based estimation of physical properties such as stellar mass, SFR, sSFR, and metallicity, there is generally a descending trend in the model performance with downgrading image resolution, implying that morphological information conveyed by images is contributory to the estimation of physical properties. These results are in line with our finding that photometry and morphology may provide comparable amount of information individually but with a large redundancy. The information lost due to the degradation of one input can be partially recovered by the other. Moving beyond this, our work points out the intrinsic synergy between photometry and morphology.

Regarding the contributions of different photometric bands, \citet{Zhong2024} found that the use of $g$-band images leads to the best performance in predicting stellar mass and sSFR among three single optical bands $g,r,z$, and adding a second band would further improve the results. Using feature importance, \citet{Hoyle2015} identified the $g$-band fiber magnitude as the most important feature for the photo-$z$ estimation among different definitions of magnitudes in optical bands. Similarly, \citet{DelliVeneri2019} analyzed the importance of different definitions of magnitudes, and found that the colors involving the $g$ band are among the most important features for estimating SFR, while other bands (e.g., the $z$ band) are less important. These findings are broadly consistent with ours in implying the important role of the $g$ band and its synergy with other bands for the galaxies observed by the SDSS.

Feature importance has also been adopted in other studies. \citet{Acquaviva2016} analyzed the importance of optical magnitudes and colors (as well as squared colors) for estimating metallicity, while stellar mass was also included in the input data and it dominates over all the photometric features. \citet{Bonjean2019} analyzed the importance of luminosities (i.e., $W1$-band and $W3$-band) and colors (i.e., $W1-W2$ and $W2-W3$) in the WISE infrared bands, finding that their contributions are divergent when estimating stellar mass, SFR, and the two properties together. \citet{Zeraatgari2024} included both the SDSS and WISE data, and found that $W2-W3$ and $W1-W2$ are most important for estimating stellar mass and SFR together, while $g-r$ is most important for estimating metallicity.

In general, we caution that feature importance may not be able to show the actual contribution of each input feature to the target due to the existence of co-linearity or dependences between features, as also noted by \citet{Euclid_Humphrey2023} and \citet{Lu2024}. Moving forward, we note that feature importance has difficulties in revealing multivariate data structures and multiple information components (especially the synergistic information). For example, \citet{Acquaviva2016}, \citet{Bonjean2019}, and \citet{DelliVeneri2019} included spec-$z$ in the analysis, but did not discuss the possible synergy between spec-$z$ and photometry. Similarly, it is difficult to tell if there is any synergy between any two bands in these studies. Additionally, \citet{Bonjean2019} found that both $W1-W2$ and spec-$z$ exhibit almost no importance, and suggested that the contribution of spec-$z$ may be absorbed by $W1$-band and $W3$-band luminosities. According to our results, we may further suggest that the contribution of $W1-W2$ would also be absorbed by the luminosities, since we found that $W1-W2$ has no significant effect unexplained by spec-$z$.

Finally, we notice a relevant study by \citet{DIsanto2018}, who proposed a greedy forward approach to select features useful for the photo-$z$ estimation and offered a physical interpretation for the selected features. Nonetheless, this approach requires re-implementing KNN to estimate the target for every combination of input features, which is computationally costly. In this regard, our causal analysis method is much more efficient in that it requires training a model only once and then any features can be tested.

\section{Discussion} \label{sec:discussions}

We discuss the limitations of this work and suggest further investigations in future endeavors. Firstly, both the causal analysis and the mutual information decomposition are data-dependent, meaning that our results may be confined to the certain data used in our analysis. In particular, although we found that the SDSS optical images can essentially make the WISE infrared photometry noncontributory to the stellar mass estimation, we do not generalize this conclusion to other data. In fact, we anticipate that the infrared data produced by upcoming surveys such as \textit{Euclid} and \textit{Roman} would provide more physical information and establish better synergy with optical data. More generally, a good data prior is the prerequisite for obtaining robust physical interpretations, involving various factors such as data coverage, data quality, and the balance of data distributions. Future surveys will produce rich data and make it possible to test how the results from causal and information-theoretical analysis would vary with different data priors.

Secondly, the list of parameters used in the causal analysis is non-exhaustive, thus there may exist undiscovered variables, causal structures, and physical effects that are relevant to stellar mass. Our results may be improved and enriched by using a more extensive list of parameters (e.g., including environmental properties). Furthermore, unobserved confounders that involve physical and systematic effects may introduce spurious associations between the known variables (e.g., the dependence between the inclination $b/a$ and spec-$z$). In this work, we only discussed the possible effects behind the effective relations between variables, and did not attempt to explicitly identify unobserved confounders. However, for robust establishment of causal structures and rigorous physical interpretations, more work would be required to differentiate confounding effects from direct causal relations.

Finally, all the results in our work may be potentially impacted by the implementations of deep learning models. In general, different deep learning-based techniques and implementations may produce inconsistent results even if they are conducted on the same data, or cannot robustly reveal the trends in data. Such problems may be due to a lack of model expressivity, insufficient training, or imperfect modeling of data distributions. We have assessed all the model implementations in this work and found no inconsistency in our qualitative causal analysis and mutual information decomposition. Nonetheless, our current methods may be refined (e.g., regarding the model expressivity, the modeling and estimation of probabilities, and the embedding of information from data) in order to perform rigorous quantitative analysis of the dependence relations within data.

\section{Conclusion} \label{sec:conclusion}

In this work, we applied the causal analysis and the mutual information decomposition to interpret the deep learning-based estimation of galaxy stellar mass, exploring the intersection of deep learning, causal learning, and information theory that breaks through the limitation of pure statistical associations in a data-driven manner. In particular, we focused on what information can be provided by galaxy images in addition to integrated photometry, attempting to dig into the mechanisms behind image-based stellar mass estimation models.

Using optical galaxy images and catalog data from the SDSS and infrared photometry from the WISE, we established a causal graph that describes the stellar mass-predicting process of end-to-end deep learning models, leveraging a framework that involves supervised contrastive learning and KNN procedures. With this approach, for a few photometry-only and image-based models, we were able to detect and compare the external variables that contain the information on stellar mass not accounted for by the model inputs. We then uncovered meaningful causal structures between the external variables and stellar mass for a photometry-only model, which offer an indication of how the variables encompassed in images in addition to integrated photometry build causal links to the target variable (i.e., stellar mass). We also performed the mutual information decomposition to give a quantification of the multicomponent contributions of different input datasets to the stellar mass estimation. The redundant, unique, and synergistic components produced by the decomposition enable us to investigate how the available information on stellar mass is distributed across multiple input sets and how these data interact when making stellar mass predictions. Our main results and their implications are summarized below.

\begin{itemize}
    \item We confirm the general conclusion from other works that morphological features contained within images can improve the estimation of galaxy physical properties complementary to integrated photometry. Our work has advanced our knowledge in this area by obtaining insightful causal structures between morphological features, stellar mass, and other galaxy properties that are not covered by optical photometry. As morphological information is encoded in images, all the causal structures involving morphological features can be captured by image-based models. We analyzed a few morphological parameters from the SDSS catalog that may contribute to the stellar mass estimation, including S\'{e}rsic indices $n$, inclinations $b/a$, and Petrosian radii $R_{50}$ and $R_{90}$ in the five optical bands, as well as their intra- and cross-band combinations. Both intra- and cross-band features have proven to be important. For star-forming galaxies, a few features such as $[b/a]_g$, $n_g/n_r$, and $R_{90,g}/R_{50,g}$ were found to be partial indicators of spec-$z$, a property that has the strongest association with stellar mass; other features such as $R_{50,g}/R_{50,r}$ provide the information on stellar mass in a way that cannot be substantially explained via spec-$z$. For passive galaxies, on the contrary, the contribution of $R_{50,g}/R_{50,r}$ can be largely explained by spec-$z$. These trends may originate from disparate systematic and physical effects. The meaningful causal structures between morphological features and stellar mass offer insights for interpreting image-based models.
    \item The aforementioned morphological parameters can convey an essential portion of the overall morphological information useful for the stellar mass estimation. Although the contribution of one single morphological feature is generally insignificant (in the presence of photometry), the overall contribution of morphology may be considerable, and even comparable to the contribution of photometry.
    \item Despite the considerable individual contributions derived from photometry and morphology on their own, we highlight the fact that images can empirically provide even more information than a simple concatenation of these two parts. This is because having them separately measured would introduce intrinsic dispersions and measurement errors, thereby diluting their intrinsic synergistic information. Therefore, it would be preferable to use images each as a whole for tasks that require both photometric and morphological features. An exhaustive collection of photometric and morphological parameters might still not perform comparably with the use of images.
    \item For the SDSS and WISE data in our analysis, the use of optical images can essentially relieve the requirement for infrared photometry. Infrared photometry would no longer make any recognizable incremental contribution to the stellar mass estimation once optical images are used, even if the $W1 - W2$ color provides extra spec-$z$ information not covered by the optical bands available. This is due to the misinformation (i.e., negative synergistic information) introduced by the inclusion of both infrared photometry and optical images that offsets the unique contribution of infrared photometry. The misinformation may be caused by noise in infrared photometry or a change in the input data distribution due to the involvement of infrared photometry. Notably, a model that uses optical images can also outperform a photometry-only model that uses both optical and infrared photometry.
    \item We noted a substantial information gain when using images in at least two optical bands, compared to using images in a single band only, suggesting that there is a boon to using multi-band images. Particularly for the SDSS data, $g$-band images strongly synergize with other bands. In tandem with $r$-band images, they are even more predictive for stellar mass than the other bands combined. This is in line with the importance of cross-band features such as $n_g/n_r$ and $R_{50,g}/R_{50,r}$, indicating the synergy between the two bands. On the contrary, all the bands other than the $g$ band do not exhibit such strong synergy. Their incremental contributions in the presence of other bands would be trivial if they do not provide a high level of unique information. Although the extra information provided by a single band could be significant when there is only one other band available, the incremental contribution would largely decline when more bands are involved. Therefore, compared to excluding the $g$ band, excluding one or two other bands would generally have a weaker impact when multiple bands are used. These results, as well as the discussions on infrared photometry, suggest that sufficient and optimized exploitation of information contained in images in a few bands may greatly reduces the necessity of using data over a broader span of bands, especially for data analysis applications faced with limited band coverage.
    \item The inclusion of spec-$z$ with photometric data produces a large amount of synergistic information, and predicts the dependences between stellar mass and most physical properties, demonstrating the predictive power of spec-$z$ in synergy with photometric data. However, the spec-$z$ data are highly insufficient in future astronomical surveys. Furthermore, we note that photo-$z$, measured with photometric data, would not provide significant synergistic information with photometric data themselves, thus it cannot act as an ideal substitute for spec-$z$ in the estimation of physical properties. Therefore, properly leveraging photometric data is still crucial.
    \item More meaningful data structures and distributions have been uncovered in our analysis. Star-forming galaxies have been found to have stronger associations between stellar mass and morphological parameters such as $[b/a]_g$, $n_g/n_r$, $R_{90,g}/R_{50,g}$, and $R_{50,g}/R_{50,r}$, as well as spec-$z$ and SFR, compared to passive galaxies. This indicates that different galaxy populations or subsamples of data could exhibit disparate behaviors. Furthermore, as a result of data imbalance, the galaxies in an overpopulated parameter space generally have a strong prior and dominate the average information level over the whole sample; whereas for those in an underpopulated parameter space, the input data would generally provide more information on stellar mass due to their large entropy (e.g., for low and high-mass ends), or conversely, result in negative mutual information (e.g., with spec-$z$ or $W1$-band photometry as input data at low or high magnitudes).
\end{itemize}

This work is an attempt to go beyond mere predictions and gain interpretable insights into the contributory information involved in the estimation of galaxy stellar mass. More broadly, the interpretability of data analysis tools would be increasingly crucial for data-driven astrophysical applications not limited to galaxy physics. For example, by resorting to the causal and information-theoretical analysis, more advanced data-driven models for astrophysical parameter estimations could be developed in the future to utilize structural causal connections, lower redundant information, and enhance synergy between variables or datasets, thereby optimizing data exploitation and improving model robustness. Furthermore, physical interpretations and insights can be obtained by uncovering meaningful data structures and distributions, facilitating the discovery and understanding of complex multivariate physical processes. With the large-scale astronomical datasets envisioned by future surveys, we anticipate to see broad impacts of integrating the predictive power of deep learning with the interpretability of various techniques, such as causal learning and information theory. This integration would play significant roles in the mining of big data in astronomy and promote more data-driven astrophysical research.

\section*{Code availability} \label{sec:code_availability}

The code used in this work is available at \url{https://github.com/QiufanLin/InterpretSM}.

\begin{acknowledgements}

This work is supported by ``The Major Key Project of PCL''.

H.X.R. acknowledges financial support from the National Natural Science Foundation of China (62201306).

This work makes use of the Sloan Digital Sky Survey (SDSS) data. Funding for SDSS-III has been provided by the Alfred P. Sloan Foundation, the Participating Institutions, the National Science Foundation, and the U.S. Department of Energy Office of Science. The SDSS-III web site is http://www.sdss3.org/. SDSS-III is managed by the Astrophysical Research Consortium for the Participating Institutions of the SDSS-III Collaboration including the University of Arizona, the Brazilian Participation Group, Brookhaven National Laboratory, Carnegie Mellon University, University of Florida, the French Participation Group, the German Participation Group, Harvard University, the Instituto de Astrofisica de Canarias, the Michigan State/Notre Dame/JINA Participation Group, Johns Hopkins University, Lawrence Berkeley National Laboratory, Max Planck Institute for Astrophysics, Max Planck Institute for Extraterrestrial Physics, New Mexico State University, New York University, Ohio State University, Pennsylvania State University, University of Portsmouth, Princeton University, the Spanish Participation Group, University of Tokyo, University of Utah, Vanderbilt University, University of Virginia, University of Washington, and Yale University.

This publication makes use of data products from the Wide-field Infrared Survey Explorer, which is a joint project of the University of California, Los Angeles, and the Jet Propulsion Laboratory/California Institute of Technology, funded by the National Aeronautics and Space Administration.

\end{acknowledgements}

\bibliographystyle{aa}
\bibliography{aa54065-25}

\begin{appendix}

\section{Details of model training and validation} \label{sec:training_dev}

The deep learning models used for the supervised contrastive learning procedure (Sect.~\ref{sec:causal_path_XY}) and the mutual information estimation (Sect.~\ref{sec:implement_info}) were both trained from scratch using the mini-batch gradient descent with the default Adam optimizer \citep{Adam}. We conducted 180\,000 iterations in total to train each model. The learning rate was initially set to $10^{-4}$, and reduced by a factor of 5 after 60\,000 training iterations. For the supervised contrastive learning procedure, a mini-batch of 64 input-stellar mass pairs was randomly selected from the training sample in each training iteration; while for the mutual information estimation, the mini-batch size was set to 128. The galaxy images, if involved in the input data, were randomly flipped and rotated with 90 deg steps before being input into the estimator, while the dimensions were fixed to $64\times64\times5$ and the galaxies were always located at the image center. We used 520 bins spanning between 6.0 and 12.5 dex to express the stellar mass probability density estimates in both the supervised contrastive learning procedure and the mutual information estimation, corresponding to a bin width of 0.0125 dex.

To assess the robustness of our default model implementations, we tested different hyperparameters, stellar mass bin widths, and network architectures. Firstly, with the default bin width and network architectures, we observed that the total loss converged with the number of training iterations above 120\,000, and the comparison between the validation sample and the training sample indicated no trend of overfitting as the training continued. We thus took 180\,000 iterations as a default choice to ensure sufficient training for all the models. Secondly, the bin width of 0.0125 dex is well within a range of binning choices in which the estimation of mutual information remains stable. We found that a bin width as large as 0.25 dex (roughly corresponding to the average width between the 16th and 84th percentiles of the PDF of log stellar mass) was already adequate to characterize the input-stellar mass dependences and produce similar results as our default bin width, despite there being a significant discretization effect in terms of expressing mutual information. On the other hand, a trend of overfitting would appear with much smaller bin widths (< 0.003 dex) or deeper network architectures compared to our default choices, degrading the results for the validation sample and the test sample. Therefore, the bin width of 0.0125 dex is a robust choice that can not only avoid overfitting but also greatly lower the discretization effect.

In summary, none of the alternative implementations could qualitatively affect the results produced by the subsequent steps in the causal analysis (Sects.~\ref{sec:detect_X} and \ref{sec:str_causal_effects_XY}), as long as the networks had sufficient expressivity and were properly trained. They also could not improve the average of mutual information estimates by a margin over statistical fluctuations ($\sim$0.02 nats), which is negligible for our results on the mutual information decomposition. These tests provide evidence that our default implementations are not limited by the binning effect, the model expressivity, or overfitting.

In addition, we note that the probability densities produced by end-to-end models may be miscalibrated, namely, not representing the actual target distributions given the input data \citep{Lin2024}, and thus may bias the mutual information estimates. However, the effect of miscalibration can be significantly alleviated as we only took into account the minus log-probability value at the given stellar mass bin (i.e., the cross-entropy loss) rather than the shape of each probability density. Furthermore, as we only considered the average of individual instance-wise mutual information estimates over any parameter space, this bias can be further suppressed.

\section{Determination of the number $k$ for selecting nearest neighbors} \label{sec:assess_knn}

\begin{figure}[H]
\centering
\includegraphics[width=1.0\linewidth]{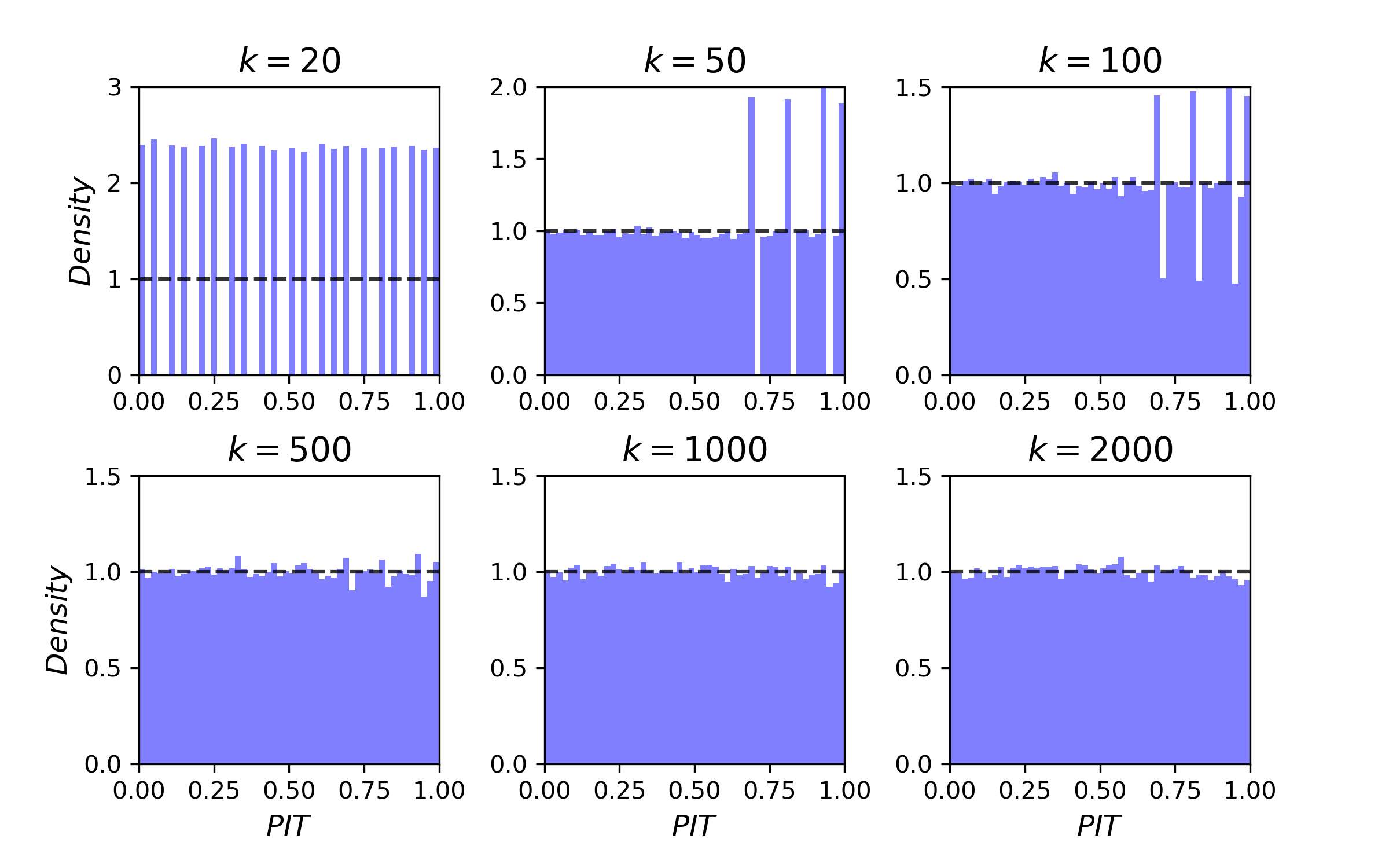}
\caption{Probability integral transform (PIT) distributions over the test sample for the photometry-only model $\mathbf{M}_{ugriz}$ defined in Table~\ref{tab:models_causal}, estimated using the nearest neighbors with different $k$ values. The black dashed lines indicate normalized uniform distributions.}
\label{fig:pit1}
\end{figure}

\begin{figure}
\centering
\includegraphics[width=1.0\linewidth]{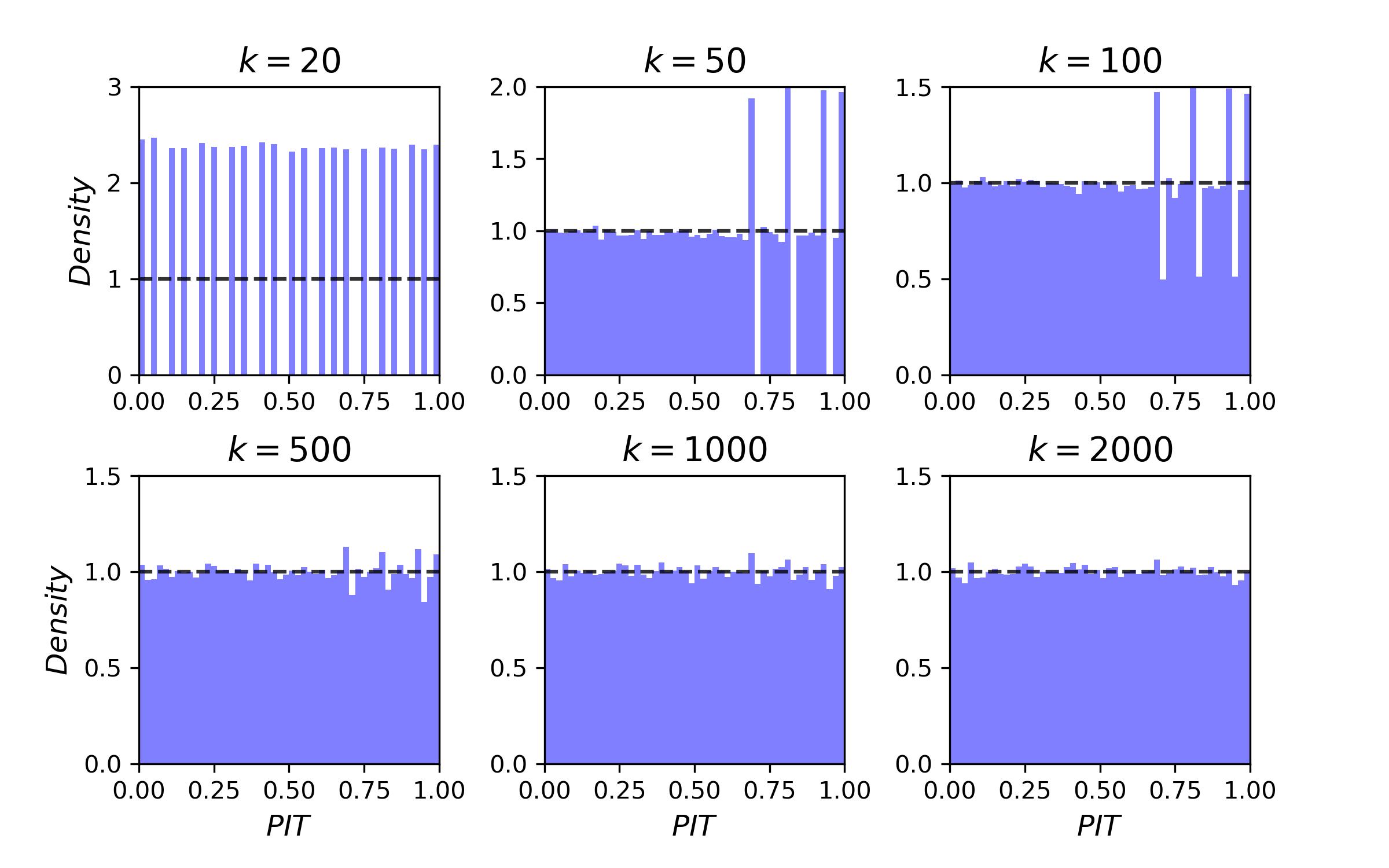}
\caption{Same as Fig.~\ref{fig:pit1}, but for the photometry-only model $\mathbf{M}_{ugrizW123}$ defined in Table~\ref{tab:models_causal}.}
\label{fig:pit2}
\end{figure}

\begin{figure}
\centering
\includegraphics[width=1.0\linewidth]{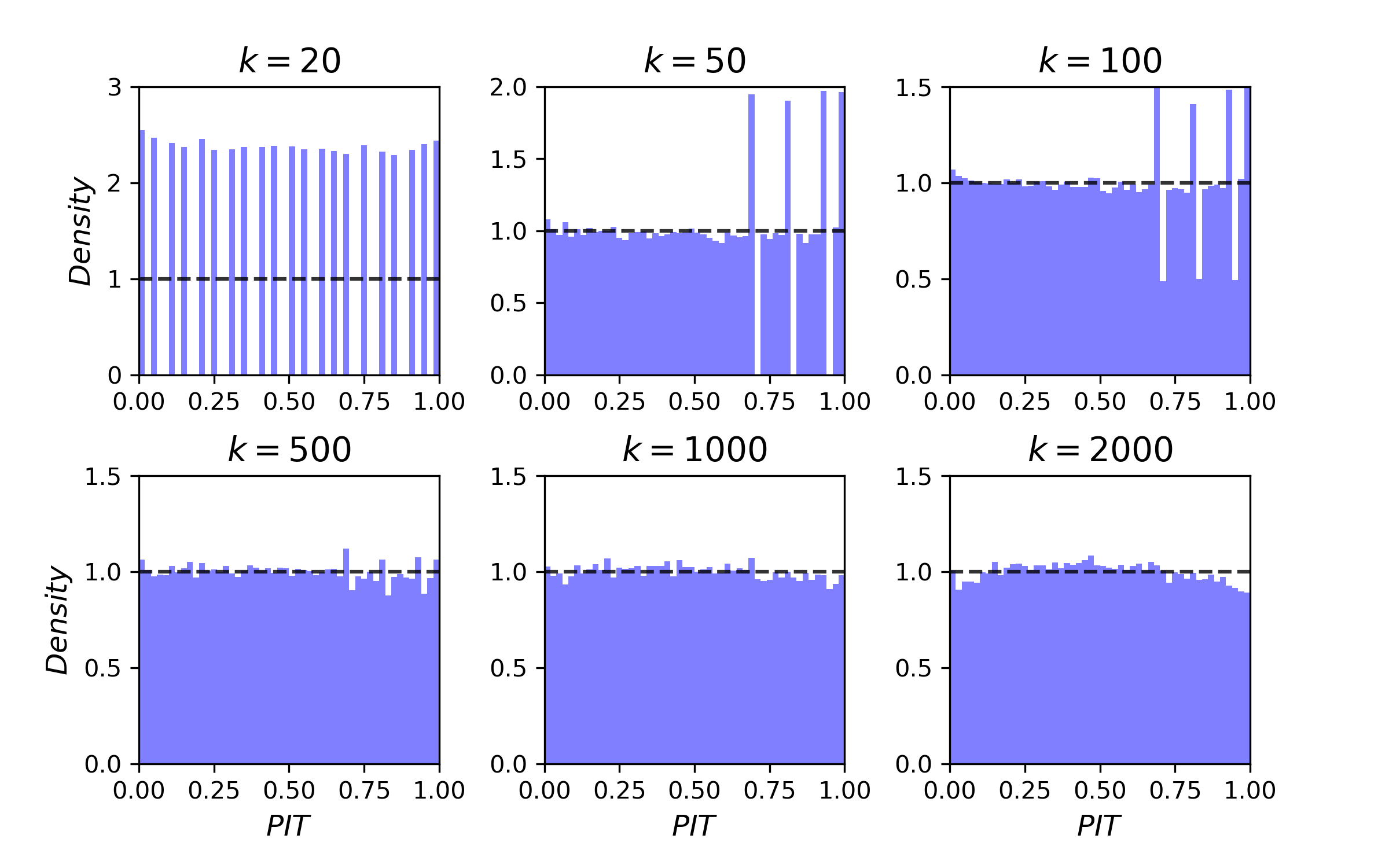}
\caption{Same as Fig.~\ref{fig:pit1}, but for the image-based model $\mathbf{I}_{ugriz}$ defined in Table~\ref{tab:models_causal}.}
\label{fig:pit3}
\end{figure}

\begin{figure}
\centering
\includegraphics[width=1.0\linewidth]{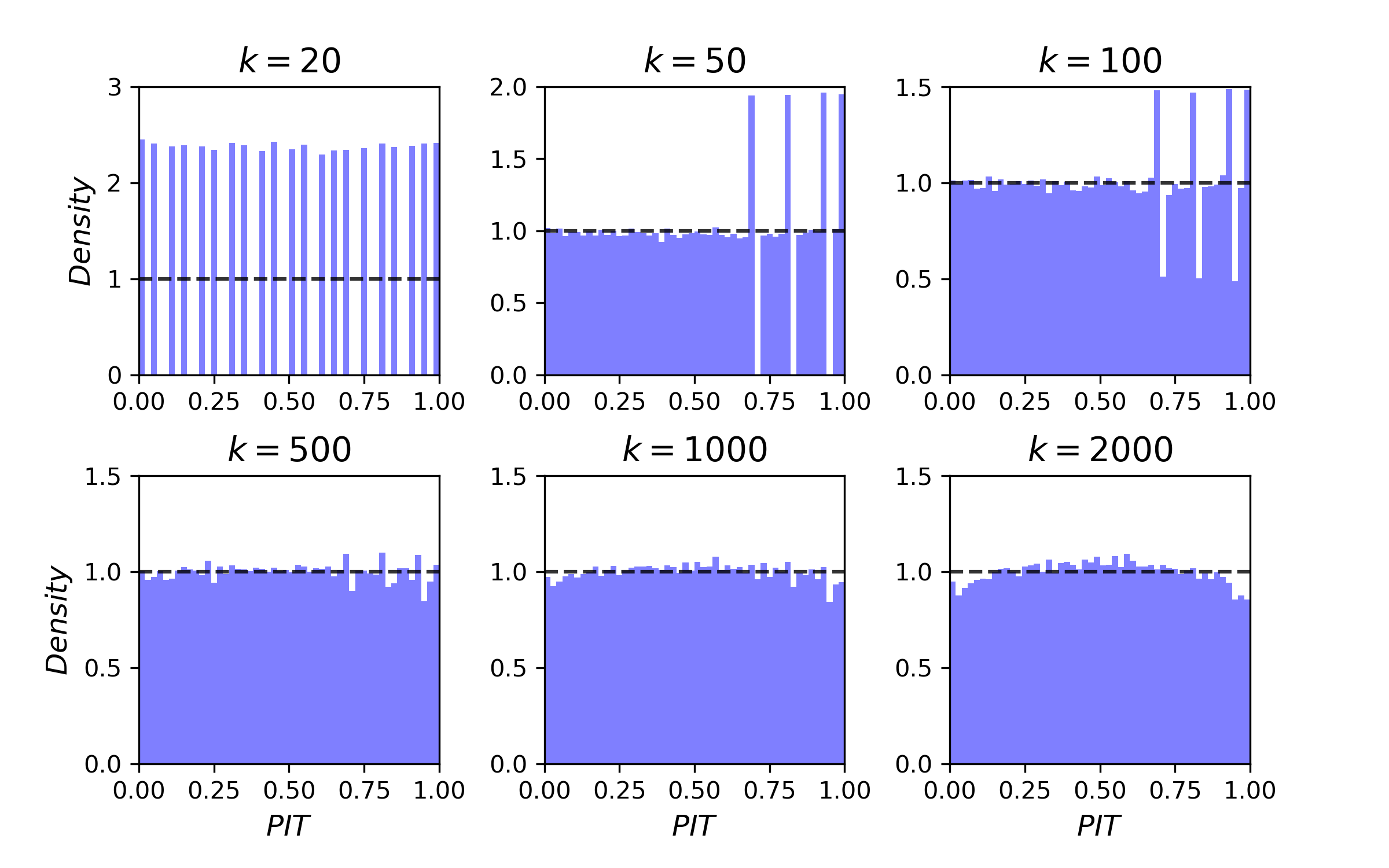}
\caption{Same as Fig.~\ref{fig:pit1}, but for the image-based model $\mathbf{I}_{ugriz} \cup \mathbf{M}_{W123}$ defined in Table~\ref{tab:models_causal}.}
\label{fig:pit4}
\end{figure}

\begin{figure}
\centering
\includegraphics[width=1.0\linewidth]{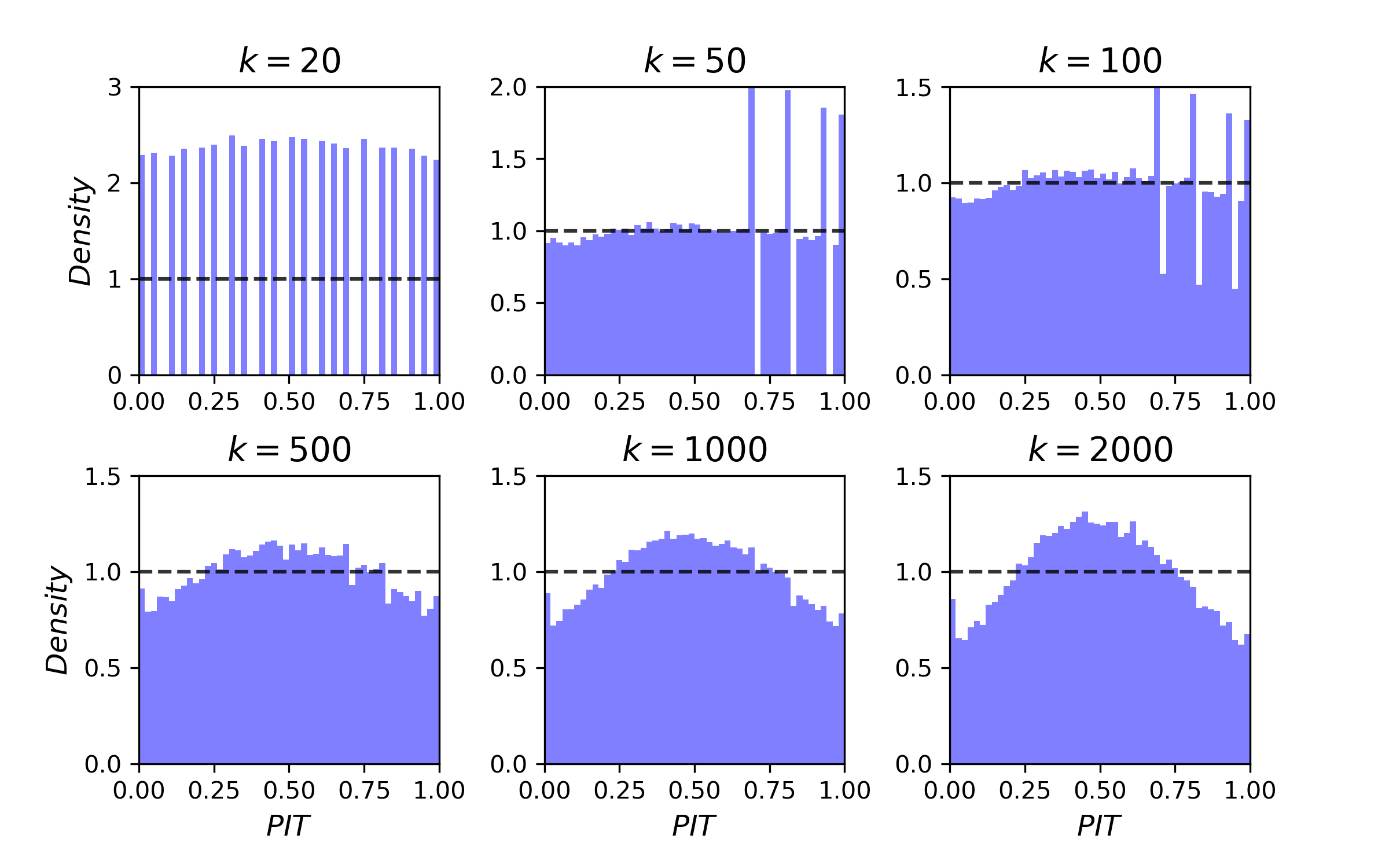}
\caption{Same as Fig.~\ref{fig:pit1}, but for the image-based model $\mathbf{I}_{ugriz} \cup \mathbf{M}_{W123} \cup z_{spec}$ defined in Table~\ref{tab:models_causal}.}
\label{fig:pit5}
\end{figure}

The conditional independence tests in Sects.~\ref{sec:detect_X} and \ref{sec:str_causal_effects_XY} both rely on KNN applied on the $S$ space. The selected nearest neighbors of each test galaxy should be neither too many nor too few in order to properly characterize the local multivariate distribution conditioned on $S$, otherwise the validity of these tests would be questioned. We adopted the probability integral transform (PIT) as a metric to determine the number $k$ for selecting nearest neighbors. The PIT for each test galaxy given a number of nearest neighbors is numerically defined as the fraction of the nearest neighbors whose stellar mass values are smaller than the stellar mass of the test galaxy. If the nearest neighbors are properly selected, the PIT distribution over the test sample should be close to uniformity between 0 and 1. For the five photometry-only or image-based models defined in Table~\ref{tab:models_causal}, we checked the PIT distributions over the test sample for several $k$ values (i.e., $k=20, 50, 100, 500, 1000, 2000$), as illustrated in Figs.~\ref{fig:pit1}, \ref{fig:pit2}, \ref{fig:pit3}, \ref{fig:pit4}, and \ref{fig:pit5}. We found that a small $k$ would result in strong shot noise; while a large $k$ would lead to concave PIT distributions, meaning that the distribution constituted by the nearest neighbors is already nonlocal. On the contrary, for all the cases, a medium value $k=100$ reaches a balance between avoiding strong shot noise and fulfilling the requirement for localization. 

\begin{figure}
\centering
\includegraphics[width=0.9\linewidth]{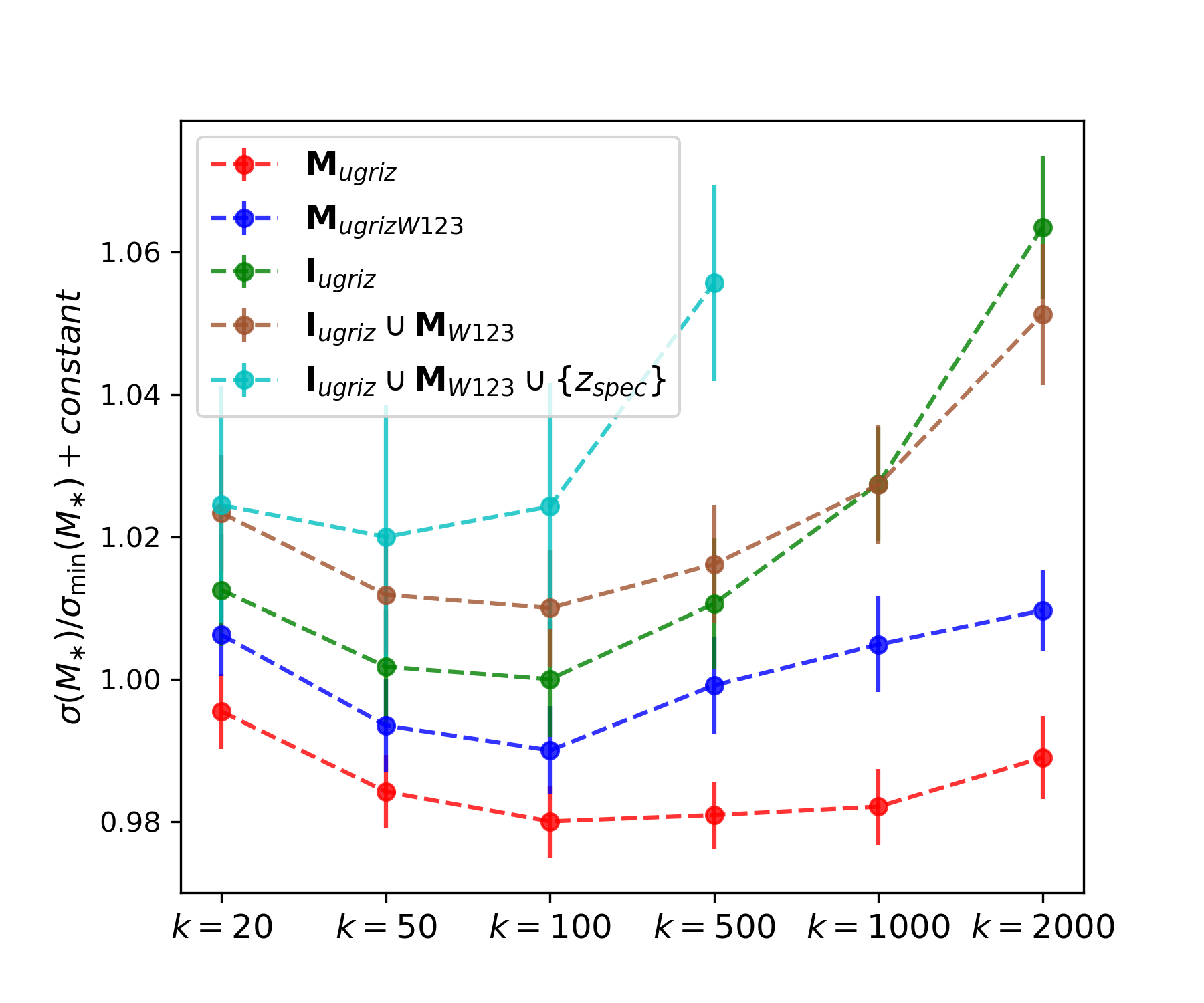}
\caption{Estimation accuracy quantified by $\sigma (M_{\ast}) / \sigma_{\min} (M_{\ast})$ as a function of $k$ for the five models defined in Table~\ref{tab:models_causal}. $\sigma (M_{\ast})$ refers to the standard deviation of log stellar mass residuals estimated using a given $k$ of nearest neighbors, and $\sigma_{\min} (M_{\ast})$ refers to the minimum value over the $k$ grid for each model. The error bars are obtained using bootstrap. The points at $k=1000$ and $k=2000$ for the model $\mathbf{I}_{ugriz} \cup \mathbf{M}_{W123} \cup z_{spec}$ are not shown due to significantly high values.}
\label{fig:sigma_k}
\end{figure}

As a cross-check, we also analyzed the evolution of stellar mass estimation accuracy with different $k$ values for the five models using the test sample. The estimation accuracy is quantified by the standard deviation of $\log(M_{\ast}^{est} / M_{\odot}) - \log(M_{\ast} / M_{\odot})$, denoted as $\sigma (M_{\ast})$, where $\log(M_{\ast} / M_{\odot})$ refers to the log stellar mass value of each test galaxy, and $\log(M_{\ast}^{est} / M_{\odot})$ refers to the mean log stellar mass estimate using a given number $k$ of nearest neighbors. We normalize $\sigma (M_{\ast})$ by the minimum value $\sigma_{\min} (M_{\ast})$ estimated over the grid of selected $k$ values for each model. As shown in Fig.~\ref{fig:sigma_k}, $\sigma (M_{\ast}) / \sigma_{\min} (M_{\ast})$ reaches the lowest point around $k=100$ for all the models. Therefore, we took $k=100$ throughout our work.

\FloatBarrier

\section{More results} \label{sec:more_res}

\subsection{Detection of external variables} \label{sec:more_corr}

\begin{figure*}
\begin{center}
\centerline{\includegraphics[width=1.1\linewidth]{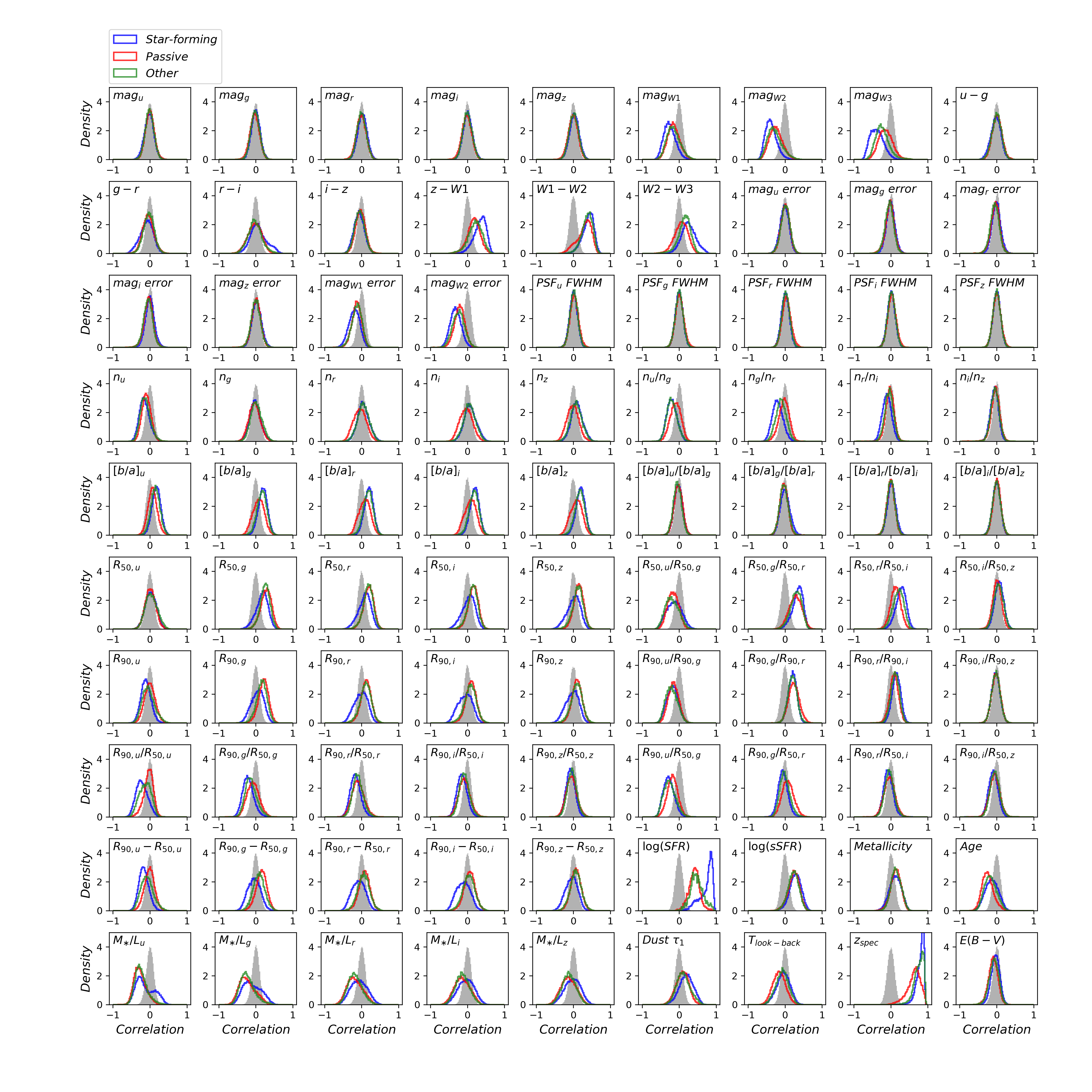}}
\caption{Distributions of local correlations between stellar mass and the parameters listed in Table~\ref{tab:data} for the photometry-only model $\mathbf{M}_{ugriz}$ defined in Table~\ref{tab:models_causal}. The original distributions are separately shown for star-forming, passive, and other galaxies from the test sample, illustrated as the colored curves. The distributions shown in grey are used as a contrast, produced by randomly permuting the stellar mass values within the nearest neighbors of each test galaxy.}
\label{fig:correlation_mag11}
\end{center}
\end{figure*}

\begin{figure*}
\begin{center}
\centerline{\includegraphics[width=1.1\linewidth]{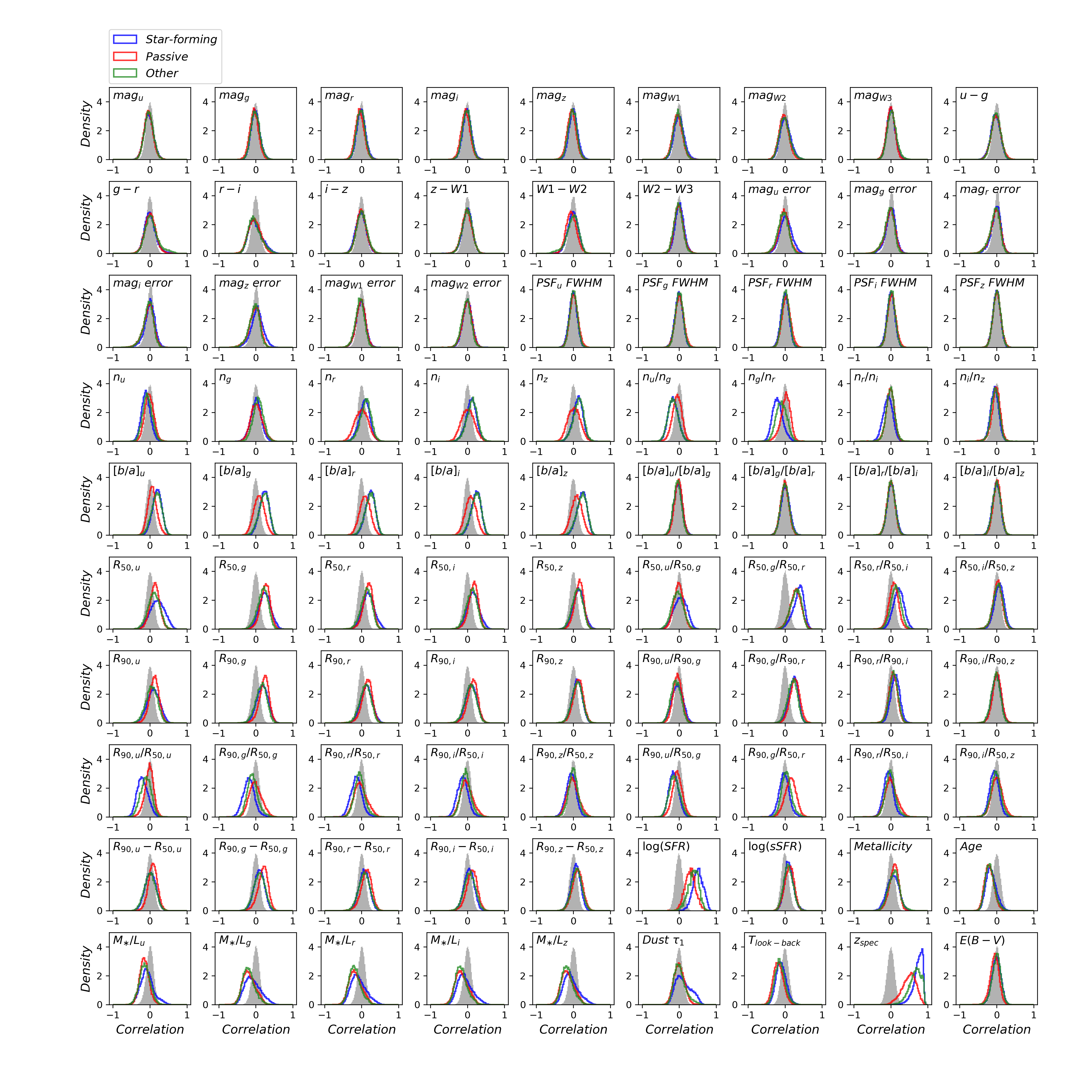}}
\caption{Same as Fig.~\ref{fig:correlation_mag11}, but for the photometry-only model $\mathbf{M}_{ugrizW123}$ defined in Table~\ref{tab:models_causal}.}
\label{fig:correlation_mag12}
\end{center}
\end{figure*}

\begin{figure*}
\begin{center}
\centerline{\includegraphics[width=1.1\linewidth]{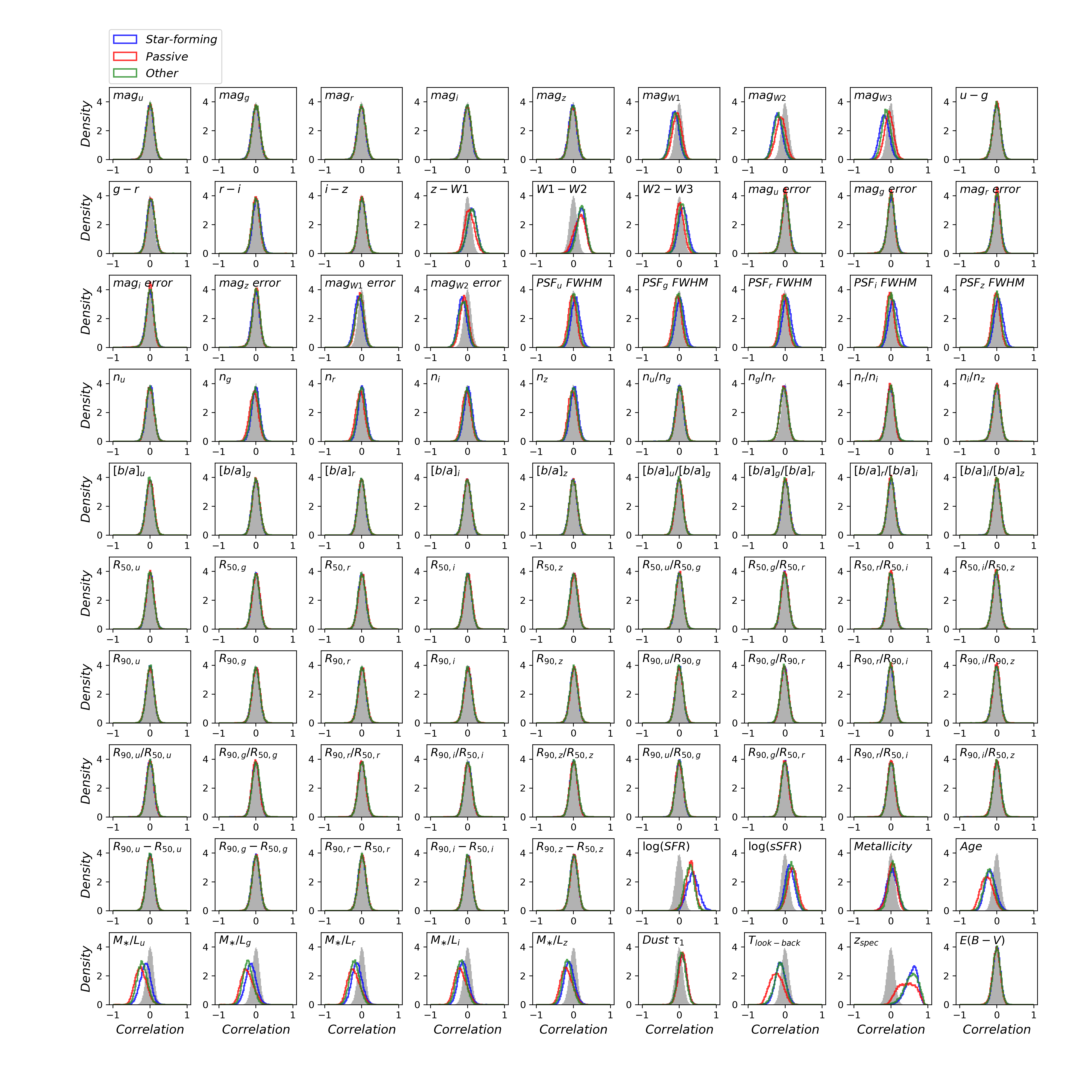}}
\caption{Same as Fig.~\ref{fig:correlation_mag11}, but for the image-based model $\mathbf{I}_{ugriz}$ defined in Table~\ref{tab:models_causal}.}
\label{fig:correlation_img21}
\end{center}
\end{figure*}

\begin{figure*}
\begin{center}
\centerline{\includegraphics[width=1.1\linewidth]{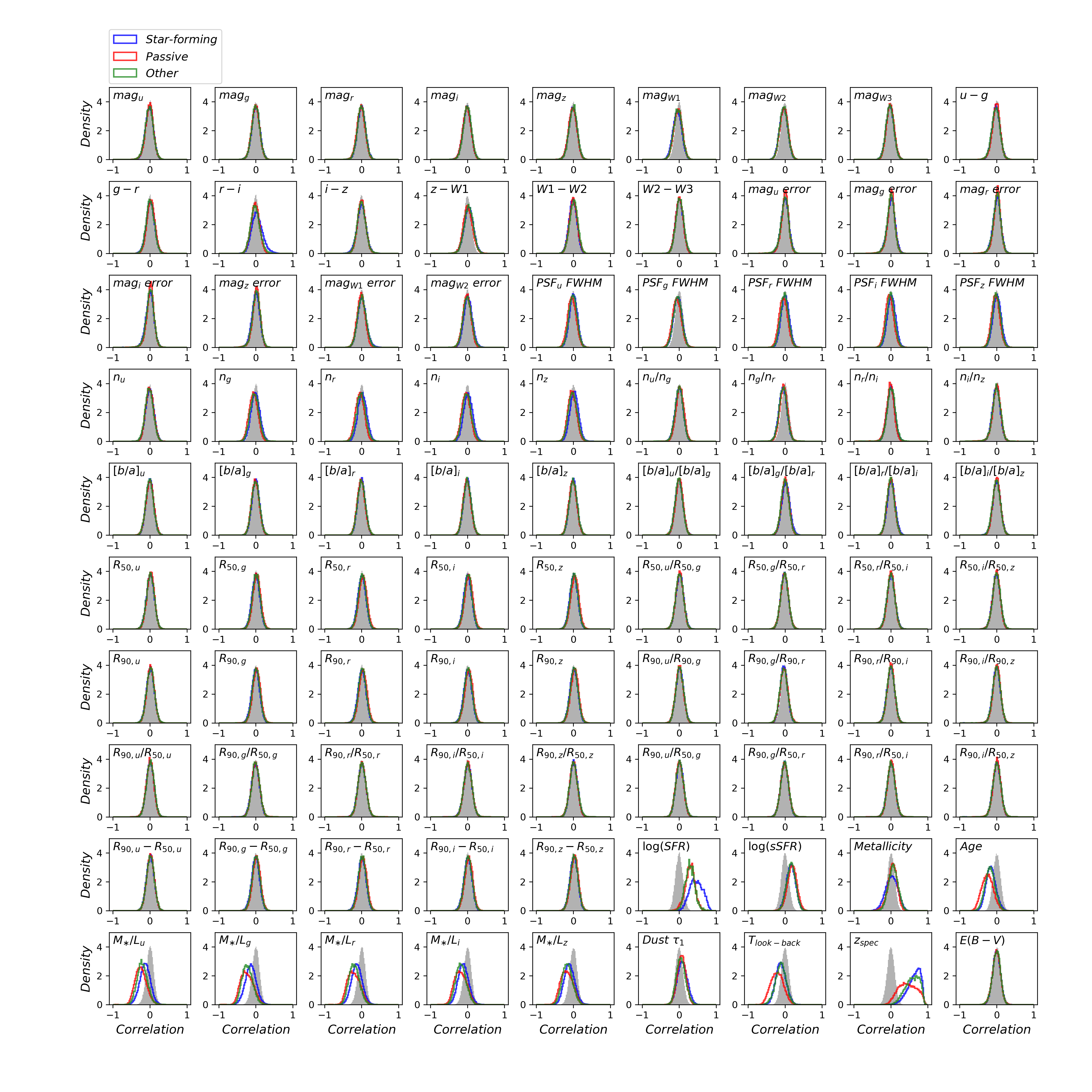}}
\caption{Same as Fig.~\ref{fig:correlation_mag11}, but for the image-based model $\mathbf{I}_{ugriz} \cup \mathbf{M}_{W123}$ defined in Table~\ref{tab:models_causal}.}
\label{fig:correlation_img22}
\end{center}
\end{figure*}

\begin{figure*}
\begin{center}
\centerline{\includegraphics[width=1.1\linewidth]{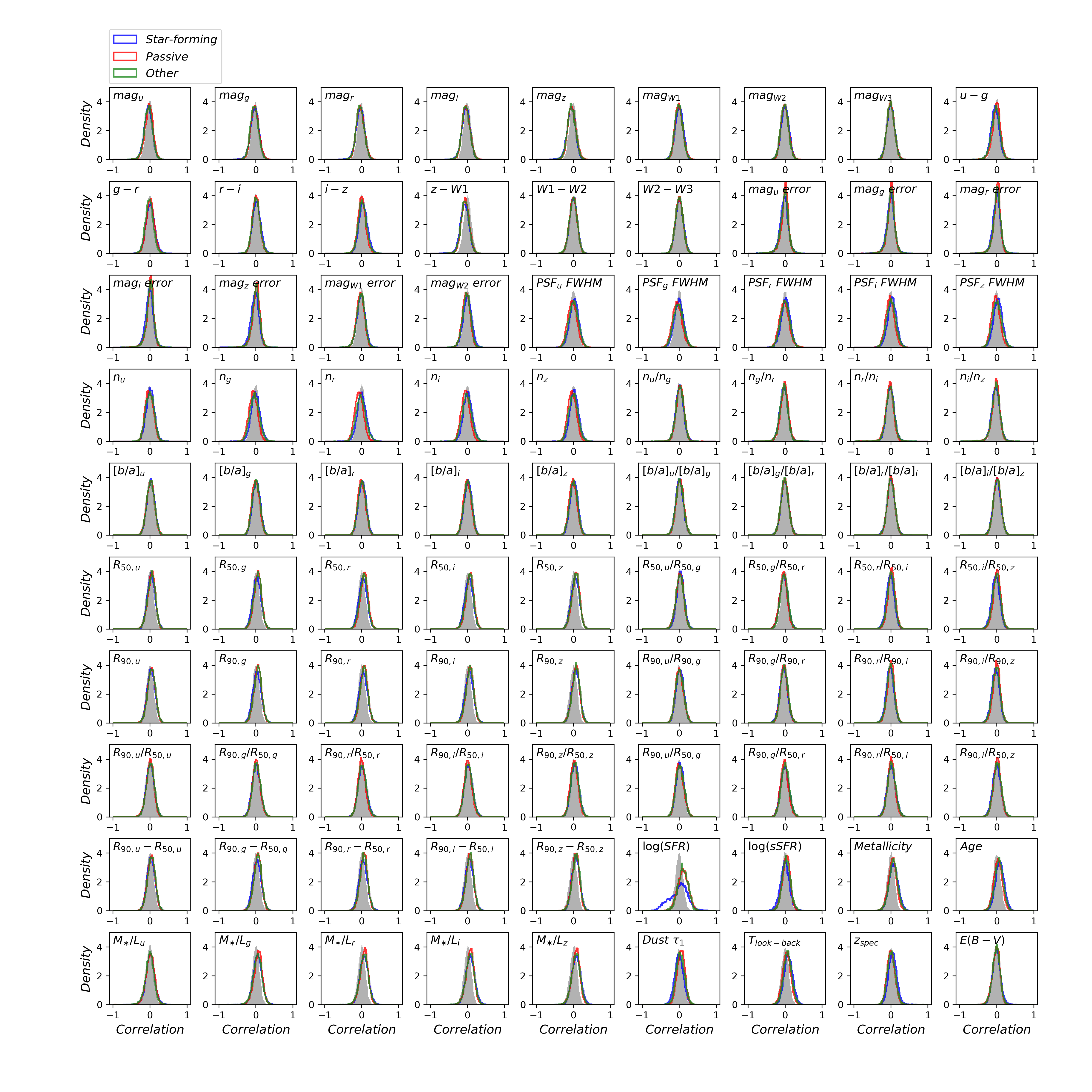}}
\caption{Same as Fig.~\ref{fig:correlation_mag11}, but for the image-based model $\mathbf{I}_{ugriz} \cup \mathbf{M}_{W123} \cup z_{spec}$ defined in Table~\ref{tab:models_causal}.}
\label{fig:correlation_img23}
\end{center}
\end{figure*}

In Figs.~\ref{fig:correlation_mag11}, \ref{fig:correlation_mag12}, \ref{fig:correlation_img21}, \ref{fig:correlation_img22}, and \ref{fig:correlation_img23}, we show more results on the local correlation distributions supplementary to Fig.~\ref{fig:correlation_selected} for the five photometry-only or image-based models defined in Table~\ref{tab:models_causal}, respectively. The list of the shown parameters, summarized in Table~\ref{tab:data}, includes various photometric properties, morphological features, physical properties, and other catalog data.

Other than the results presented in Fig.~\ref{fig:correlation_selected}, we found that there exist clear residual correlations between stellar mass and optical colors for the photometry-only models, illustrated as the inconsistency between the original and reference correlation distributions. This is because optical magnitudes rather than colors are the direct model inputs. The possible misinformation due to noise or data imbalance makes magnitudes and colors not exploited simultaneously. Although not shown, the impact of such residual correlations on the stellar mass estimation indicated by the predictive efficiency is insignificant. For image-based models, there are generally no strong residual correlations for either optical magnitudes or colors, probably because the inclusion of morphology covers or synergizes with the color information not fully exploited by the photometry-only models. In addition, for the models not involving infrared photometry, there are correlations between stellar mass and infrared magnitude errors, which is due to the fact that the WISE infrared magnitude errors are a strong function of infrared magnitudes.

While PSFs may contribute in shaping the dependence relations between measured morphological features and stellar mass, they do not have apparently direct contributions to the stellar mass estimation for the SDSS data. Therefore, we did not include PSF FWHMs in the input data for any model. On the contrary, we always used galactic reddening $E(B-V)$ as an additional input. Although not shown, the impact of galactic reddening $E(B-V)$ on the stellar mass estimation cannot be fully covered by photometry or images. The results illustrated in the figures confirm that our models fed with galactic reddening $E(B-V)$ have learned to leverage this information for the stellar mass estimation.

\subsection{Analysis of causal structures between external variables and stellar mass} \label{sec:more_cond_corr}

\begin{figure*}
\begin{center}
\centerline{\includegraphics[width=1.1\linewidth]{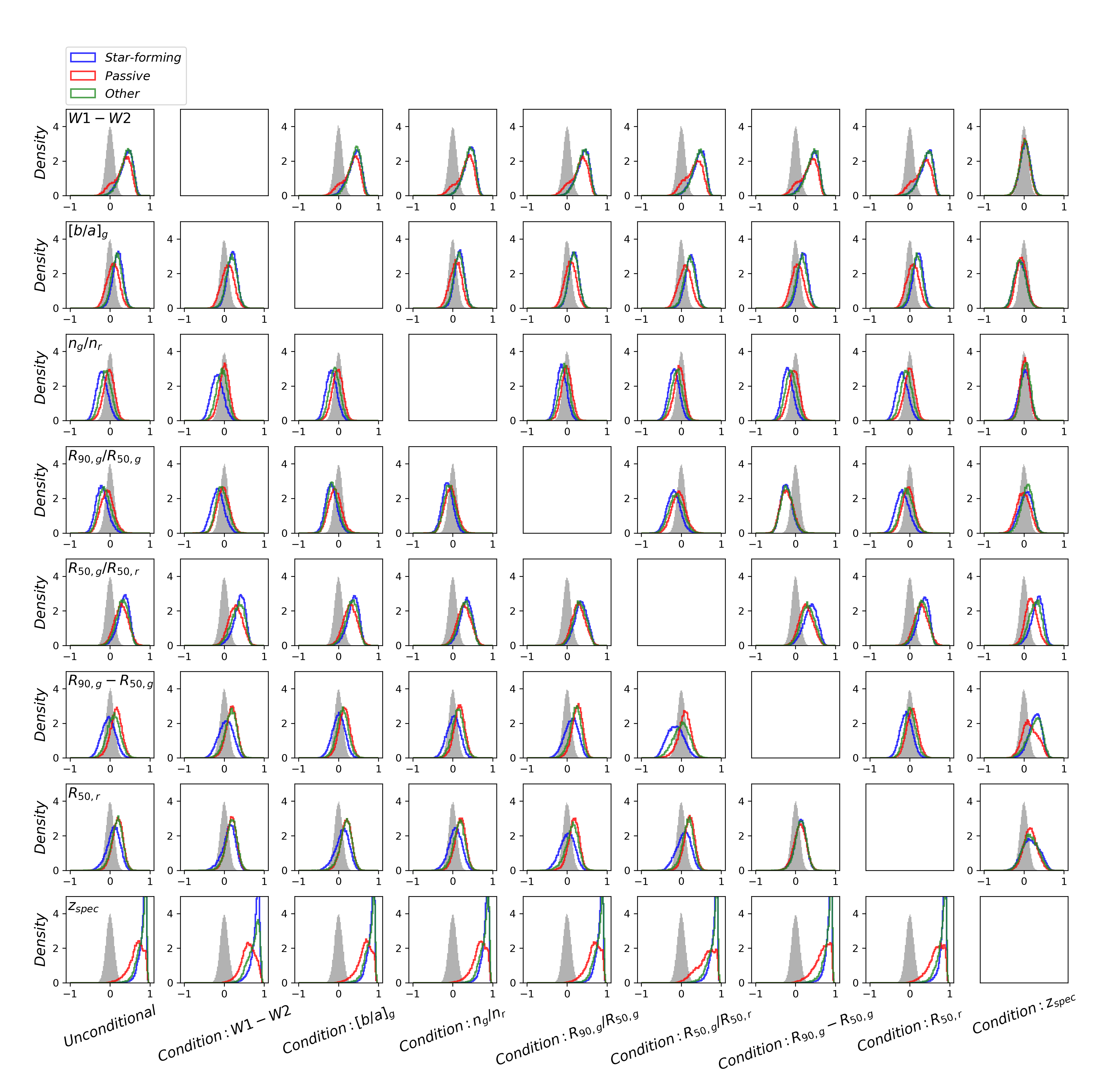}}
\caption{Distributions of conditional local correlations between stellar mass and representative parameters for the photometry-only model $\mathbf{M}_{ugriz}$ defined in Table~\ref{tab:models_causal}. Each row corresponds to a parameter labeled on the first column. Similar to Fig.~\ref{fig:pred_efficiency_cond}, for each parameter, the first column shows the unconditional correlation distributions to be compared with, and each of the remaining columns shows the conditional correlation distributions with the conditional variable labeled on the bottom. All the parameters including stellar mass are first conditioned on the $g-r$ color of the nearest neighbors of each test galaxy before computing the (conditional) correlations. The original distributions are separately shown for star-forming, passive, and other galaxies from the test sample, illustrated as the colored curves. The distributions shown in grey are used as a contrast, produced by randomly permuting the stellar mass values within the nearest neighbors of each test galaxy.}
\label{fig:correlation_cond}
\end{center}
\end{figure*}

To complement the discussions in Sect.~\ref{sec:res_pred_efficiency_cond} in which the conditional predictive efficiency was adopted, we applied the correlation metric for the model $\mathbf{M}_{ugriz}$, with both stellar mass and query variables quadratically regressed on conditional variables before estimating correlations. In Fig.~\ref{fig:correlation_cond}, we present the distributions of such conditional correlations between stellar mass and the representative parameters shown in Fig.~\ref{fig:pred_efficiency_cond}. All the parameters including stellar mass are first quadratically regressed on the $g-r$ color of the nearest neighbors of each test galaxy, then the conditional correlations for every shown parameter are estimated by conditioning on every other parameter, compared to the unconditional correlations in the first column. As shown, there are clear changes in the conditional correlations in contrast to the unconditional ones for some parameters, such as the correlations for $W1 - W2$ conditioned on spec-$z$. These trends are consistent with those shown in Fig.~\ref{fig:pred_efficiency_cond}. Nonetheless, compared to the conditional predictive efficiency, the correlation is less indicative of the impacts of certain variables on the stellar mass estimation.

\subsection{Decomposition of contributions of different photometric bands} \label{sec:more_decompose2}

\begin{figure*}
\begin{center}
\centerline{\includegraphics[width=1.0\linewidth]{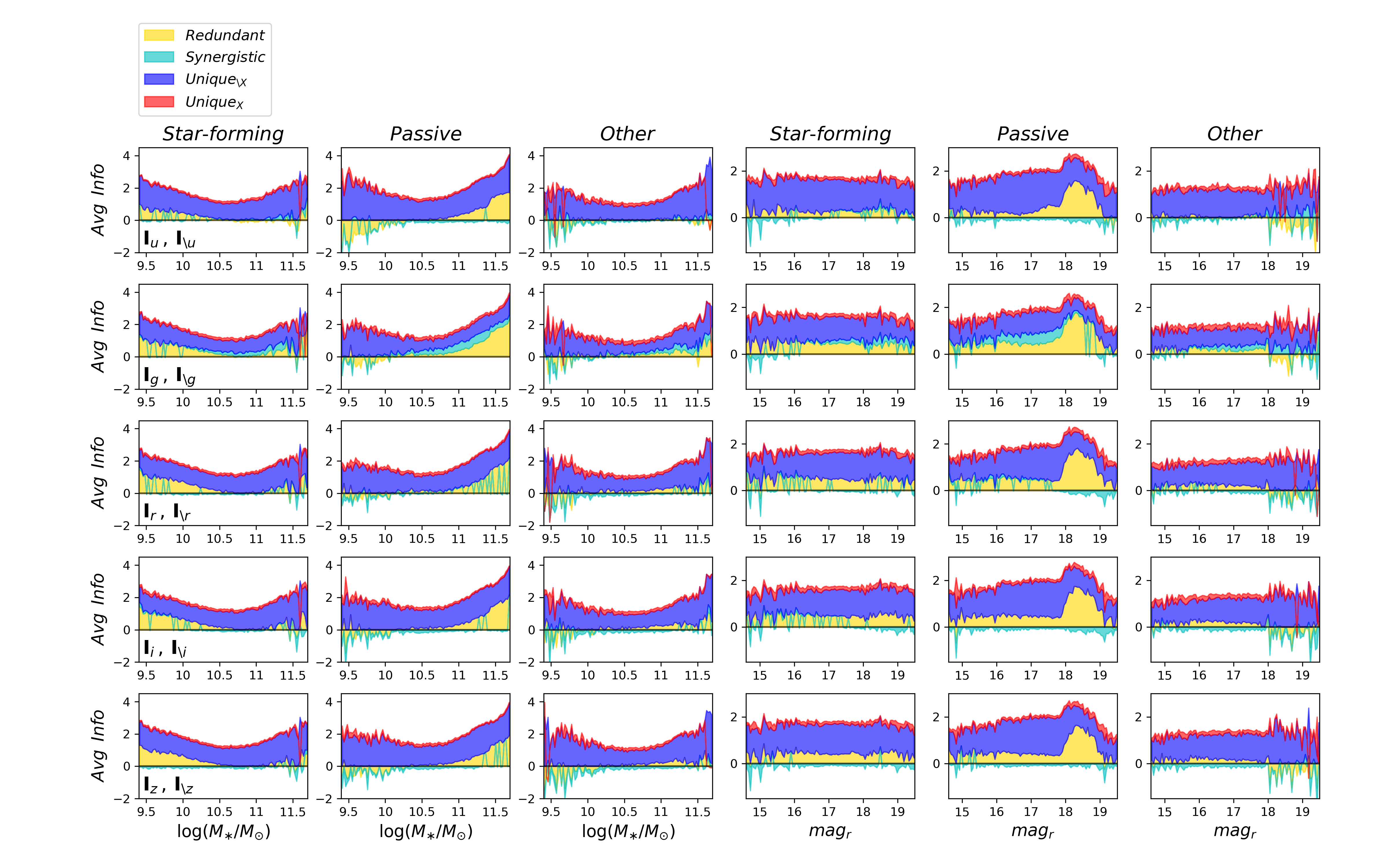}}
\caption{Stack plots of the redundant, unique, and synergistic information components (in units of nats) as a function of stellar mass or $r$-band magnitude for <$\mathbf{I}_{X} \,,\, \mathbf{I}_{\backslash X}$> defined in Table~\ref{tab:info_bands}. Each row corresponds to a case in which the images in one band are separated out, shown for star-forming, passive, and other galaxies from the test sample. The label $X$ refers to a single band that is separated out, running over all the optical bands, and $\backslash X$ refers to the remaining bands, distinguished for the unique information (shown in red and blue, respectively).}
\label{fig:info_imgdrop3}
\end{center}
\end{figure*}

\begin{figure*}
\begin{center}
\centerline{\includegraphics[width=1.0\linewidth]{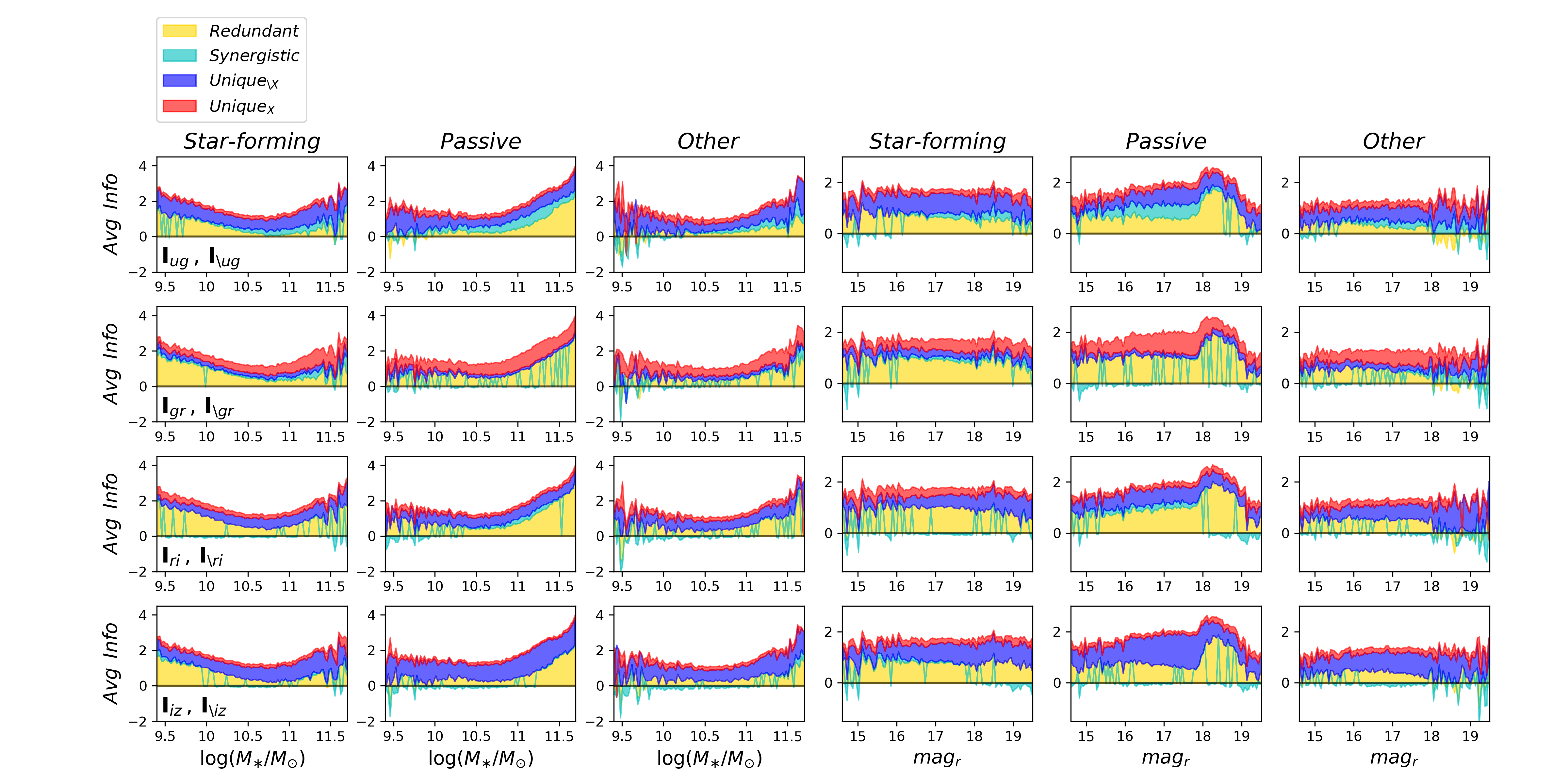}}
\caption{Same as Fig.~\ref{fig:info_imgdrop3}, but with two adjacent bands separated out in each case.}
\label{fig:info_imgdrop4}
\end{center}
\end{figure*}

\begin{figure*}
\begin{center}
\centerline{\includegraphics[width=1.0\linewidth]{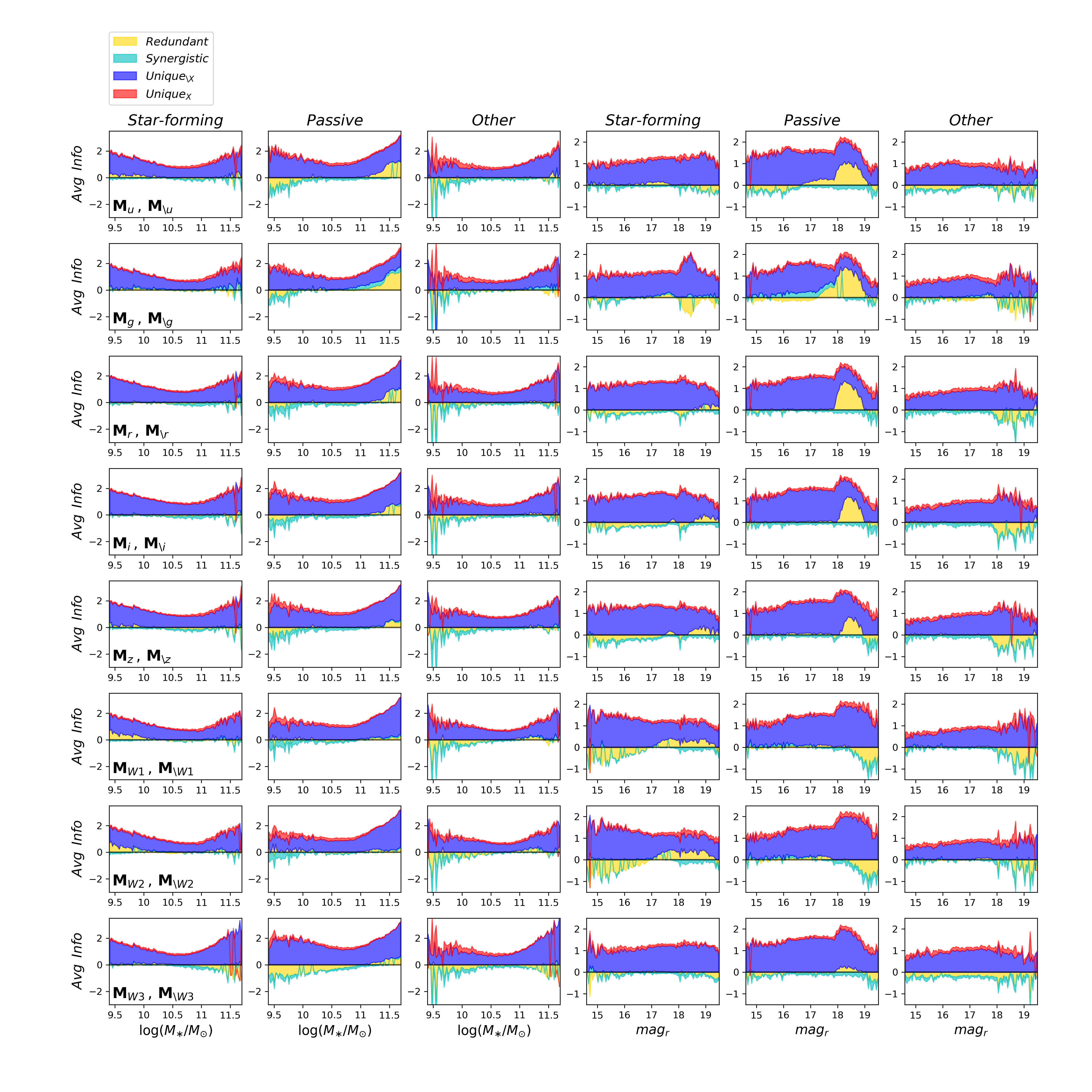}}
\caption{Same as Fig.~\ref{fig:info_imgdrop3}, but for <$\mathbf{M}_{X} \,,\, \mathbf{M}_{\backslash X}$> defined in Table~\ref{tab:info_bands}. Each row corresponds to a case in which the photometry in one band is separated out. The label $X$ runs over all the optical and infrared bands.}
\label{fig:info_magdrop0}
\end{center}
\end{figure*}

\begin{figure*}
\begin{center}
\centerline{\includegraphics[width=1.0\linewidth]{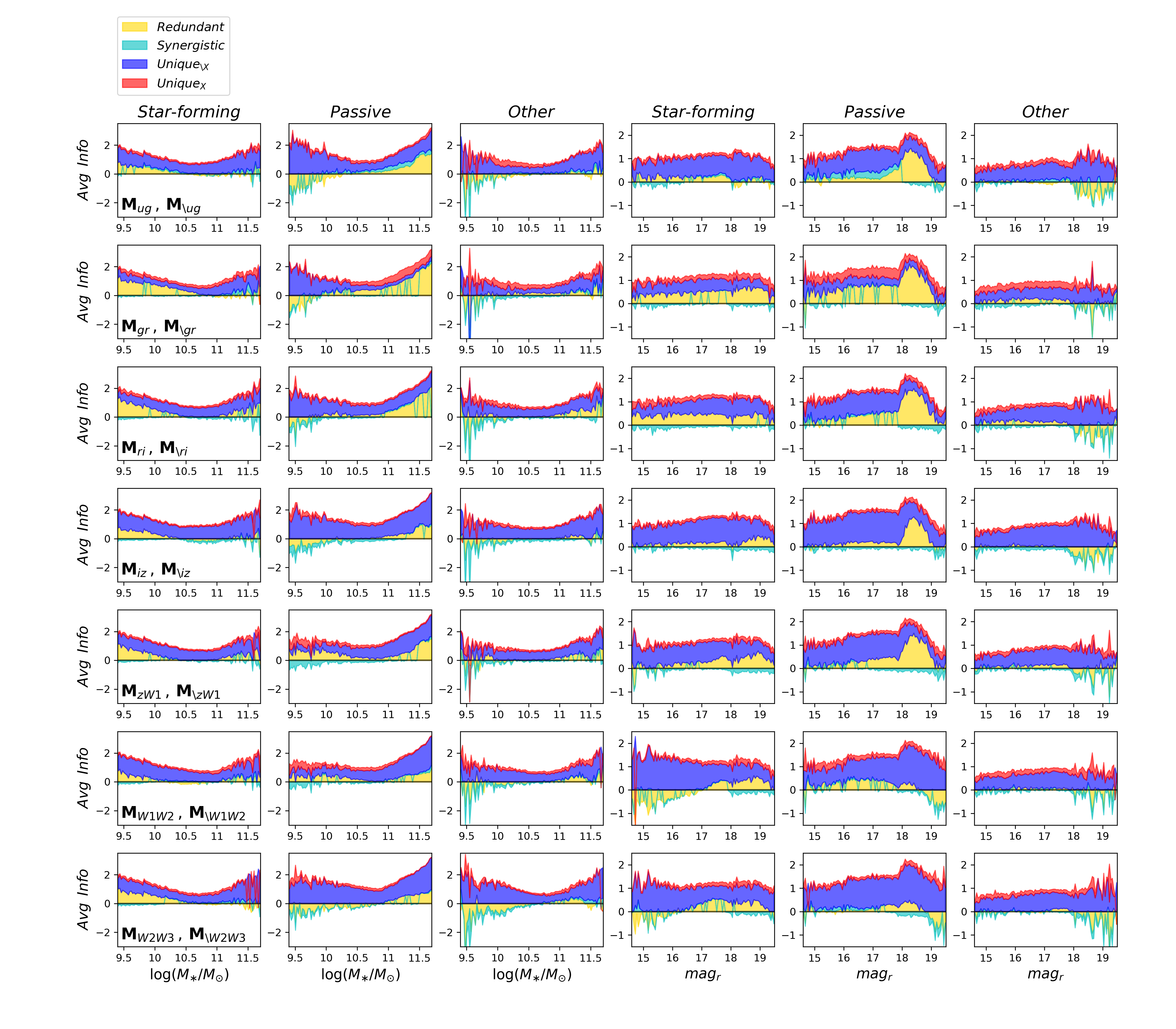}}
\caption{Same as Fig.~\ref{fig:info_magdrop0}, but with two adjacent bands separated out in each case.}
\label{fig:info_magdrop1}
\end{center}
\end{figure*}

In Figs.~\ref{fig:info_imgdrop3}, \ref{fig:info_imgdrop4}, \ref{fig:info_magdrop0}, and \ref{fig:info_magdrop1}, we show the stack plots supplementary to Figs.~\ref{fig:info_globmean_gw1drop} and \ref{fig:info_globmean_grw1w2drop} for all the cases defined in Table~\ref{tab:info_bands}, decomposing the contributions of different photometric bands to the stellar mass estimation. In general, other than the synergistic effects exhibited by the $g$ band and the behaviors of data imbalance discussed in Sect.~\ref{sec:res_decompose2}, we found no sharp discrepancy between different bands.

\end{appendix}

\end{CJK*}
\end{document}